\documentclass[a4paper]{article}

\usepackage[top=2cm,bottom=2cm,left=3cm,right=3cm]{geometry}

\usepackage{amsmath}
\usepackage{amsthm}
\usepackage{amssymb}
\usepackage{mathtools}
\usepackage{enumitem}
\usepackage[colorlinks=true, allcolors=blue]{hyperref}
\usepackage{todonotes}
\usepackage{xspace}
\usepackage[capitalize]{cleveref}

\usepackage{soul} 
\usepackage{tikz}
\usetikzlibrary{positioning}
\usetikzlibrary{calc}
\usetikzlibrary{graphs}
\usetikzlibrary{shapes,arrows,decorations.markings,decorations.text,decorations.pathreplacing}
\usepackage{caption}
\usepackage[subrefformat=parens]{subcaption}
\usepackage{blkarray}
\usepackage{booktabs}

\usepackage{thmtools}
\usepackage{thm-restate}

\usepackage{authblk}

\usepackage{orcidlink}

\usepackage[ruled]{algorithm2e}
\SetArgSty{} 
\SetTitleSty{textsc}{normalsize}
\SetNlSty{normalsize}{}{}
\DontPrintSemicolon 

\newtheorem{definition}{Definition}
\newtheorem{problem}[definition]{Problem}
\newtheorem{theorem}[definition]{Theorem}
\newtheorem{lemma}[definition]{Lemma}
\newtheorem{proposition}[definition]{Proposition}
\newtheorem{corollary}[definition]{Corollary}

\crefname{algocfline}{alg.}{algs.}
\Crefname{algocfline}{Algorithm}{Algorithms}

\DeclareMathOperator\supp{supp}
\DeclareDocumentCommand\transpose{m}{#1^{\intercal}}
\DeclareDocumentCommand\zerovec{o}{\IfNoValueTF{#1}{\mathbb{O}}{\mathbb{O}_{#1}}}
\DeclareDocumentCommand\T{}{\mathcal{T}}
\DeclareDocumentCommand\V{}{\mathcal{V}}\DeclareDocumentCommand\E{}{\mathcal{E}}
\DeclareDocumentCommand\spqrNonvirtual{}{E^{\textup{reg}}}
\DeclareDocumentCommand\spqrVirtual{}{E^{\textup{virt}}}
\DeclareDocumentCommand\orderO{o}{\mathcal{O}\IfValueTF{#1}{\left(#1\right)}{}}

\DeclareDocumentCommand\algoReduceTree{}{\ensuremath{\hyperref[algo_reducetree]{\textsc{ReduceTree}}}\xspace}
\DeclareDocumentCommand\algoProcessTree{}{\ensuremath{\hyperref[algo_processtree]{\textsc{ProcessTree}}}\xspace}
\DeclareDocumentCommand\algoSplitSkeleton{}{\ensuremath{\hyperref[algo_splitskeleton]{\textsc{SplitSkeleton}}}\xspace}
\DeclareDocumentCommand\algoMergeTree{}{\ensuremath{\hyperref[algo_mergetree]{\textsc{MergeTree}}}\xspace}
\DeclareDocumentCommand\algoFindSplittableVertices{}{\ensuremath{\hyperref[algo_findsplittablevertices]{\textsc{FindSplittableVertices}}}\xspace}
\DeclareDocumentCommand\algoFindTreeSplittableVertices{}{\ensuremath{\hyperref[algo_findtreesplittablevertices]{\textsc{FindTreeSplittableVertices}}}\xspace}
\DeclareDocumentCommand\algoBipartiteSplit{}{\ensuremath{\hyperref[algo_bipartitesplit]{\textsc{BipartiteSplit}}}\xspace}
\DeclareDocumentCommand\algoExtendSeries{}{\ensuremath{\hyperref[algo_extendseries]{\textsc{ExtendSeries}}}\xspace}
\DeclareDocumentCommand\algoReduceSeries{}{\ensuremath{\hyperref[algo_reduceseries]{\textsc{ReduceSeries}}}\xspace}
\DeclareDocumentCommand\algoReduceParallel{}{\ensuremath{\hyperref[algo_reduceParallel]{\textsc{ReduceParallel}}}\xspace}
\DeclareDocumentCommand\algoGraphicRowAugmentation{}{\ensuremath{\hyperref[algo_graphicrowaugmentation]{\textsc{GraphicRowAugmentation}}}\xspace}

\title{A Row-wise Algorithm for Graph Realization}

\author{
Rolf van der Hulst \orcidlink{0000-0002-5941-3016},
Matthias Walter \orcidlink{0000-0002-6615-5983}
}
\affil{
University of Twente, Enschede, The Netherlands\\
\sffamily
rolf.vanderhulst@mosek.com\footnote{The first author is now affiliated with Mosek ApS.}, m.walter@utwente.nl
\rmfamily
}

\newcounter{equivalenceproofcounter}
\DeclareDocumentCommand\refEquivalenceProofCounter{m}{\hyperref[eq_equivalenceproof]{(\theequivalenceproofcounter)\ensuremath{_{\text{#1}}}}}

\tikzset{main/.style={circle, draw=black, very thick, inner sep = 0.5mm}}
\tikzset{tree/.style={red, ultra thick}}
\tikzset{cotree/.style={thick, blue}}
\tikzset{virtualtree/.style={red, ultra thick, dashed}}
\tikzset{virtualcotree/.style={blue, thick, dashed}}
\tikzset{marked/.style={blue, thick, postaction={decorate, decoration={markings,
        mark=between positions 0.45 and 0.55 step 0.1 with {\draw [-,blue,ultra thick,solid] (0,-0.7mm) -- (0,0.7mm);}
        }}}}
\tikzset{virtualmarked/.style={blue, dashed,postaction={decorate, decoration={markings,
        mark=between positions 0.45 and 0.55 step 0.1 with {\draw [-,blue,ultra thick,solid] (0,-0.7mm) -- (0,0.7mm);}
        }}}}
\tikzset{behindrect/.style={draw=black!50,fill=black!02}}

\tikzset{dimmedmarked/.style={blue, thick, postaction={decorate, decoration={markings, mark=between positions 0.45 and 0.55 step 0.1 with {\draw [-,blue!20,ultra thick,solid] (0,-0.7mm) -- (0,0.7mm);}
}}}}

\tikzset{colormarked/.style={yellow!80!orange, thick, postaction={decorate, decoration={markings, mark=between positions 0.45 and 0.55 step 0.1 with {\draw [-,yellow!50!orange!100,ultra thick,solid] (0,-0.7mm) -- (0,0.7mm);}
}}}}

\providecommand{\keywords}[1]
{
  \small
  \textbf{{Keywords:}} #1
}

\begin{document}

\maketitle

\begin{abstract}
  Given a $\{0,1\}$-matrix $M$, the graph realization problem for $M$ asks if there exists a spanning forest such that the columns of $M$ are incidence vectors of paths in the forest.
  The problem is closely related to the recognition of network matrices, which are a large subclass of totally unimodular matrices and have many applications in mixed-integer programming.
  Existing efficient algorithms for graph realization grow a submatrix in a column-wise fashion whilst maintaining a graphic realization. In the context of mixed-integer linear programming, this limits the set of  submatrices of the constraint matrix that can efficiently be determined to be network matrices to network submatrices that span all rows and a subset of the columns.
  This paper complements the existing work by providing an algorithm that works in a row-wise fashion and uses similar data structures, and enables the detection of arbitrary graphic submatrices.
  The main challenge in designing efficient algorithms for the graph realization problem is ambiguity as there may exist many graphs realizing $M$.
  The key insight for designing an efficient row-wise algorithm is that a graphic matrix is uniquely represented by an SPQR-tree, a graph decomposition that stores all graphs with the same set of cycles.
  The developed row-wise algorithm uses data structures that are compatible with the column-wise algorithm and can be combined with the latter to detect maximal graphic submatrices.
\end{abstract}


\section{Introduction}

Graphs are important objects in mathematics, and occur in many different fields and practical applications. Representing graphs using matrices, such as the adjacency matrix or the node-edge incidence matrix, can provide powerful insights into the structure and facilitate solving problems involving the graph.
In this work, we consider the \emph{representation matrix} of a graph, defined as follows.

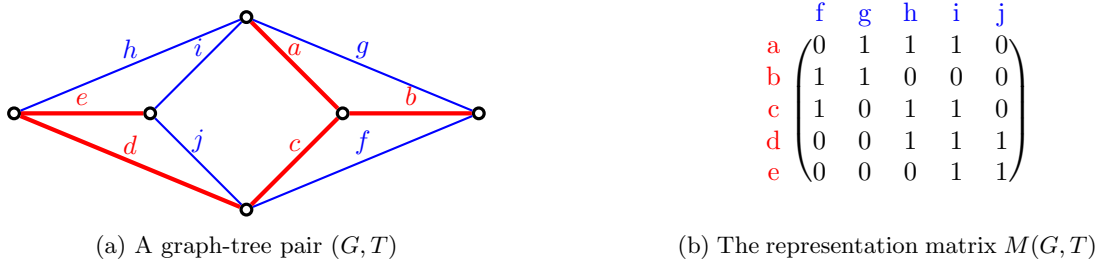
\begin{figure}[htpb]
\begin{subfigure}{.5\textwidth}
            \centering
            \begin{tikzpicture}[node distance = 1.8 cm]
                \node[main] (1) {};
                \node[main] (2) [left of=1] {};
                \node[main] (3) [below left of=2] {};
                \node[main] (4) [above left of=2] {};
                \node[main] (5) [above left of=3] {};
                \node[main] (6) [left of=5]{};
                \draw[tree] (1) -- (2) node [midway,above=-0.04cm] {$b$};
                \draw[cotree] (1) -- (3) node [midway,above=-0.04cm] {$f$};
                \draw[cotree] (1) -- (4) node [midway,above=-0.04cm] {$g$};
                \draw[tree] (2) -- (4) node [midway,above=-0.01cm] {$a$};
                \draw[tree] (3) -- (2) node [midway,above=-0.01cm] {$c$};

                \draw[cotree] (3) -- (5) node [midway,above=-0.01cm] {$j$};
                \draw[tree] (6) -- (3) node [midway,above=-0.04cm] {$d$};
                \draw[tree] (5) -- (6) node [midway,above=-0.04cm] {$e$};
                \draw[cotree] (4) -- (6) node [midway,above=-0.04cm] {$h$};
                \draw[cotree] (5) -- (4) node [midway,above=-0.01cm] {$i$};
            \end{tikzpicture}
            \caption{A graph-tree pair $(G,T)$}
\end{subfigure}
\hfill
\begin{subfigure}{.37\textwidth}
            \centering
            \begin{blockarray}{cccccc}
             & \color{blue}{f} & \color{blue}{g} & \color{blue}{h} & \color{blue}{i} & \color{blue}{j} \\
            \begin{block}{c(ccccc)}
              \color{red}{a} & 0 & 1 & 1 & 1 & 0  \\
              \color{red}{b} & 1 & 1 & 0 & 0 & 0 \\
              \color{red}{c} & 1 & 0 & 1 & 1 & 0  \\
              \color{red}{d} & 0 & 0 & 1 & 1 & 1 \\
              \color{red}{e} & 0 & 0 & 0 & 1 & 1  \\
            \end{block}
            \end{blockarray}
            \caption{The representation matrix $M(G,T)$}
\end{subfigure}

\caption{A graph-tree pair $(G,T)$ and its representation matrix. Edges in $T$ are marked bold and red, edges in $E\setminus T$ are marked blue.}
\label{fig_representation_example}
\end{figure}

Given a connected multigraph $G$ with vertex set $V(G)$, edge set $E(G)$ and a spanning tree $T \subseteq E(G)$ of G, the \emph{fundamental path} $P_e(T) \coloneqq P_{u,w}(T) \subseteq T$ of an edge $e = \{u,w\} \in E \setminus T$ is defined as the unique path in $T$ connecting the two end-vertices of $e$.
For such a \emph{graph-tree pair} $(G,T)$, let $M(G,T) $ be a binary $|T| \times |E\setminus T|$ matrix with rows indexed by edges in $T$ and columns indexed by edges in $E\setminus T$.
For a pair of edges $(e,f) \in T \times (E \setminus T)$ we let $M_{e,f} = 1$ if $e \in P_f(T)$ and $M_{e,f} = 0$ otherwise.
The matrix $M(G,T)$ constructed in this fashion is called the \emph{representation matrix} of $(G,T)$. \Cref{fig_representation_example} shows an example of a graph-tree pair and its representation matrix.

Our central problem, the \emph{graph realization problem} asks for the reverse operation:

\begin{problem}\label{prob_graph_realization}
Given a binary matrix $B$, is there a multigraph $G$ with spanning tree $T \subseteq E(G)$ such that $B = M(G,T)$ holds?
\end{problem}

In the affirmative case, $B$ is said to be \emph{graphic}, and $(G,T)$ is said to \textit{realize} $B$.

\paragraph{Related problems.}
Our main motivation for investigating \cref{prob_graph_realization} comes from \emph{network matrices}, which are closely related to graphic matrices. Given a directed multigraph $G = (V,A)$, with some (not necessarily rooted) spanning forest $T$, the network matrix $N(G,T)$ is a $|T|\times |A\setminus T|$ matrix where for edge pair $(e,f)\in T \times (A\setminus T)$ satisfies
\[N(G,T)_{e,f} = \begin{cases}
    +1 & \text{if $e$ occurs in $P_f(T)$ forwardly,}\\
    -1 & \text{if $f$ occurs in $P_f(T)$ backwardly,}\\
    0  & \text{otherwise.}
\end{cases}\]

The graph realization problem is closely connected to the problem of determining whether a given $\{-1,0,1\}$-matrix $B$ is a network matrix.
After solving the graph realization problem on the binary support of $B$, one only needs to determine the direction of each arc in order to determine whether the matrix is a network matrix.
Finding the arc directions is relatively straightforward to do, as first outlined by Camion~\cite{Camion1963}. Bixby and Cunningham~\cite{Bixby1980} go into more detail (see Algorithm 7 of~\cite{Bixby1980}).

Network matrices are a large class of \textit{totally unimodular} matrices~\cite{Tutte1965}.
This makes them of practical interest for (mixed)-integer linear optimization, because problems with a totally unimodular constraint matrix and integer right hand side can be solved in polynomial time using linear optimization~\cite{Hoffman1956} with, say, the Ellipsoid method~\cite{GroetschelLS81,KarpP80,PadbergR81} or interior-point methods~\cite{Karmarkar84}.
Even if only part of the constraint matrix of an integer linear optimization program is a network matrix, a large network \emph{submatrix} can be useful to reduce the solution time.
For instance, branch-and-cut algorithms can use stronger cutting planes that exploit network-design substructures~\cite{Achterberg2010}.
In recent work, the authors of~\cite{Aprile2024} consider integer programs with $\Delta$-modular constraint matrices that contain a large transposed network matrix and only a constant number of rows that do not belong to the transposed network matrix, and show that these integer programs are solvable in time polynomial in $\Delta$ and the size of the constraint matrix.

Additionally, the detection of total unimodularity requires the solution of graph realization problems~\cite{Seymour80,Truemper1990}.
Although detecting whether a matrix is a network matrix can be done in polynomial time~\cite{Tutte1960}, both the problem of finding the largest graphic submatrix (using various definitions of `largest') and the problem of finding the largest network submatrix are NP-hard~\cite{Bartholdi1982}.

The graph realization problem is closely connected to graphic matroids, and can be reformulated in terms of matroids.
A binary matrix $B$ defines a \emph{linear matroid} that has the columns of $\begin{bmatrix}
    I \mid B
\end{bmatrix}$ as a ground set, where subsets are independent if and only if the corresponding column vectors are linearly independent over $\mathbb{F}_2$.
Every graph $G = (V,E)$ also has an associated graphic matroid, which has the set $E$ of edges as its ground set and the forests of $G$ as its independent sets.
The graph realization problem can then be reformulated as follows:
given a binary matrix $B$, is the linear matroid given by $B$ isomorphic to a graphic matroid?
Although we will not use the matroid perspective throughout this work, it may be useful for readers that are familiar with matroid theory.

\paragraph{Known methods.}
Numerous methods have been proposed to solve the graph realization problem.
Tutte gave a first polynomial-time algorithm~\cite{Tutte1960,Tutte1964}.
Many other polynomial-time algorithms were later developed by a large variety of authors~\cite{Bixby1980,BixbyWagner1988,Cunningham1982,Fujishige1980,Gavril1983,Iri1968,Tomizawa1976}, and the books \cite{Seshu1961,Truemper1998} explain some of these in further detail.
The most impressive results for graph realization were obtained by Bixby and Wagner~\cite{BixbyWagner1988} and Fujishige~\cite{Fujishige1980}.
Let $k$ be the number of nonzeros of the input matrix $B \in \{0,1\}^{m \times n}$.
Both papers achieve an `almost linear' running time of $\orderO(k\alpha(k,m))$, where $\alpha$ denotes the extremely slowly growing \emph{inverse Ackermann function} \cite{Tarjan1984}.

The algorithms described in both papers work in a similar column-wise fashion, namely by growing a graphic submatrix by one column in each step.
First, they determine some initial graphic matrix $M$ given by a subset of columns of $B$, typically by a single column.
If $M$ has a single column, $M$ represents a graph that is a cycle.
Then they augment $M$ with a new column $c$ and (efficiently) determine if the matrix $[ M \mid c]$ is graphic.
In the affirmative case, they set $M\coloneqq[ M \mid c ]$ and repeat the augmentation, terminating only when $M = B$ or when $[ M \mid c]$ is found to be non-graphic.

It is well known that there may be many graph-tree pairs $(G,T)$ sharing the same representation matrix $M(G,T)$~\cite{Whitney33}.
This ambiguity is one of the main challenges in designing algorithms for \cref{prob_graph_realization}.
The algorithms due to Bixby and Wagner and due to Fujishige both maintain complicated data structures in order to represent the graphic matrix $M$, and efficiently check if the new column $c$ can be augmented whilst preserving graphicness.
Bixby and Wagner use a so-called \emph{t-decomposition}, whereas Fujishige uses the more complicated \emph{PQ-trees}.

In this work, we will use \emph{SPQR-trees}, which are very similar to the \emph{t-decomposition} used by Bixby and Wagner.

\paragraph{Research gap.}
As there is an interest in determining large graphic submatrices, we observe a gap in previous research.
Although the existing algorithms are very efficient, there is no algorithm that grows a graphic (sub)matrix in a \emph{row-wise} fashion.
In the context of mixed-integer programming, this limits the set of submatrices that can be efficiently determined to be graphic to submatrices that always contain all rows.
An algorithm that could detect submatrices in a row-wise fashion is useful, as it enables the detection of graphic submatrices and network submatrices in many applications.
Many models in practice contain graphic submatrices given by a set as rows.
Examples include models with edge inequalities $x_u + x_v \leq 1$ for edges $\{ u, v \}$ that model a stable set polytope, or models that contain network matrices formed by a subset of rows and all columns, such as those formed by flow conservation constraints that occur in many production planning and network design problems, such as the lot sizing problem~\cite{Karimi2003} and fixed-charge network flow problems~\cite{Hirsch68}.
Furthermore, there also exist models that contain transposed network submatrices given by a subset of the columns, which often appear in a setting where the variables describe an ordering.
Examples of these submatrices occur in the natural dates formulation of the single machine scheduling polytope~\cite{Balas1985} and in several formulations of the periodic event scheduling problem~\cite{Liebchen2006}.

The lack of a row-wise algorithm can be explained by its difficulty. In the column augmentation algorithms the new column is added by reversing the deletion of an edge in a represented graph, which is equivalent to the addition of the column edge.
Augmenting a row amounts to reversing the contraction of an edge in the graph, which is a more complicated operation than addition.
As we would like to grow a graphic matrix in a row-wise fashion, we are interested in solving the \emph{graphic row augmentation problem}:

\begin{problem}
  \label{prob_graphic_row_augmentation}
  Given a graphic matrix $M$ and a binary vector $b$, determine graphicness of the matrix
  \begin{equation}
    M' \coloneqq \begin{bmatrix} M \\ \transpose{b} \end{bmatrix}. \label{eq_augmented_matrix}
  \end{equation}
\end{problem}

By repeatedly solving \cref{prob_graphic_row_augmentation}, we can determine if a matrix is graphic.
Note that similarly to the column case, any binary matrix with a single row is graphic.
It is realized by a graph $G$ that has two vertices that are connected by a set of parallel edges.
Moreover, if the augmentation of a row $\transpose{b}$ does not preserve graphicness, one could also continue and ignore that row, which leads to an algorithm for greedily building an inclusion-wise maximal graphic submatrix of $B$.

Given an efficient algorithm to solve \cref{prob_graphic_row_augmentation}, we could additionally combine row and column augmentation to efficiently determine graphicness of arbitrary graphic submatrices, and not just of all-row or all-column submatrices.
In order to facilitate the practical implementation of such an algorithm, we use data structures that are highly similar to the ones used by Bixby and Wagner~\cite{BixbyWagner1988}.

\paragraph{Contribution and outline.}
For a (graphic) $m\times n$ matrix $M$, we formulate an algorithm to solve the graphic row augmentation problem given by
\cref{prob_graphic_row_augmentation} in $\orderO(\alpha(m+n, m+n)(m+n))$ time and $\orderO(m+n)$ space. By repeatedly applying this algorithm, we derive an algorithm for \cref{prob_graph_realization} that runs in $\orderO(\alpha(m+n, m+n)(m^2+mn))$ time.
In \cref{sec_separations_connectivity} we show how to deal with a block structure of $M$ and reduce the graphic row augmentation problem to the case in which $M$ is connected.
\Cref{sec_general_row_augmentation} characterizes the graphic row augmentation problem in terms of graphs.
In \cref{sec_spqr_trees} we introduce the SPQR-tree data structure and show how it encodes graphic matrices.
In \cref{sec_high_level}, we provide a high-level overview of the proposed algorithm by combining the results from previous sections. The following sections present the algorithm in further detail. In \cref{sec_augmentation_reductions} we present three reductions that shrink the SPQR-tree.
In \cref{sec_augmentation_singleskeleton} and \cref{sec_augmentation_merging}, we show how the SPQR-tree can be efficiently updated to reflect a graphic row augmentation.
In \cref{sec_algorithm} we present and discuss the complete algorithm and provide worst-case bounds for space and time complexity.
Finally, we discuss our results in \cref{sec_discussion}.

\section{Separations and connectivity}
\label{sec_separations_connectivity}

It is well-known that there can be multiple graphs $G$ and trees $T$ with the same representation matrix $M(G,T)$ \cite{Whitney33}.
In other words, the graph that is represented by a certain matrix is not unique.
To formalize the corresponding ambiguity, let us define a \emph{$k$-separation} as a partition of $E(G)$ into $E_1$ and $E_2$ such that $|E_1|,|E_2| \geq k$ holds and such that the \emph{corresponding graphs} $G_1$ and $G_2$ have exactly $k$ \emph{separating vertices} in common, where $G_i$ is the graph with edge set $E_i$ and vertex set $V_i \coloneqq \bigcup_{e \in E_i} e$.
A graph is called \emph{$k$-connected} if it is connected and it has no $\ell$-separation for all $\ell \in \{1,2,\dotsc,k-1\}$.
Our definition of $k$-connected is equivalent to the one used by Tutte~\cite{Tutte1966}.
Tutte $k$-connectivity implies the more commonly used $k$-vertex connectivity.
A vertex that is common to two parts of a $1$-separation is called an \emph{articulation vertex}.

Consider a multigraph $G=(V,E)$. For a vertex $v\in V$, we use $\delta(v)$ to denote the edges incident to $v$. Occasionally, we will write $\delta_G(v)$ to clarify in which graph we are considering the neighborhood of $v$.
For a subset of edges $F\subseteq E$, we use the notation 
$G\setminus F$ to denote the graph $G=(V,E\setminus F)$ where the edges in $F$ are removed and we use $G\slash F$ to denote the graph obtained by contracting each edge $e\in F$ into a single vertex.
For a subset of vertices $U\subseteq V$, we use $G-U$ to denote the induced subgraph $G[V\setminus U]$, where the vertices $U$ and all incident edges have been removed. For a singular vertex $v\in V$, we occasionally abuse notation and write $G-v$, instead.

Although \cref{prob_graphic_row_augmentation} asks for the addition of a new row, the reverse operations of deleting a row (or a column) from a graphic matrix $M=M(G,T)$ provides important intuition. In particular, if $M'$ is the matrix obtained by deleting a column $c$ from $M$, this corresponds to the deletion of the corresponding edge in $E\setminus T$, and for $G' = (V, E \setminus \{c\})$ it holds that $M'=M(G',T)$.
If $M'$ is the matrix obtained by deleting a row $r$ from $M$, then it can be obtained by contracting the corresponding tree edge, and we observe that $M' = M(G \slash \{r\},T \slash \{r\})$ holds.
Since graphicness is maintained under the deletion of rows and columns, it follows that graphicness is maintained under taking submatrices. For more details, we refer to \cite[Chapter~3]{Truemper1998}.

We start with a reduction to the case in which we can assume that $M$ in~\eqref{eq_augmented_matrix} is \emph{connected}, meaning that the graph with the adjacency matrix
$
  \begin{bmatrix}
    0 & M \\
    \transpose{M} & 0
  \end{bmatrix}
$
is connected.
If our input matrix $A$ is connected, one can, via a breadth-first search, reorder the rows in such a way that $M$ remains connected in consecutive updates.
However, this sequential connectivity property is lost if we may later skip rows whose addition does not preserve graphicness.
Consequently, we show how to actually treat matrices $M$ that are not connected.

To this end, we consider the case in which the matrix $M$ consists of $k$ block submatrices $M_1, M_2, \dotsc, M_k$ each of which is connected, and that $\transpose{b} = (\transpose{b_1}, \transpose{b_2}, \dotsc, \transpose{b_{k-1}}, \transpose{b_k})$ is the corresponding partition of the new row $b$:
\begin{subequations}
  \label{eq_matrix_blocks}
  \begin{align}
    M' &= \begin{bmatrix}
      M_1    &      0 & \hdots &   0 &   0 \\
        0    &    M_2 &        &   0 &   0 \\
      \vdots &        & \ddots &     & \vdots \\
        0    &      0 &        & M_{k-1} &   0 \\
        0    &      0 & \hdots &   0 & M_k \\
      \transpose{b_1} & \transpose{b_2} & \hdots & \transpose{b_{k-1}} & \transpose{b_k}
    \end{bmatrix}, ~b_i \neq \zerovec
    \label{eq_matrix_blocks_layout} \\
    M'_i &\coloneqq \begin{bmatrix} M_i \\ \transpose{b_i} \end{bmatrix} \label{eq_matrix_blocks_parts}
  \end{align}
\end{subequations}
Note that we require each $b_i$ to contain at least one $1$-entry.
Moreover, we allow that any such submatrix $M_i$ has no rows, in which case it must have a single column in order to be connected.
Although this is clearly unusual, the representation matrix of a graph consisting of one vertex and one loop is a $0$-by-$1$ matrix since its unique spanning tree has no edges.
However, we will only deal with this extreme case if $\transpose{b}$ has a nonzero in a column in which $M$ only has $0$-entries.
The following theorem ensures that we can assume connectivity in subsequent sections.

\begin{theorem}
  \label{thm_combine_blocks}
  Let $M \in \{0,1\}^{m \times n}$ be a graphic matrix and let $b \in \{0,1\}^n$ be such that the matrix $M'$ is of the form~\eqref{eq_matrix_blocks_layout}.
  Then $M'$ is graphic if and only if all matrices $M'_i$ as in~\eqref{eq_matrix_blocks_parts} are graphic for $i=1,2,\dotsc,k$.
  Moreover, every pair $(G',T')$ with $M' = M(G',T')$ can be obtained as follows.
  For $i=1,2,\dotsc,k$, let $(G'_i,T'_i)$ be such that $M'_i = M(G'_i,T'_i)$ holds and let $e'_i \in T'_i$ be the edge corresponding to the last row of $M'_i$.
  Let $G'_0$ be a graph with two vertices and exactly $k+1$ edges $e_0,e_1,\dotsc,e_k$ that connect these two vertices.
  Obtain $G'$ (resp.\ $T'$) from $G'_0, G'_1, \dotsc, G'_k$ (resp.\ from $G'_0, G'_1, \dotsc, G'_k$) by identifying $e_i$ with $e_i'$ for every $i=1,2,\dotsc,k$, and then removing these $2k$ edges.
  The edge $e_0$ then corresponds to the last row of $M'$.
\end{theorem}

\begin{proof}
  Since graphicness is maintained under taking submatrices, graphicness of $M'$ implies graphicness of $M'_i$ for all $i=1,2,\dotsc,k$.
  In order to prove the reverse direction of the first statement, assume $M'_i = M(G'_i,T'_i)$ for $i=1,2,\dotsc,k$.
  Let $G'_0$, $G'$, $T'$ as well as all $e'_i$ and $e_i$ be as in the theorem.
  By construction, $T'$ consists of the union of the sets $T'_i \setminus \{e'_i\}$, augmented by $e_0$.
  We now show $M(G',T') = M'$.
  Consider any column of $M'$, which is also a column of $M'_i$ for some $i \in \{1,2,\dotsc,k\}$ and let $e \in E(G_i') \setminus T_i'$ be the corresponding edge.
  By construction of $G'$ and $T'$, the fundamental path $P_e(T')$ contains the same edges as $P_e(T_i')$, except that $e_0$ in $T'$ is exchanged for $e'_i \in P_e(T')$.
  Since the row vector of $M'_i$ corresponding to $e'_i$ and the row vector of $M'$ corresponding to $e_0$ are both $\transpose{b}_i$, this shows that the column vectors of $M'$ and of $M(G',T')$ are identical.
  We conclude that $M'$ is indeed graphic and that $M' = M(G',T')$ holds.

  It remains to show that all realizations of $M'$ are of that type.
  To this end, consider any $(G^\star,T^\star)$ with $M' = M(G^\star,T^\star)$.
  Let $e^\star \in T^\star$ denote the edge corresponding to the last row of $M'$.
  Let, for every $i \in \{1,2,\dotsc,k\}$, $G'_i$ and $T'_i$ be obtained from $G^\star$ and $T^\star$, respectively, by contracting all edges of $T^\star$ corresponding to rows that do not belong to $M'_i$ and (for $G'_i$) deleting all edges of $E(G^\star) \setminus T^\star$ corresponding to columns that do not belong to $M'_i$.
  By construction we have $M(G'_i,T'_i) = M'_i$.
  The edge $e^\star$ remains present in each of the constructed graphs, and we denote this edge by $e'_i$ in the corresponding graph and tree.
  Let $G_i \coloneqq G'_i / e'_i$ and $T_i \coloneqq T'_i / e'_i$, respectively.
  Clearly, the edge sets of the $G_i$ form a partiton of $E(G^\star) \setminus \{e^\star\}$.
  Hence, $G^\star$ must be so that the contraction of $e^\star$ yields a graph that consists of the $G_i$, all having a single vertex $v^\star \in V(G_1) \cap V(G_2) \cap \dotsb \cap V(G_k)$ in common.
  Moreover, since each $b_i$ contains at least one $1$-entry, every $G_i$ has some edge $f$ such that $e^\star \in P_f(T^\star)$ holds.
  Hence, the joint vertex $v^\star$ of the parts of $G^\star / e^\star$ must be the one to which $e^\star$ was contracted.
  This shows that $(G^\star,T^\star)$ is equal to the pair $(G',T')$ (with respect to the $(G'_i,T'_i)$ and $e'_i$ constructed above) as stated in the theorem.
\end{proof}

In our introduction, we assumed that $G$ was a connected multigraph. For any disconnected graph $G$ and any spanning forest $T$, $M(G,T)$ has two or more disconnected blocks. As we can process each block individually, it is clear that the assumption that $G$ is connected was not necessary and can be easily dealt with. Moreover, from now on we can assume that the current matrix $M$ is connected.
As a consequence, we can focus on $2$-connected graphs due to the following result.

\begin{proposition}[Truemper~\protect{\cite[Prop.~3.2.31]{Truemper1998}}]
  \label{thm_matrix_connected_graph_biconnected}
  Let $G$ be a connected graph such that $M = M(G,T)$ holds.
  Then $G$ is 2-connected if and only if $M$ is connected.
\end{proposition}

\section{Graphic row augmentation}
\label{sec_general_row_augmentation}

In the previous section, we observed that the graphic row augmentation problem reduces to the case in which $M' = \begin{bmatrix} M \\ \transpose{b} \end{bmatrix}$ holds, where $M$ is connected. In this section, we do not explicitly use the connectivity property of $M$, but derive more general results for when $M$ is graphic.
We denote by $Y \coloneqq \supp(b)$ the edges corresponding to the subset of columns with a $1$-entry in $\transpose{b}$.
To gain some intuition, let us assume that we are given some graph $G$ with tree $T$ such that $M = M(G,T)$.
We then want to find operations on $G$ that add a new edge $r^\star$ (indexing $\transpose{b}$) such that for all $y\in Y$ the paths $P_y(T)$ are elongated with $r^\star$ and the paths $P_e(T)$ are unaltered for all edges $e$ that correspond to columns $c$ of $M$ with $b_c = 0$.

Suppose that $M'$ is indeed graphic such that $M' = M(G',T')$ for some graph $G'$ with tree $T'$.
If we were to remove the last row $r^\star$ of $M'$, we would obtain $M$ back again.
Now, we consider what happens to $G'$ and $T'$ in this case.
The removal of the row $r^\star$ corresponds to a contraction of the edge corresponding to $r^\star$ in $G'$.
Clearly, such a contraction shortens precisely those fundamental paths of the edges $e \in E \setminus T$ for which $r^\star\in P_e(T)$ holds.
Thus, after performing the contraction, we find a graph $G$ with tree $T = T'\setminus \{r^\star\}$ such that $M = M(G,T)$ holds.

An algorithm that solves \cref{prob_graphic_row_augmentation} would need to find the `reverse' of such a contraction.
Hence, we can intuitively determine that we need to take two steps.
First, we identify a vertex $v \in V(G)$ that is to be split into vertices $v_1$ and $v_2$.
Second, we check if there exists a bipartition of the edges in $\delta(v)$, so that we can reassign the edges in $\delta(v)$ to either $\delta(v_1)$ or $\delta(v_2)$. In particular, this bipartition should ensure that the fundamental paths of the column edges $Y$ are elongated and the fundamental paths of the column edges $C \setminus Y$ remain unchanged.

If there is such a bipartition, we can split $v$ into two new vertices $v_1$ and $v_2$, reassign the edges $\delta(v)$ to $\delta(v_1)$ and $\delta(v_2)$, and add the new edge $r^\star$ between $v_1$ and $v_2$. This way, we create a new graph-tree pair $(G',T')$ with $T'=T\cup\{r^\star\}$ such that $M'=M(G',T')$.

We will now formalize the above intuition.
The following definitions were inspired by Truemper's work on recognizing total unimodularity, which contains an algorithm for solving the problem in case the involved graphs are $3$-connected (see case~2 of the TEST-C subroutine in~\cite{Truemper1990}).
However, Truemper's algorithm lacks a detailed explanation and proofs.
In fact, we identified a minor mistake and later provide a fix.

\begin{definition}[$Y$-reduced graph, auxiliary graph, $Y$-splittable vertices]
  \label{SetupDefinition}
  Let $G$ be a multigraph and let $Y \subseteq E(G)$ be a subset of its edges.
  We define the \emph{$Y$-reduced graph of $G$} as the graph $G_Y \coloneqq (V(G), E(G) \setminus Y)$.
  Moreover, by $G^v_Y \coloneqq G_Y[V \setminus \{v\}]$ we denote the graph obtained by removing edges $Y$ and vertex $v$ (along with its incident edges).
  The corresponding \emph{auxiliary graph} $H^v_Y \coloneqq H^v_Y(G)$ is the graph having a vertex for each connected component of $G^v_Y$ and with edges $\{h_1,h_2\} \in E(H^v_Y)$ if and only if there is an edge $\{v_1,v_2\} \in Y$ with $v_1 \in V(h_1)$ and $v_2 \in V(h_2)$.
  Finally, we say that a vertex $v \in G$ is \emph{$Y$-splittable} with respect to $G$ if $H^v_Y$ is bipartite.
\end{definition}

Note that the auxiliary graph $H^v_Y$ is not bipartite if it contains loops, i.e., if an edge from $Y$ has both end-vertices in the same connected component of $G^v_Y$.

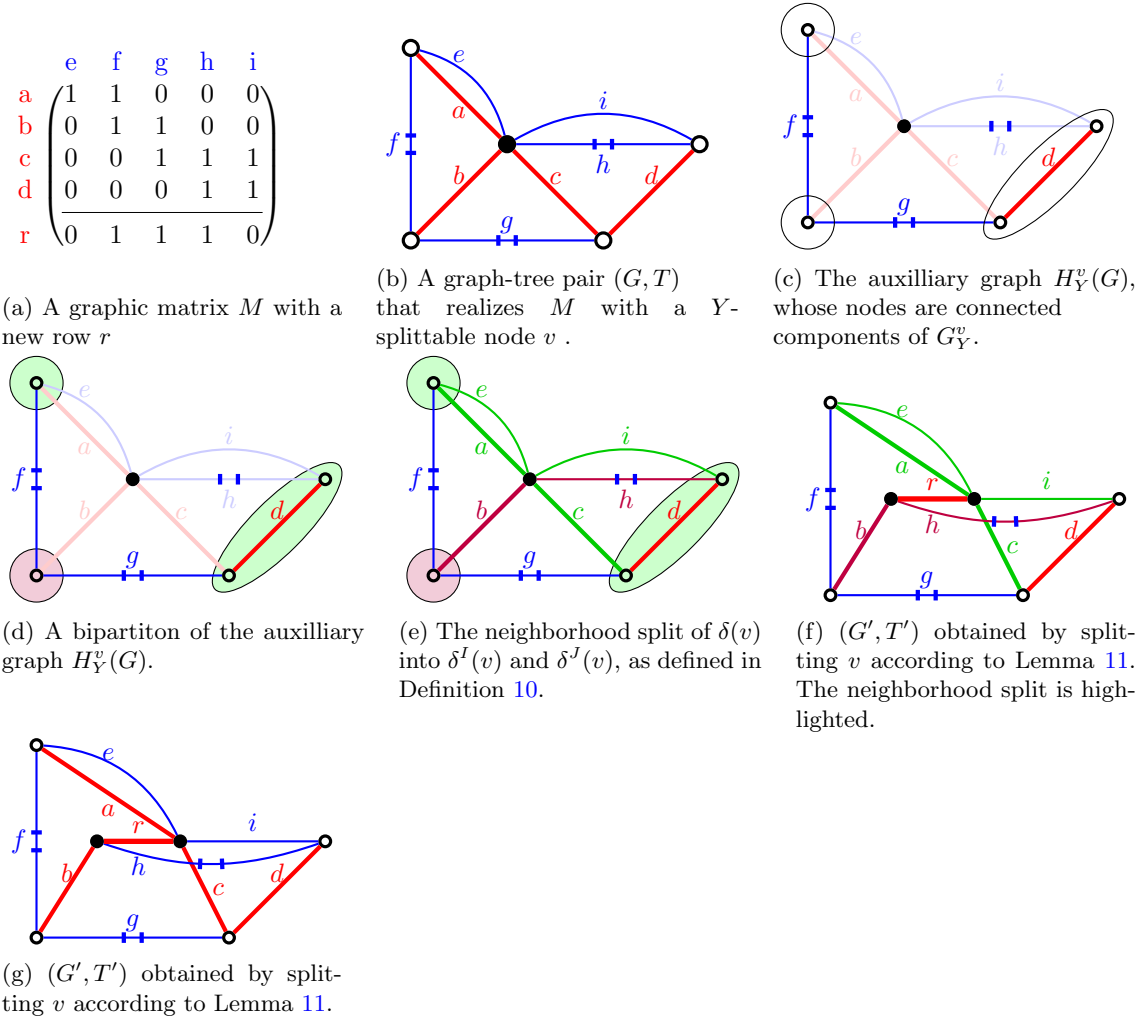
\begin{figure}[htpb]
  \begin{subfigure}[b]{0.3\textwidth}
    \begin{blockarray}{cccccc}
      & \color{blue}{e} & \color{blue}{f} & \color{blue}{g} & \color{blue}{h} & \color{blue}{i} \\
    \begin{block}{c(ccccc)}
      \color{red}{a} & 1 & 1 & 0 & 0 & 0  \\
      \color{red}{b} & 0 & 1 & 1 & 0 & 0 \\
      \color{red}{c} & 0 & 0 & 1 & 1 & 1  \\
      \color{red}{d} & 0 & 0 & 0 & 1 & 1 \\
      \cmidrule(lr){2-6}
      \color{red}{r} & 0 & 1 & 1 & 1 & 0  \\
    \end{block}
    \end{blockarray}
    \subcaption{A graphic matrix $M$ with a new row $r$.\label{fig_splittable_example_a}}
  \end{subfigure}
\hfill
  \begin{subfigure}[b]{.32\textwidth}
    \begin{tikzpicture}[node distance= 1.8cm]
      \node[main, inner sep= 2pt] (1) {};
      \node[main, inner sep= 2pt, fill=black] (2) [below right of=1] {};
      \node[main, inner sep= 2pt] (3) [below left of=2] {};
      \node[main, inner sep= 2pt] (4) [below right of=2] {};
      \node[main, inner sep= 2pt] (5) [above right of=4] {};
      \draw[tree] (1) -- (2) node [midway,below=0.0cm] {$a$} node[midway, above= 0.30cm, blue] {$e$};
      \draw[tree] (2) -- (3) node [midway,above=-0.04cm] {$b$};
      \draw[tree] (2) -- (4) node [midway,above=-0.04cm] {$c$};
      \draw[tree] (4) -- (5) node [midway,above=-0.04cm] {$d$};
      \draw[marked] (1) -- (3) node [midway,left=-0.00cm] {$f$};
      \draw[marked] (3) -- (4) node [midway,above=-0.04cm] {$g$};
      \draw[marked] (2) -- (5) node [midway,below= 0.0cm] {$h$} node [midway,above= 0.33cm] {$i$};
      \draw[cotree] (1) to [bend left] (2);
      \draw[cotree] (2) to [bend left] (5);
    \end{tikzpicture}
    \subcaption{%
      A graph-tree pair $(G,T)$ \\
      that realizes $M$ with a \\
      $Y$-splittable vertex $v$.
      \label{fig_splittable_example_b}
    }
  \end{subfigure}
\hfill
  \begin{subfigure}[b]{.32\textwidth}
    \begin{tikzpicture}[node distance= 1.8cm]
      \node[main] (1) {};
      \node[main, fill=black] (2) [below right of=1] {};
      \node[main] (3) [below left of=2] {};
      \node[main] (4) [below right of=2] {};
      \node[main] (5) [above right of=4] {};
      \draw[tree,red!20] (1) -- (2) node [midway,below=0.0cm] {$a$} node[midway, above= 0.30cm, blue!20] {$e$};
      \draw[tree,red!20] (2) -- (3) node [midway,above=-0.04cm] {$b$};
      \draw[tree,red!20] (2) -- (4) node [midway,above=-0.04cm] {$c$};
      \draw[tree] (4) -- (5) node [midway,above=-0.04cm] {$d$};
      \draw[marked] (1) -- (3) node [midway,left=-0.00cm] {$f$};
      \draw[marked] (3) -- (4) node [midway,above=-0.04cm] {$g$};
      \draw[dimmedmarked,blue!20] (2) -- (5) node [midway,below= 0.0cm] {$h$} node [midway,above= 0.33cm] {$i$};
      \draw[cotree,blue!20] (1) to [bend left] (2);
      \draw[cotree,blue!20] (2) to [bend left] (5);
      \draw (0,0) ellipse (10pt and 10pt);
      \draw (0,-2.5456) ellipse (10pt and 10pt);
      \draw[rotate around={45:(3.18198,-1.90919)}] (3.18198,-1.90919) ellipse (33pt and 10pt);
    \end{tikzpicture}
    \subcaption{%
      The auxiliary graph $H^v_Y(G)$, whose vertices are connected \\
      components of $G^v_Y$.
      \label{fig_splittable_example_c}
    }
  \end{subfigure}

  \begin{subfigure}[t]{.32\textwidth}
    \begin{tikzpicture}[node distance= 1.8cm]
      \draw[fill=green!60] (0,0) ellipse (10pt and 10pt);
      \draw[fill=yellow!80!orange!60] (0,-2.5456) ellipse (10pt and 10pt);
      \draw[rotate around={45:(3.18198,-1.90919)},fill=green!60] (3.18198,-1.90919) ellipse (33pt and 10pt);
      \node[main] (1) {};
      \node[main, fill=black] (2) [below right of=1] {};
      \node[main] (3) [below left of=2] {};
      \node[main] (4) [below right of=2] {};
      \node[main] (5) [above right of=4] {};
      \draw[tree,red!20] (1) -- (2) node [midway,below=0.0cm] {$a$} node[midway, above= 0.30cm, blue!20] {$e$};
      \draw[tree,red!20] (2) -- (3) node [midway,above=-0.04cm] {$b$};
      \draw[tree,red!20] (2) -- (4) node [midway,above=-0.04cm] {$c$};
      \draw[tree] (4) -- (5) node [midway,above=-0.04cm] {$d$};
      \draw[marked] (1) -- (3) node [midway,left=-0.00cm] {$f$};
      \draw[marked] (3) -- (4) node [midway,above=-0.04cm] {$g$};
      \draw[dimmedmarked,blue!20] (2) -- (5) node [midway,below= 0.0cm] {$h$} node [midway,above= 0.33cm] {$i$};
      \draw[cotree,blue!20] (1) to [bend left] (2);
      \draw[cotree,blue!20] (2) to [bend left] (5);
    \end{tikzpicture}
    \subcaption{%
      A bipartition of the auxiliary graph $H^v_Y(G)$.
    }
    \label{fig_splittable_example_d}
  \end{subfigure}
\hfill
  \begin{subfigure}[t]{.32\textwidth}
    \begin{tikzpicture}[node distance= 1.8cm]
      \draw[fill=green!60] (0,0) ellipse (10pt and 10pt);
      \draw[fill=yellow!80!orange!60] (0,-2.5456) ellipse (10pt and 10pt);
      \draw[rotate around={45:(3.18198,-1.90919)},fill=green!60] (3.18198,-1.90919) ellipse (33pt and 10pt);
      \node[main] (1) {};
      \node[main, fill=black] (2) [below right of=1] {};
      \node[main] (3) [below left of=2] {};
      \node[main] (4) [below right of=2] {};
      \node[main] (5) [above right of=4] {};
      \draw[tree,green!80!black] (1) -- (2) node [midway,below=0.0cm] {$a$} node[midway, above= 0.30cm, green!80!black] {$e$};
      \draw[tree,yellow!50!orange!100] (2) -- (3) node [midway,above=-0.04cm] {$b$};
      \draw[tree,green!80!black] (2) -- (4) node [midway,above=-0.04cm] {$c$};
      \draw[tree] (4) -- (5) node [midway,above=-0.04cm] {$d$};
      \draw[marked] (1) -- (3) node [midway,left=-0.00cm] {$f$};
      \draw[marked] (3) -- (4) node [midway,above=-0.04cm] {$g$};
      \draw[colormarked,yellow!50!orange!100] (2) -- (5) node [midway,below= 0.0cm] {$h$} node [midway,above= 0.33cm, green!80!black] {$i$};
      \draw[cotree,green!80!black] (1) to [bend left] (2);
      \draw[cotree,green!80!black] (2) to [bend left] (5);
    \end{tikzpicture}
    \subcaption{%
      The neighborhood split of $\delta(v)$ into $\delta^I(v)$ and $\delta^J(v)$, as defined in \cref{def_SplitConstruction}.
    }
    \label{fig_splittable_example_e}
  \end{subfigure}
\hfill
  \begin{subfigure}[t]{.3\textwidth}
    \begin{tikzpicture}[node distance= 1.8cm]
      \node[main] (1) {};
      \node[main, fill=black] (2a) at (0.8,-1.2728) {};
      \node[main, fill=black] (2b) at (1.9,-1.2728) {};
      
      \node[main, draw=none] (2) [below right of=1] {};
      \node[main] (3) [below left of=2] {};
      \node[main] (4) [below right of=2] {};
      \node[main] (5) [above right of=4] {};
      
      \draw[tree,green!80!black] (1) -- (2b) node [midway,below=0.0cm] {$a$} node[midway, above= 0.30cm, green!80!black] {$e$};
      \draw[tree,yellow!50!orange!100] (2a) -- (3) node [midway,above=-0.04cm] {$b$};
      \draw[tree,green!80!black] (2b) -- (4) node [midway, right =-0.04cm] {$c$};
      \draw[tree] (4) -- (5) node [midway,above=-0.04cm] {$d$};
      \draw[marked] (1) -- (3) node [midway,left=-0.00cm] {$f$};
      \draw[marked] (3) -- (4) node [midway,above=-0.04cm] {$g$};
      \draw[cotree,green!80!black] (2b) -- (5) node [midway,above= 0.0cm] {$i$};
      \draw[cotree,green!80!black] (1) to [bend left] (2b);
      \draw[colormarked,yellow!50!orange!100] (2a) to [bend right=19] (5);
      \draw[tree,line width=2pt] (2a) -- (2b) node [midway, above=-0.04cm] {$r$} node [midway, below = 0.05cm, yellow!50!orange!100] {$h$};
    \end{tikzpicture}
    \subcaption{%
      $(G',T')$ obtained by splitting $v$ according to \cref{thm_SplitConstructionProof}.
      The neighborhood split is highlighted.
    }
    \label{fig_splittable_example_f}
  \end{subfigure}
\hfill
  \begin{subfigure}[t]{0.3\textwidth}
    \begin{tikzpicture}[node distance= 1.8cm]
      \node[main] (1) {};
      \node[main, fill=black] (2a) at (0.8,-1.2728) {};
      \node[main, fill=black] (2b) at (1.9,-1.2728) {};

      \node[main, draw=none] (2) [below right of=1] {};
      \node[main] (3) [below left of=2] {};
      \node[main] (4) [below right of=2] {};
      \node[main] (5) [above right of=4] {};

      \draw[tree,red] (1) -- (2b) node [midway,below=0.0cm] {$a$} node[midway, above= 0.30cm, blue] {$e$};
      \draw[tree,red] (2a) -- (3) node [midway,above=-0.04cm] {$b$};
      \draw[tree,red] (2b) -- (4) node [midway, right =-0.04cm] {$c$};
      \draw[tree] (4) -- (5) node [midway,above=-0.04cm] {$d$};
      \draw[marked] (1) -- (3) node [midway,left=-0.00cm] {$f$};
      \draw[marked] (3) -- (4) node [midway,above=-0.04cm] {$g$};
      \draw[cotree] (2b) -- (5) node [midway,above= 0.0cm] {$i$};
      \draw[cotree,blue] (1) to [bend left] (2b);
      \draw[marked,blue] (2a) to [bend right=19] (5);
      \draw[tree,line width=2pt] (2a) -- (2b) node [midway, above=-0.04cm] {$r$} node [midway, below = 0.05cm, blue] {$h$};
    \end{tikzpicture}
    \subcaption{%
      $(G',T')$ obtained by splitting $v$ according to \cref{thm_SplitConstructionProof}.
    }
    \label{fig_splittable_example_g}
  \end{subfigure}
  \caption{%
    An example of a graph-tree pair $(G,T)$ with a $Y$-splittable vertex $v$ (marked in black), its auxiliary graph $H^v_Y(G)$ and the updated graph $G'$ obtained after splitting $v$ into two vertices, reassigning the neighboring edges and adding the new row edge.
    Edges in $T$ are marked red and bold, all other edges are marked in blue.
    The edges in $Y = \{f,g,h\}$ are marked with two stripes.
    In \subref{fig_splittable_example_d}, the bipartition of $H^v_Y(G)$ is shown.
    In \subref{fig_splittable_example_e}, the neighborhood split into $\delta^I(v)$ and $\delta^J(v)$ is highlighted using green and yellow edges.
    In \subref{fig_splittable_example_f} and \subref{fig_splittable_example_g}, the neighborhood split of \subref{fig_splittable_example_d} is applied using \cref{thm_SplitConstructionProof}, splitting up $\delta^I(v)$ and $\delta^J(v)$ into two new vertices, that are both marked black, which are connected by a newly added row edge $r$.
  }
  \label{fig_splittable_example}
\end{figure}

\Cref{fig_splittable_example} shows an example of a graph-tree pair with a $Y$-splittable vertex, and highlights how $v$ and its neighboring edges can be split so that we can obtain $(G',T')$.

Based on \cref{SetupDefinition}, a few basic but insightful results follow.
Given a subset of edges $Y \subseteq E$, we say that $Y$ is a \emph{star centered at $v$} if
$Y \subseteq \delta(v)$, such that all edges $y \in Y$ have the vertex $v$ in common.

\begin{proposition}
  \label{thm_yStarLemma}
  Let $G$ be a multigraph and let $Y \subseteq E(G)$ be a subset of edges such that $Y$ is a star centered at $v$.
  Then $v$ is $Y$-splittable.
\end{proposition}

\begin{proof}
  From $Y \subseteq \delta(v)$ it follows that the auxiliary graph $H^v_Y$ has no edges, and thus $H^v_Y$ is bipartite.
\end{proof}

\begin{corollary}\label{thm_yEmptyLemma}
  Every vertex of a multigraph is $\emptyset$-splittable.
\end{corollary}

\begin{proof}
  Apply \cref{thm_yStarLemma} to each vertex.
\end{proof}

We can also derive a few necessary conditions for a  $Y$-splittable vertex $v$ when $Y$ is not a star.
\begin{lemma}
  \label{thm_splittableNotStarArticulationLemma}
  Let $G$ be a multigraph and let $Y \subseteq E(G)$ be an edge subset.
  If $v \in V(G)$ is $Y$-splittable and $Y$ is not a star centered at $v$, then $v$ is an articulation vertex of $G_Y$.
\end{lemma}

\begin{proof}
  As $Y$ is not a star centered at $v$, there must exist an edge $e \in Y$ that is not incident to $v$.
  Since $e \in Y\setminus\delta(v)$, it induces an edge $\{h_1,h_2\}$ in the auxiliary graph $H_Y^v$. Since $H^v_Y$ is bipartite, $h_1 \neq h_2$ holds, which implies that $G^v_Y$ has at least two connected components.
\end{proof}

\begin{lemma}
  \label{thm_MustLieOnPath}
  Let $G$ be a multigraph with spanning tree $T \subseteq E(G)$ and a set of edges $Y \subseteq E(G) \setminus T$.
  If $v$ is $Y$-splittable, then $v$ must lie on the fundamental path $P_y(T)$ for each $y \in Y$.
\end{lemma}

\begin{proof}
  Assume, for the sake of contradiction, that there exists an edge $y \in Y$ such that $v$ is not on $P_y(T)$.
  Since $P_y(T) \subseteq T \setminus \delta(v)$ and $T\cap Y = \emptyset$ hold, $P_y(T)$ must lie in a single connected component of $G^v_Y$.
  This implies that $H^v_Y$ has a loop for this component, and hence $H^v_Y$ is not bipartite.
  This contradicts the assumption that $v$ was $Y$-splittable, which concludes the proof.
\end{proof}

\cref{thm_yStarLemma,thm_splittableNotStarArticulationLemma,thm_MustLieOnPath} provide necessary conditions for a vertex $v$ to be $Y$-splittable.
Now, we consider the next step where we split $v$ into two new vertices and reassign the edges incident to $v$ to these vertices. This step is formalized in \cref{def_SplitConstruction} and \cref{thm_SplitConstructionProof}, where \cref{thm_SplitConstructionProof} proves that for a $Y$-splittable vertex $v$ a certain reassignment of incident edges elongates exactly the fundamental paths of edges $Y$ by splitting $v$ into two vertices $v_1$ and $v_2$ and subsequently setting the new tree edge $r^\star$ to connect $v_1$ and $v_2$.

\begin{definition}[neighborhood split]
  \label{def_SplitConstruction}
  Let $G$ be a multigraph with a set of edges $Y \subseteq E(G)$.
  Let $v$ be a $Y$-splittable vertex and let $I$ and $J$ be the two sides of a corresponding bipartition of $H^v_Y$.
  Then the associated \emph{neighborhood split} of $v$ is the partition of $\delta(v)$ into $\delta^I(v)$ and $\delta^J(v)$ defined via
  \[
    \delta^I(v) \coloneqq \{\{u,v\}\in \delta(v) \mid \text{ either } \{u,v\}\in Y \text{ or there exists } h \in I \text{ with } u \in h \text{ (but not both)} \}.
  \]
\end{definition}

\begin{lemma}
  \label{thm_SplitConstructionProof}
  Let $G$ be a multigraph with spanning tree $T \subseteq E(G)$, and let $Y \subseteq E(G) \setminus T$ be a subset of the non-tree edges.
  Let $v$ be a $Y$-splittable vertex and let $I,J\subseteq V(H^v_Y)$ denote the bipartition of $H^v_Y$. Construct the multigraph $G'$ from $G$ by splitting $v$ into vertices $i$ and $j$, adding the new edge $r^\star = \{i,j\}$, and replacing the edges $\{u,v\} \in \delta(v)$ by $\{u,i\}$ (resp.\ by $\{u,j\}$) if $u \in \delta^I(v)$ (resp.\ $u \in \delta^J(v)$).
  Construct the spanning tree $T' \coloneqq T \cup \{ r^\star \}$ of $G'$.
  Then for each $e \in E \setminus T$ we have
  \[
    P_e(T') = \begin{cases}
      P_e(T) \cup \{ r^\star \} &\text{if } e \in Y, \\
      P_e(T) &\text{otherwise}.
    \end{cases}
  \]
\end{lemma}

\begin{proof}
  First note that $T'$ is indeed a spanning tree of $G'$.

  Second, consider an edge $e \in E \setminus (T \cup Y)$.
  Denote by $C \coloneqq P_e(T) \cup \{e\}$ its fundamental cycle and observe that $C \cap Y = \emptyset$ holds.
  If $P_e(T) \cap \delta(v) = \emptyset$ then none of the edges in the path are changed by changing $v$ and $\delta(v)$ and thus $P_e(T) = P_e(T')$ holds.
  Otherwise, let $f,f' \in \delta(v) \cap C$ be the two cycle edges incident to $v$.
  From $C \cap Y = \emptyset$ it follows that $C \setminus \delta(v)$ is a path in the reduced graph $G_Y^v$ and hence, $f$ and $f'$ connect to the same connected component of $G_Y^v$.
  Again from $C \cap Y = \emptyset $ we have $f,f' \notin Y$, and thus $f$ and $f'$ are reassigned to the same vertex from $\{i,j\}$ in $G'$. This implies that in $G'$ the path $P_e(T)$ remains unchanged, so we conclude that $P_e(T) = P_e(T')$ holds.

  Third, consider an edge $y \in Y$.
  Once again, denote by $C \coloneqq P_y(T) \cup \{y\}$ the fundamental cycle.
  By \cref{thm_MustLieOnPath} we know that $v$ must lie on $P_y(T)$, and we obtain $C \cap \delta(v) = \{f,f'\}$ for suitable edges $f,f'$.
  We claim that these edges are reassigned to different end-vertices, and we distinguish two cases to prove it.

\medskip
\noindent
  \textbf{Case 1: $y\in \{f,f'\}$.}
  We can assume $y = f$ without loss of generality.
  Then $f' \in T$ and $P_y(T) \setminus \{f'\}$ is a path that lies in a single connected component of the reduced graph $G^v_Y$. Both $y$ and $f'$ must connect to this component.
  Since $f\in Y$ and $f' \notin Y$ hold, they are indeed reassigned to different end-vertices from $\{i,j\}$ in $G'$.

\medskip
\noindent
  \textbf{Case 2: $y \notin \{f,f'\}$.}
  In this case, $P_y(T) \setminus \{f,f'\}$ is no longer a path and thus disconnected.
  Let $P^1$ and $P^2$ be the two (maximal) subpaths formed by $P_y(T) \setminus \{f,f'\}$, which may consist of just a single vertex.
  Each such path $P^k$ ($k=1,2$) lies in a single connected component $h_k$ of $G^v_Y$.
  Since $v$ is $Y$-splittable, $H^v_Y$ is bipartite.
  Moreover, since $y$ connects a vertex from $P^1$ with one from $P^2$, $h_1$ and $h_2$ must be on different sides of the bipartition.
  Due to $f,f' \notin Y$, they are indeed reassigned to different end-vertices from $\{i,j\}$ in $G'$.

  In both cases, $f$ and $f'$ were reassigned to different end-vertices from $\{i,j\}$ in $G'$.
  Thus, the new edge $r^\star = \{i,j\}$ lies on the fundamental path of $P_y(T')$, which yields $P_y(T') = P_y(T) \cup \{ r^\star \}$.
  This concludes the proof.
\end{proof}

Now, we have the tools to show the main result of this section, which characterizes the graphic row augmentation problem for a matrix $M$ in terms of $Y$-splittability of the graphs represented by $M$.

\begin{theorem}
  \label{thm_graphic_splittable}
  Let $M$ and $M'$ be binary matrices and $b$ be a binary vector as in~\eqref{eq_augmented_matrix} and define
  $Y \coloneqq \supp(\transpose{b})$.
  Then $M'$ is graphic if and only if there exists a graph $G = (V,E)$ and a tree $T \subseteq E$ with $M = M(G,T)$ such that $G$ has a $Y$-splittable vertex.
\end{theorem}

\begin{proof}
We first prove that graphicness of $M'$ implies existence of the graph-tree as stated in the theorem.
To this end, assume that $M'$ has a realization $M' = M(G',T')$.
Let $r^\star$ index the last row given by $\transpose{b}$.
Removal of row $r^\star$ from the matrix $M'$ corresponds to the contraction of the edge $r^\star \in T'$ with end-vertices $v_1,v_2$ into a new vertex $v$.
Hence, $M = M(G,T)$ holds, where $T = T' / r^\star$ and $G = G' / r^\star$.

We claim that $v$ is $Y$-splittable in $G$.
If $Y$ is a star centered at $v$, the statement follows from \cref{thm_yStarLemma}.
Otherwise, $Y$ is not a star centered at $v$ so there must exist some edge $y \in Y$ that is not incident to $v$.
Deletion of $r^\star$ from the tree $T'$ yields two trees $T_1$ and $T_2$ with vertex sets $K_1 = V(T_1)$ and $K_2 = V(T_2)$, respectively, where $v_1\in K_1$ and $v_2\in K_2$. Note that the edge $y$ must connect $K_1$ with $K_2$, since its fundamental path $P_y(T')$ must contain $r^\star$ because $M'$ is graphic.
As $y$ is not incident to $v$ in $G$, there must exist vertices $w_1\in K_1$, $w_1\neq v_1$ and $w_2\in K_2$, $w_2\neq v_2$. This shows that $|K_i|\geq 2$ for $i=1,2$.

Consider any edge $e \in E(G') \setminus T'$ that connects $K_1$ with $K_2$.
Since $r^\star \in P_e(T')$ holds, we must have $e \in Y$, and all edges in $Y$ must connect $K_1$ to $K_2$.
Hence, no such edge belongs to $G_Y$.
Then the vertex sets $K_i \setminus \{v_i\}$ for $i=1,2$ are non-empty (since $|K_i|\geq 2$) and belong to $G$.
Since $K_1\setminus\{v_1\}$ is not connected to $K_2\setminus\{v_2\}$ by any edge in $G_Y$, we conclude that $v$ is an articulation vertex of $G_Y$.
Now, let us show that $H_Y^v$ is bipartite.
For $i=1,2$, we define $H_i \coloneqq \{ h \in V(H_Y^v) : V(h) \subseteq K_i\setminus\{v_i\} \}$.
Clearly, $H_1$ and $H_2$ are disjoint since $ K_1 $ and $ K_2 $ are.
The vertices of any connected component of $G^v_Y$ are a subset of either $K_1\setminus\{v_1\}$ or $K_2\setminus\{v_2\}$ as these vertex sets are disconnected in $G^v_Y$.
Then $H_1\cup H_2 = V(H^v_Y)$ holds since these vertex sets cover all vertices of $G^v_Y$, i.e., $(K_1\setminus\{v_1\}) \cup (K_2\setminus\{v_2\}) = V(G^v_Y)$.
Thus, $H_1$ and $H_2$ form a bipartition of the components $V(H^v_Y)$.
As argued above, every edge $y\in Y$ must connect $K_1$ to $K_2$ as $M'$ is graphic.
This implies that every edge $(h_i,h_j)\in E(H^v_Y)$ has $h_i\in H_1$ and $h_j\in H_2$ (or vice versa).
This shows that $H_Y^v$ is bipartite with bipartition $H_1$, $H_2$.
We conclude that $v$ is $Y$-splittable.

We now prove the reverse direction.
We consider a graph $G$ with spanning tree $T \subseteq E(G)$ with $M = M(G,T)$ that has a $Y$-splittable vertex $v$ and show that $M'$ is graphic.
We obtain $G'$ and $T'$ according to \cref{thm_SplitConstructionProof}.
Then \cref{thm_SplitConstructionProof} shows that $G'$ correctly elongates the fundamental paths of all edges in $Y$ and does not modify those of other edges.
This shows $M' = M(G',T')$ and in particular that $M'$ is graphic.
\end{proof}

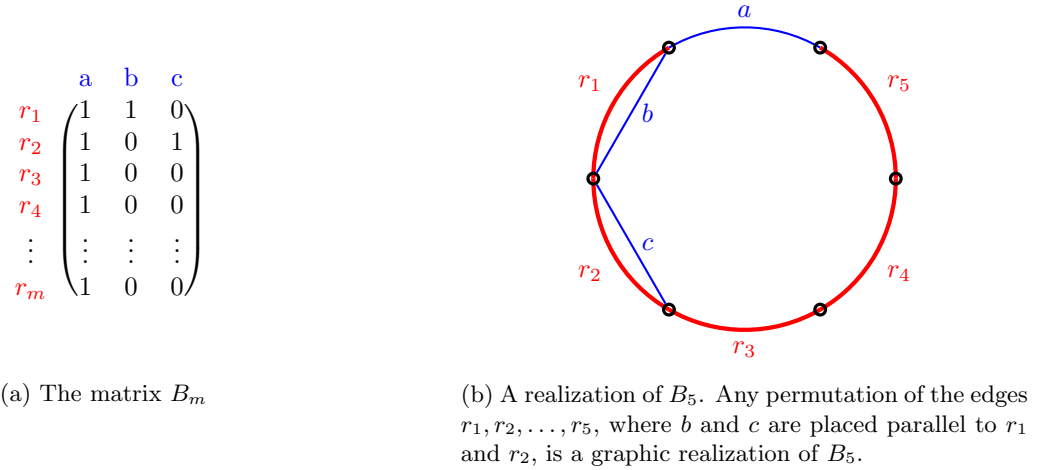
\begin{figure}[htpb]
\begin{subfigure}[t]{.37\textwidth}
            \vspace{-4.0cm}
            \centering
            \begin{blockarray}{cccc}
             & \color{blue}{a} & \color{blue}{b} & \color{blue}{c} \\
            \begin{block}{c(ccc)}
              \color{red}{$r_1$} & 1 & 1 & 0  \\
              \color{red}{$r_2$} & 1 & 0 & 1 \\
              \color{red}{$r_3$} & 1 & 0 & 0  \\
              \color{red}{$r_4$} & 1 & 0 & 0  \\
              \vdots & \vdots & \vdots & \vdots \\
              \color{red}{$r_{m}$} & 1 & 0 & 0 \\
            \end{block}
            \end{blockarray}
            \vspace{0.45cm}

            \subcaption{The matrix $B_m$}
\end{subfigure}
\hfill
\begin{subfigure}[t]{.5\textwidth}
            \centering
            \begin{tikzpicture}
            \draw[cotree] (120:2) arc (120:60:20mm)  node [midway,above = 0.01cm] {$a$};
            \draw[tree] (180:2) arc (180:120:20mm) node [midway,above left = 0.01cm] {$r_1$};
            \draw[tree] (240:2) arc (240:180:20mm) node [midway,below left = 0.01cm] {$r_2$};
            \draw[tree] (300:2) arc (300:240:20mm) node [midway, below = 0.01cm] {$r_3$};
            \draw[tree] (360:2) arc (360:300:20mm) node [midway, below right = 0.01cm] {$r_4$};
            \draw[tree] (420:2) arc (420:360:20mm) node [midway, above right = 0.01cm] {$r_5$};

            \node[main] (v_1) at (360/6 * 1:2cm) {};
            \node[main] (v_2) at (360/6 * 2:2cm) {};
            \node[main] (v_3) at (360/6 * 3:2cm) {};
            \node[main] (v_4) at (360/6 * 4:2cm) {};
            \node[main] (v_5) at (360/6 * 5:2cm) {};
            \node[main] (v_6) at (360/6 * 6:2cm) {};

            \draw[cotree] (v_2) -- (v_3) node [midway,right=0.01cm] {$b$};
            \draw[cotree] (v_3) -- (v_4) node [midway,right=0.01cm] {$c$};

   \end{tikzpicture}
   \subcaption{A realization of $B_5$. Any permutation of the edges $r_1,r_2,\dotsc,r_5$, where $b$ and $c$ are placed parallel to $r_1$ and $r_2$, is a  graphic realization of $B_5$.}
  \end{subfigure}
  \caption{%
    A series of matrices $B_m$ with $m\geq 2$ rows where each matrix $B_m$ has $\frac{m!}{2}$ realizations.
    Only $(m-1)!$ realizations contain a $\{b,c\}$-splittable vertex (when $b$ and $c$ share an adjacent vertex).
    Thus, the fraction of graphs that realizes $B_m$ and contains a $\{b,c\}$-splittable vertex is $2/m$.
  }
  \label{fig_example_many_graphs_few_realizations}
\end{figure}

\Cref{thm_graphic_splittable} provides a good characterization of when the matrix $M'$ is graphic. However, it is difficult to turn into a full algorithm.
In particular, it does not tell us how to find $G$ explicitly.
This is problematic, as there may exist a large number of graphs that each realize $M$, of which only a few may have a $Y$-splittable vertex.
\Cref{fig_example_many_graphs_few_realizations} shows an example with a series of graphic matrices with $m$ rows and a given set $Y$, for which only a fraction of $2/m$ of all represented graphs contains a $Y$-splittable vertex.

In the next section, we will explain how one can use an \emph{SPQR-tree}, which is a graph decomposition that represents all graphs that realize $M$.
Then our goal will become to update the SPQR-tree of $M$ to an SPQR-tree for $M'$, by efficiently finding the $Y$-splittable vertices across all graphs realizing $M$.

\section{Representing graphic matrices using SPQR-trees}
\label{sec_spqr_trees}

In the previous section we showed that the augmentation problem can be reduced to the search for a $Y$-splittable vertex in case the matrix $M$ in~\eqref{eq_augmented_matrix} represents a unique graph.
However, this does not need to be the case.
Note that we can assume that $M$ is connected by~\cref{thm_combine_blocks} and by~\cref{thm_matrix_connected_graph_biconnected} that every realization $G$ (with spanning tree $T$ such that $M = M(G,T)$ holds) is $2$-connected.

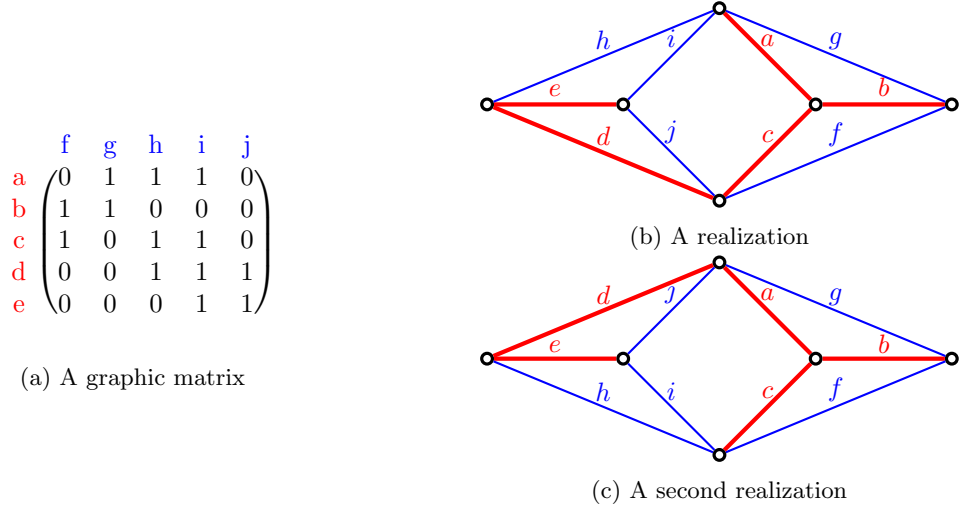
\begin{figure}[htpb]
  \begin{subfigure}{.37\textwidth}
    \centering
    \begin{blockarray}{cccccc}
      & \color{blue}{f} & \color{blue}{g} & \color{blue}{h} & \color{blue}{i} & \color{blue}{j} \\
      \begin{block}{c(ccccc)}
        \color{red}{a} & 0 & 1 & 1 & 1 & 0  \\
        \color{red}{b} & 1 & 1 & 0 & 0 & 0 \\
        \color{red}{c} & 1 & 0 & 1 & 1 & 0  \\
        \color{red}{d} & 0 & 0 & 1 & 1 & 1 \\
        \color{red}{e} & 0 & 0 & 0 & 1 & 1  \\
      \end{block}
    \end{blockarray}
    \caption{%
      A graphic matrix.
    }%
    \label{fig_2isomorphic_example_a}
\vspace{1.5cm}
  \end{subfigure} \hspace{1.0cm}
  \begin{subfigure}{.5\textwidth}
    \centering
    \begin{tikzpicture}[node distance = 1.8 cm]
      \node[main] (1) {};
      \node[main] (2) [left of=1] {};
      \node[main] (3) [below left of=2] {};
      \node[main] (4) [above left of=2] {};
      \node[main] (5) [above left of=3] {};
      \node[main] (6) [left of=5]{};
      \draw[tree] (1) -- (2) node [midway,above=-0.04cm] {$b$};
      \draw[cotree] (1) -- (3) node [midway,above=-0.04cm] {$f$};
      \draw[cotree] (1) -- (4) node [midway,above=-0.04cm] {$g$};
      \draw[tree] (2) -- (4) node [midway,above=-0.01cm] {$a$};
      \draw[tree] (3) -- (2) node [midway,above=-0.01cm] {$c$};

      \draw[cotree] (3) -- (5) node [midway,above=-0.01cm] {$j$};
      \draw[tree] (6) -- (3) node [midway,above=-0.04cm] {$d$};
      \draw[tree] (5) -- (6) node [midway,above=-0.04cm] {$e$};
      \draw[cotree] (4) -- (6) node [midway,above=-0.04cm] {$h$};
      \draw[cotree] (5) -- (4) node [midway,above=-0.01cm] {$i$};
    \end{tikzpicture}
    \caption{%
      A realization.
    }%
    \label{fig_2isomorphic_example_b}
    \begin{tikzpicture}[main,node distance = 1.8 cm]
      \node[main] (1) {};
      \node[main] (2) [left of=1] {};
      \node[main] (3) [below left of=2] {};
      \node[main] (4) [above left of=2] {};
      \node[main] (5) [above left of=3] {};
      \node[main] (6) [left of=5]{};
      \draw[tree] (1) -- (2) node [midway,above=-0.04cm] {$b$};
      \draw[cotree] (1) -- (3) node [midway,above=-0.04cm] {$f$};
      \draw[cotree] (1) -- (4) node [midway,above=-0.04cm] {$g$};
      \draw[tree] (2) -- (4) node [midway,above=-0.01cm] {$a$};
      \draw[tree] (3) -- (2) node [midway,above=-0.01cm] {$c$};

      \draw[cotree] (4) -- (5) node [midway,above=-0.01cm] {$j$};
      \draw[tree] (6) -- (4) node [midway,above=-0.04cm] {$d$};
      \draw[tree] (5) -- (6) node [midway,above=-0.04cm] {$e$};
      \draw[cotree] (3) -- (6) node [midway,above=-0.04cm] {$h$};
      \draw[cotree] (5) -- (3) node [midway,above=-0.01cm] {$i$};
    \end{tikzpicture}
    \caption{%
      Another realization.
    }%
    \label{fig_2isomorphic_example_c}
  \end{subfigure}
  \caption{%
    A graphic matrix with two $2$-isomorphic realizations.
    The realization in~\subref{fig_2isomorphic_example_c} can be obtained from \subref{fig_2isomorphic_example_b} using a reversal on the 2-separation given by $E_1 = \{d,e,h,i,j\}$ and $E_2 = E \setminus E_1$.
  }%
  \label{fig_2isomorphic_example}
\end{figure}

Suppose that $G$ has a 2-separation $(E_1,E_2)$ with separating vertices $\{u,v\} = V(E_1) \cap V(E_2)$.
Let $G'$ be a graph with the same edge labels but with the sets of incident edges of $u$ and $v$ in $E_2$ swapped, i.e., setting $\delta_{G'}(u) \coloneqq (\delta_G(u) \cap E_1) \cup (\delta_G(v) \cap E_2)$ and $\delta_{G'}(v) \coloneqq (\delta_G(v) \cap E_1) \cup (\delta_G(u) \cap E_2)$, and leaving all other edges unchanged. We say that $G'$ is obtained from $G$ by \emph{reversing} $E_2$.
Moreover, two graphs are called \emph{$2$-isomorphic} if one can be obtained from the other by a sequence of reversals (for arbitrary 2-separations).
An example of two $2$-isomorphic graphs can be found in \Cref{fig_2isomorphic_example}.
Whitney~\cite{Whitney32,Whitney33} showed that two graphs $G$ and $G'$ are $2$-isomorphic if and only if they have the same cycles.
In particular, $3$-connected graphs are uniquely determined by their cycles.

This is highly relevant for the graph realization problem since the columns of a matrix $M = M(G,T)$ encode the set of \emph{fundamental cycles} $P_e(T) \cup \{e\}$ for all $e \in E(G) \setminus T$.
Consequently, $2$-isomorphic graphs $G$ and $G'$ (with a spanning tree $T$) satisfy $M(G,T) = M(G',T)$.
Despite the fact that $M(G,T)$ only encodes a (usually small) subset of $G$'s cycles, the reverse implication also holds.
This follows from the fact that these fundamental cycles form a basis of the cycle vector space~\cite[Section 1.9]{Diestel12}.

\begin{proposition}[Consequence of Whitney~\cite{Whitney32,Whitney33}]
  \label{thm_realizations_2isomorphic}
  Let~$G$ and~$G'$ be $2$-connected graphs with identical edge labels $E(G) = E(G')$ and let~$T$ be a spanning tree of~$G$.
  Then $M(G,T) = M(G',T)$ holds if and only if~$G$ and~$G'$ are $2$-isomorphic.
  In particular, in the affirmative case $T$ is also a spanning tree of $G'$.
\end{proposition}

Thus, in order to represent the set of graphs represented by $M$ we could, in principle, maintain one graph $G$ that realizes $M$ and maintain all its 2-separations, and thus its possible reversals.
This is done by the fast column-wise algorithms for graph realization proposed by Bixby and Wagner~\cite{BixbyWagner1988} and Fujishige~\cite{Fujishige1980} by means of so-called \emph{t-decompositions} and \emph{PQ-graphs}, respectively.
The former relies on fundamental work of Cunningham and Edmonds~\cite{Cunningham1980}.
A variant of the t-decomposition appears in the more modern literature under the name \emph{SPQR-tree} or \emph{SPQR-decomposition}~\cite{DiBattistaT89,DiBattistaT90,GutwengerM01,PatrizioBR13}. Most of the recent literature considering SPQR-trees seems to only consider the setting of computing or using an SPQR-tree for a specific graph $G$. In contrast, only the column augmentation algorithms~\cite{BixbyWagner1988, Fujishige1980}, and a more recent work by Angelini, Bl{\"a}sius and Rutter on the mutual duality problem for planar graphs~\cite{PatrizioBR13}, use SPQR-trees to represent multiple graphs at once and use Whitney's reversals.

We will stick to that notion, but augment the represented $2$-isomorphic graphs $G$ with a spanning tree $T \subseteq E(G)$.
Moreover, in order to keep the presentation simple, our notion of an SPQR-tree is undirected.

\begin{figure}[htpb]
\usetikzlibrary{positioning}
    \usetikzlibrary{calc}
    \begin{subfigure}[b]{0.45\textwidth}
    \centering
    $\begin{blockarray}{cccccccccc}
                 & \color{blue}{g} & \color{blue}{h} & \color{blue}{i} & \color{blue}{j} & \color{blue}{k} & \color{blue}{l} & \color{blue}{m} & \color{blue}{n} & \color{blue}{o}\\
    \begin{block}{c(ccccccccc)}
    \color{red}{a} & 1 & 1 & 0 & 1 & 0 & 0 & 0 & 0 & 0 \\
    \color{red}{b} & 1 & 1 & 1 & 1 & 0 & 0 & 0 & 0 & 0 \\
    \color{red}{c} & 1 & 1 & 1 & 0 & 1 & 0 & 0 & 0 & 0 \\
    \color{red}{d} & 1 & 1 & 1 & 0 & 0 & 1 & 0 & 0 & 0 \\
    \color{red}{e} & 1 & 1 & 1 & 0 & 0 & 0 & 1 & 1 & 1 \\
    \color{red}{f} & 0 & 0 & 0 & 1 & 0 & 0 & 0 & 0 & 0 \\
    \end{block}
    \end{blockarray}$
    \vspace{47pt}
    \subcaption{The graphic matrix $M$.}
    \end{subfigure}
    \begin{subfigure}[b]{0.19\textwidth}
    \centering
        \begin{tikzpicture}
            \draw[behindrect] (-0.5,0.5) rectangle (2.15, -3.0);
    \node[main] (A) {};
    \node[main] (B) [below = of A]{};
    \draw[cotree] (A)  to [bend right] (B) ;
    \draw[cotree] (A) -- (B) node[midway, rotate=0, right=-0.1cm] {$h$} node[midway, left=0.1cm, rotate=0] {$g$};

    \node[main] (5) [below right = 0.707cm of A]{};
    \node[main] (6) [right = of B]{};

    \draw[cotree] (B) -- (5) node[midway,below right=-0.15cm] {$i$};
    \draw[tree] (5) -- (6) node[midway,above right=-0.15cm] {$b$};
    \draw[tree] (A) -- (5) node[midway,above right=-0.15cm] {$a$};

    \node[main] (19) [right = of A]{};
    \draw[tree] (6) -- (19) node[midway,right=-0.1cm] {$f$};
    \draw[cotree] (A) -- (19) node[midway,above=-0.1cm] {$j$};

    \node[main] (7) [below = of B]{};
    \node[main] (8) [right = of 7]{};

    \draw[cotree] (7)  to [bend left] (B) ;
    \draw[tree] (7) -- (B) node[midway, rotate=0,right=-0.1cm] {$c$} node[midway, left=0.1cm, rotate=0,blue] {$k$};

    \draw[cotree] (6)  to [bend left] (8) ;
    \draw[cotree] (6)  to [in=0,out=0,looseness = 1.2] (8) ;
    \draw[cotree] (6)  to [in=0,out=0,looseness = 2.0] (8) ;

    \draw[tree] (6) -- (8) node[midway, rotate=0,left=-0.1cm] {$e$} node[midway, right= 0.02cm, blue] {$m$} node[midway, right=0.37cm, blue] {$n$} node[midway, right=0.64cm, blue] {$o$};

    \draw[cotree] (7) to [bend right] (8);
    \draw[tree] (7) -- (8) node[midway, rotate=0,above=-0.1cm] {$d$} node[midway, below=0.1cm, rotate=0,blue] {$l$};

    \end{tikzpicture}
    \vspace{1.0cm}
    \subcaption{A realization\\ $(G,T)$ of $M$.}
    \end{subfigure}
    \begin{subfigure}[b]{0.35\textwidth}
        \centering
        \begin{tikzpicture}
                \draw[behindrect] (-0.6,0.5) rectangle (4.6, -5.3);
    \node[main] (A) {};
    \node[main] (B) [below = of A]{};
    \draw[cotree] (A)  to [bend right] (B) ;
    \draw[cotree] (A) -- (B) node[midway, rotate=0,left=-0.15cm] {$h$} node[midway, left=0.12cm, rotate=0] {$g$};
    \draw[virtualtree] (A) to [bend left] (B);

    \node[main] (3) [right = of A]{};
    \node[main] (4) [below = of 3]{};
    \node[main] (5) [below right = 0.707cm of 3]{};
    \node[main] (6) [right = of 4]{};

    \draw[virtualcotree] (3) -- (4);
    \draw[virtualtree] (4) -- (6);
    \draw[cotree] (4) -- (5) node[midway,below right =-0.2cm] {$i$};
    \draw[tree] (5) -- (6) node[midway,below left=-0.2cm] {$b$};
    \draw[tree] (3) -- (5) node[midway,below left=-0.2cm] {$a$};
    \draw[virtualcotree] (3) to [bend left] (6);

        \node[main] (17) [right = of 6]{};
    \node[main] (18) [above = of 17]{};
    \node[main] (19) [below right = 0.707cm of 18]{};
    \draw[virtualtree] (17) to (18);
    \draw[tree] (17) -- (19) node[midway,below right=-0.2cm] {$f$};
    \draw[cotree] (18) -- (19) node[midway,above right=-0.1cm] {$j$};

    \node[main] (7) [below = of 4]{};
    \node[main] (8) [right = of 7]{};
    \node[main] (9) [below = of 8]{};
    \node[main] (10) [left = of 9]{};
    \draw[virtualcotree] (7) -- (8);
    \draw[virtualtree] (8) -- (9);
    \draw[virtualtree] (9) -- (10);
    \draw[virtualtree] (10) -- (7);

    \node[main] (11) [left = of 7]{};
    \node[main] (12) [below = of 11]{};
    \draw[cotree] (11)  to [bend right] (12) ;
    \draw[tree] (11) -- (12) node[midway, rotate=0, left=-0.11cm] {$c$} node[midway, left=0.07cm, rotate=0,blue] {$k$};
    \draw[virtualcotree] (11) to [bend left] (12);

    \node[main] (13) [right = of 8]{};
    \node[main] (14) [below = of 13]{};
    \draw[cotree] (13)  to [bend left] (14) ;
    \draw[cotree] (13)  to [in=0,out=0,looseness = 1.2] (14) ;
    \draw[cotree] (13)  to [in=0,out=0,looseness = 2.0] (14) ;

    \draw[tree] (13) -- (14) node[midway, rotate=0,left=-0.1cm] {$e$} node[midway, right= 0.02cm, blue] {$m$} node[midway, right=0.37cm, blue] {$n$} node[midway, right=0.64cm, blue] {$o$};

    \draw[virtualcotree] (13) to [bend right] (14);

    \node[main] (15) [below = of 10]{};
    \node[main] (16) [right = of 15]{};
    \draw[cotree] (15) to [bend right] (16);
    \draw[tree] (15) -- (16) node[midway, rotate=0,above=-0.1cm] {$d$} node[midway, below=0.05cm, rotate=0,blue] {$l$};
    \draw[virtualcotree] (15) to [bend left] (16);

    \draw[black] ($(A.north)+(-0.4,0.3)$) rectangle ($(B.south)+(0.4,-0.3)$);
    \draw[black] ($0.5*(A.north)+0.5*(B.south) + (0.4,0.0)$) -- ($0.5*(3.north)+0.5*(4.south)+ (-0.4,0.0)$) ;
    \draw[black] ($(3.north)+(-0.4,0.3)$) rectangle ($(6.south)+(0.4,-0.3)$);
    \draw[black] ($0.5*(4.south)+0.5*(6.south) + (0.0,-0.3)$) -- ($0.5*(7.north)+0.5*(8.north)+ (0.0,0.3)$) ;

    \draw[black] ($0.5*(9.south)+0.5*(10.south) + (0.0,-0.3)$) -- ($0.5*(15.north)+0.5*(16.north)+ (0.0,0.4)$) ;

    \node[black] (Box1) at ($(A.north) + 0.5*(0.0,0.3)$) {\bf{P}};
    \node[black] (Box2) at ($(3.north) + 0.5*(2.2,0.3)$) {\bf{R}};
    \node[black] (Box3) at ($(8.north) + 0.5*(0.0,0.3)$) {\bf{S}};
    \node[black] (Box4) at ($(11.north) + 0.5*(0.0,0.3)$) {\bf{P}};
    \node[black] (Box5) at ($(13.north) + (0.5,0.15)$) {\bf{P}};
    \node[black] (Box6) at ($(16.north) + 0.5*(0.0,0.5)$) {\bf{P}};

    \draw[black] ($(7.north)+(-0.4,0.3)$) rectangle ($(9.south)+(0.4,-0.3)$);
    \draw[black] ($0.5*(11.north)+0.5*(12.south) + (0.4,0.0)$) -- ($0.5*(7.north)+0.5*(10.south)+ (-0.4,0.0)$) ;
    \draw[black] ($0.5*(11.north)+0.5*(12.south) + (0.4,0.0)$) -- ($0.5*(7.north)+0.5*(10.south)+ (-0.4,0.0)$) ;
    \draw[black] ($0.5*(8.north)+0.5*(9.south) + (0.4,0.0)$) -- ($0.5*(13.north)+0.5*(14.south)+ (-0.4,0.0)$) ;
    \draw[black] ($(11.north)+(-0.4,0.3)$) rectangle ($(12.south)+(0.4,-0.3)$);

    \draw[black] ($(13.north)+(-0.4,0.3)$) rectangle ($(14.south)+(0.95,-0.3)$);

    \draw[black] ($(15.north)+(-0.4,0.4)$) rectangle ($(16.south)+(0.4,-0.4)$);

    \draw[black] ($0.5*(17.south)+0.5*(18.north) -(0.4,0.0)$) -- ($0.5*(3.north)+0.5*(4.south)+ (1.6,0.0)$) ;
    \draw[black] ($(18.north)+(-0.4,0.3)$) rectangle ($(17.south)+(0.95,-0.3)$);
    \node[black] (Box7) at ($(18.north) + (0.5,0.15)$) {\bf{S}};

    \end{tikzpicture}
    \subcaption{The unique minimal SPQR-tree $\T$ associated to $M$.}
    \end{subfigure}

    \caption{%
      A graphic matrix with a realization and its unique minimal SPQR-tree.
      Tree edges are marked in bold and red, co-tree edges are marked in blue and virtual edges are dashed.
    }
    \label{SPQRTreeExampleFigure}
\end{figure}
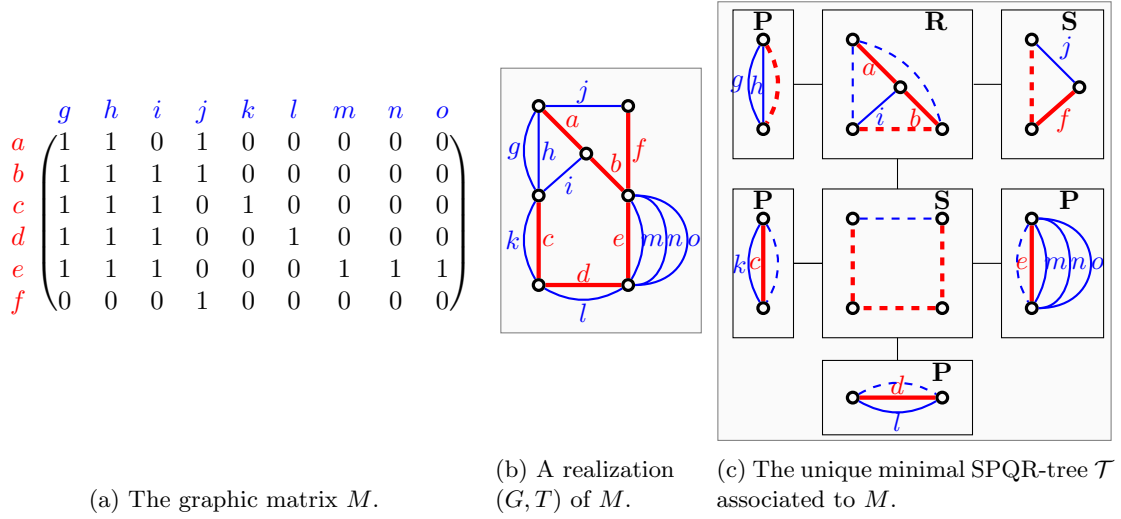

\begin{definition}[SPQR-tree, virtual edges, regular edges]
\label{def_SPQR_tree}
  An \emph{SPQR-tree} is a tree $\T = (\V, \E)$ that is supported by a map $\pi$.
  For every node $\mu \in \mathcal{V}$ there is an associated connected multi-graph $G_\mu$ without loops, called the \emph{skeleton}, and a spanning tree $T_\mu$ of $G_\mu$.
  We assume that the vertex sets of all skeletons are pairwise disjoint.
  We denote by $E(\T)$ the union of the edges $E(G_\mu)$ over all $\mu \in \mathcal{V}$, and by $V(\T)$ the union of the vertices  $V(G_\mu)$ over all $\mu \in \mathcal{V}$.

  A subset $\spqrVirtual(\T) \subseteq E(\T)$ of size $|\spqrVirtual(\T)| = 2 |\E|$ of the edges is called \emph{virtual}:
  for every edge $\{\mu,\nu\} \in \E$ of the SPQR-tree there is a pair of virtual edges $e \in E(G_\mu)$ and $f \in E(G_\nu)$.
  This relationship is specified by means of the map $\pi : E(\T) \to E(\T) \cup \{\emptyset\}$, where $\pi(e) = f$ and $\pi(f) = e$ indicates that $e \in E(G_\mu)$ and $f \in E(G_\nu)$ are the two virtual edges corresponding to the edge $\{\mu,\nu \} \in \E$, and where $\pi(e) = \emptyset$ holds whenever $e$ is not a virtual edge.
  Every virtual edge corresponds to exactly one edge in $\E$.
  The remaining edges $\spqrNonvirtual(\T) \coloneqq E(\T) \setminus \spqrVirtual(\T)$ are called \emph{regular}.
  By $E^{\textup{reg}}_\mu(\T)$ we denote all regular edges $E(G_\mu) \cap \spqrNonvirtual(\T)$.
  For each virtual edge pair $e,f$, either $e \in T_\mu$ or $f \in T_\nu$ holds, i.e., exactly one of them is part of the respective spanning trees.
  Moreover, each skeleton $G_\mu$ is of exactly one of four types:
  \begin{enumerate}[label=(\Alph*),ref=\Alph*]
  \setcounter{enumi}{18}
  \item
    \label{node_S}
    $G_\mu$ is a cycle of length at least 3; $\mu$ is called \emph{series}.
  \setcounter{enumi}{15}
  \item
    \label{node_P}
    $G_\mu$ has exactly two vertices and at least 3 edges; $\mu$ is called \emph{parallel}.
  \setcounter{enumi}{16}
  \item
    \label{node_Q}
    $G_\mu$ has at most~2 vertices and at most 2 edges, and $G_\mu$ is the only skeleton, i.e., $\mathcal{V} = \{\mu\}$.
  \setcounter{enumi}{17}
  \item
    \label{node_R}
    $G_\mu$ is simple, $3$-connected and has at least $4$ edges; $\mu$ is called \emph{rigid}.
  \end{enumerate}
  An SPQR-tree is called \emph{minimal} if its~\eqref{node_S}-nodes are pairwise non-adjacent and its \eqref{node_P}-nodes are pairwise non-adjacent.
\end{definition}
An example of a graphic matrix and its associated SPQR-tree is depicted in \cref{SPQRTreeExampleFigure}.

To avoid confusion, we use \emph{node} when referring to some $\mu\in\V(\T)$ for some SPQR-tree $\T$, and we use \emph{vertex} for the represented graphs.
The above definition differs slightly from that of Gutwenger and Mutzel~\cite{GutwengerM01}.
In particular, we do not consider each non-virtual edge to have its unique leaf node of type \eqref{node_Q}, but rather only use \eqref{node_Q} when the graph has at most two edges, which is closer to the definition of $t$-decompositions as used by Bixby and Wagner~\cite{BixbyWagner1988}.

To see how an SPQR-tree $\T$ encodes a family of $2$-isomorphic graphs, consider one such graph $G$.
First, $E(G)$ consists of all the regular edges (of all skeletons) of $\T$.
Second, an edge $\{\mu,\nu\}$ with virtual edges $e \in E(G_\mu)$ and $f \in E(G_\nu)$ represents a 2-separation $(E_\mu,E_\nu)$ of $G$, where $E_\mu$ (resp.\ $E_\nu$) consists of all regular edges of skeletons that are closer to $\mu$ than to $\nu$ (resp.\ closer to $\nu$ than to $\mu$).
In \cref{SPQRTreeExampleFigure}, an example of such a 2-separation is the one defined by $E_1 = \{a,b,f,g,h,i,j\}$ and $E_2 = E\setminus E_1$.

Although the edges of the SPQR-tree encode many of the possible 2-separations of its represented graphs, there may still be additional 2-separations that exist within nodes of type~\eqref{node_S} and~\eqref{node_P}.
We will call a 2-separation $(E_1,E_2)$ of any graph represented by $\T$ \emph{local} if $E_i \subseteq \spqrNonvirtual_\mu(\T)$ holds for some $\mu \in \V(\T)$ and some $i \in \{1,2\}$.
In other words, local 2-separations are those for which one side is contained entirely in a single skeleton of the SPQR-tree. In \cref{SPQRTreeExampleFigure}, an example of such a 2-separation is the one defined by $E_1 = \{n,o\}$ and $E_2 = E\setminus E_1$.

Note that every pair of non-adjacent vertices of a cycle are the separating vertices of a 2-separation.
A corresponding reversal yields some reordering of the cycle edges.
In fact, any permutation of a cycle's edges yields a $2$-isomorphic cycle.
These observations motivate the following definition.

\begin{definition}[Graphs and trees represented by an SPQR-tree]
  \label{def_spqr_represented}
  Let $\T$ be an SPQR-tree supported by the map $\pi$.
  A graph-tree pair $(G,T)$ is \emph{represented by $\T$} if it can be obtained by the following steps:
  \begin{enumerate}
  \item
    \label{def_spqr_represented_permute}
    For each $\mu \in \V(\T)$ of type~\eqref{node_S}, permute the edges of the cycle $G_\mu$ arbitrarily.
  \item
    \label{def_spqr_represented_merge}
    Contract each edge $\{\mu,\nu\} \in \E(\T)$ (in any order) to a new node $\xi$.
    Let $e \in E(G_\mu)$ and $f \in E(G_\nu)$ be the two corresponding virtual edges.
    The skeleton $G_\xi$ is the graph obtained from $G_\mu$ and $G_\nu$ by identifying the two end-vertices of $e$ with the two end-vertices of $f$ (via an arbitrary bijection) and then removing $e$ and $f$.
    Define $T_\xi$ to be $(T_\mu \cup T_\nu) \setminus \{e,f\}$.
    Define $\Phi_G : V(\T)\to V(G)$ to be the mapping from the skeleton vertices to the resulting graph vertices, where initially $\Phi_G(v) = v$ holds.
    For the vertices $w_e \in e$ and $w_f \in f$ that are identified with one another into a new vertex $v'$, we update $\Phi_G$ for all $u$ where $\Phi_G(u) = w_e$ or $\Phi_G(u) = w_f$ holds to $\Phi_G(u) = v'$, instead.
  \end{enumerate}
\end{definition}

Note that in Step 2 of the definition, since either $e \in T_\mu$ or $f \in T_\nu$ holds, exactly one of the edges $e,f$ is removed from $T_\mu \cup T_\nu$.
This (inductively) ensures that $T_\xi$ is a spanning tree of $G_\xi$.
By construction, any two graphs $G, G'$ that are represented by $\T$ are $2$-isomorphic.
Moreover, every 2-separation of a represented graph either corresponds to an edge of $\T$ or to a 2-separation of an~\eqref{node_S}- or a \eqref{node_P}-node. Cunningham and Edmonds have shown this more formally:
\begin{proposition}[Consequence of Cunningham and Edmonds~\cite{Cunningham1980}, Theorem 3]
\label{thm_spqr_tree_2sep_local_edge}
Let $G$ be a graph represented by $\T=(\V,\E)$, and let $(E_1,E_2)$ be a 2-separation of $G$. Then, $(E_1,E_2)$ is either a local 2-separation for some $\mu\in\V(\T)$, or there exists an SPQR-tree edge $\epsilon=\{\mu,\nu\}\in \E$ such that the subtrees $\T_\mu$ and $\T_\nu$ formed by removing $\epsilon$ from $\T$ satisfy $E_1 = \spqrNonvirtual(\T_\mu)$ and $E_2 = \spqrNonvirtual(\T_\nu)$.
\end{proposition}

In the literature, SPQR-trees are defined without the spanning tree $T_\mu$ of each skeleton $G_\mu$.
However, the last statement of \cref{thm_realizations_2isomorphic} shows that the SPQR-tree structure is independent of the spanning tree.
By results from Hopcroft and Tarjan~\cite[Lemma~2]{HopcroftT73} and Cunningham and Edmonds~\cite[Theorem~1]{Cunningham1980}, minimal SPQR-trees are unique.
The discussion above and \cref{thm_realizations_2isomorphic} together are summarized in the main result of this section.

\begin{theorem}
  \label{thm_spqr_realizations}
  Let $A$ be a connected graphic matrix.
  Then there exists a unique minimal SPQR-tree $\T$ for which the set of represented graph-tree pairs $(G,T)$ is equal to the set of graph-tree pairs $(G,T)$ for which $A = M(G,T)$ holds.
\end{theorem}

The theorem in particular shows that the resulting SPQR-tree $\T$ is independent of the choice of $(G,T)$ (among those represented by $\T$).
Consequently, if we construct a minimal SPQR-tree $\T$ from any graph-tree pair $(G,T)$ for which $A = M(G,T)$ holds, then we can (in principle) obtain all other graph-tree pairs $(G',T')$ for which $A = M(G',T')$ holds by enumerating all graph-tree pairs that are represented by $\T$.

In the next sections we will explain how to update an SPQR-tree for a matrix $M$ when augmenting it with a new row.

\paragraph{Data structures.}
We now introduce the basic data structures that we use to represent the SPQR-tree and some basic results that are necessary for the running time analysis.
As we aim to make our row-wise algorithm compatible with the column-wise algorithm formulated by Bixby and Wagner~\cite{BixbyWagner1988}, the merging of two skeletons $G_\mu$ and $G_\nu$ should happen in a minimum amount of time.
In particular, Bixby and Wagner use \emph{disjoint set} data structures~\cite{Tarjan75} (also referred to as \emph{union-find}) and doubly linked-lists in order to represent the skeletons $G_\mu$.
A sequence of $k$ disjoint set operations on a set with $n$ elements can be done in $\orderO(m \alpha(k,n))$ time~\cite{Tarjan1984}.
This can be seen as being `almost linear' in $k$ since in practice, $\alpha(k,n)\leq 4$ holds for any realistic input that can be described on a computer.

We remark that Fink and Rutter~\cite{FinkR23} investigate an expansion operation of a vertex into a graph, which generalizes the splitting of a single vertex, and show that it can be performed in time linear to the degree of the vertex. 
We do not use this result since their definition of expansion does not apply immediately to the vertex splitting represented here. 
We believe that their results may be adapted to also be sufficient for this work, but refrained from doing so to streamline our presentation.

In an implementation of the algorithms described in this paper, we use the following data structures:
\begin{itemize}
\item
  The matrix is assumed to be a compressed sparse matrix given in row-major format, so that we can access each row with $k$ non-zero entries in $\orderO(k)$ time.
\item
  There is an explicit mapping from matrix rows and columns to edge labels.
\item
  The SPQR-tree node labels $\V(\T)$ are given by a disjoint set data structure.
\item
  The skeleton vertex labels $V(G_\mu)$ are given by a disjoint set data structure over all vertices of the SPQR-tree $\T$.
\item
  The neighboring edges $\delta(v)$ of vertices $v\in G_\mu$ link to each other using a doubly-linked list.
\item
  Each edge stores the SPQR-tree node label it is initially located in.
  In order to find the edge's current skeleton, a \emph{find}-operation on the disjoint set data structure of the SPQR-tree node labels is carried out.
\item
  Each edge stores its initial end-vertex labels.
  Similarly to the above, a \emph{find}-operation is necessary to find the correct adjacent vertices.
\item
  For each skeleton, we store its type (among~\eqref{node_S},~\eqref{node_P},~\eqref{node_Q} and~\eqref{node_R})
\item
  Each skeleton stores its edges in $Y$ using a doubly-linked list.
\item We distinguish different SPQR-trees in the SPQR-forest using a disjoint set data structure over the SPQR-tree node labels.
\end{itemize}

Moreover, we pick an arbitrary root node and represent $\T$ as a rooted arborescence.
Each SPQR-tree node $\mu$, except for the root,
stores its parent $\nu$, and each `child' virtual edge $e$ stores the corresponding virtual edge $f$.
The corresponding child member can be found by using a \emph{find}-operation on the disjoint set data structure of SPQR-tree node labels.

Additionally, we note that it is not always necessary to explicitly maintain the skeleton graph.
In particular, the skeletons of type~\eqref{node_S}, \eqref{node_P} or~\eqref{node_Q} always have the same structure, which means that it is unnecessary to track the end-vertices of all edges. This is particularly convenient for~\eqref{node_S}-nodes, whose edges may be arbitrarily permuted.

Let~$m$ and~$n$ be the number of rows and columns of the matrix respectively.
First, we show that the size of an  SPQR-tree is linear with respect to the size of the graph that it represents. We can easily extend this notion to SPQR-forests, and show that the SPQR-forest representing a $m\times n$ matrix always has linear size.
Note that any realization of a graphic $m \times n$ matrix is a graph $G = (V,E)$ with $|E| = m + n$ edges.

\begin{proposition}[Hopcroft \& Tarjan~\cite{HopcroftT73}]
  \label{thm_SPQRTreeEdgeBound}
  Let $\T = (\V, \E)$ be a minimal SPQR-tree of a 2-connected graph $G = (V, E)$ with $|E|\geq 3$.
  Then, the total number $\sum_{\mu \in \V } |E_\mu|$ of edges in the skeletons of $\T$ is at most $3|E| - 6$.
\end{proposition}

\begin{proposition}
  \label{thm_SPQRTreeNodeAndSkeletonBound}
  Let $\T= (\V, \E)$ be a minimal SPQR-tree of a $2$-connected graph $G=(V,E)$ with $|E|\geq 3$. Then, we have $|\V| \leq |E|-2$ for the number of skeletons of $\T$. For the total number of vertices in the skeletons of $\T$, $\sum_{\mu\in\V} |V_\mu|\leq 3|E| - 6$ holds.
\end{proposition}

\begin{proof}
  Since $|E|\geq 3$ holds, $\T$ consists only of nodes that are not of type \eqref{node_Q}. Thus, each skeleton $G_\mu = (V_\mu, E_\mu)$ then requires $|E_\mu|\geq 3$ by definition.
  For the first point, \cref{thm_SPQRTreeEdgeBound} implies that $|\V| \leq \frac{3|E|-6}{3} = |E| -2$.
  For the second point, we have that $|V_\mu|\leq |E_\mu|$ since $G_\mu$ is $2$-connected and $|E_\mu|\geq 3$. Then, \cref{thm_SPQRTreeEdgeBound} implies that
  $\sum_{\mu\in\V} |V_\mu| \leq \sum_{\mu\in\V} |E_\mu| \leq 3|E|-6$.
\end{proof}

Using the above-mentioned data structures and bounds on the SPQR-tree size, we consider basic operations and their time complexities:
\begin{itemize}
\item
  Split a vertex $v$ into two vertices and reassign its edges in $\orderO(|\delta(v)|)$ time.
\item
  Perform $k$ merges of two vertices into one (identifying them) in $\orderO(k\alpha(k,m+n))$ time.
\item
  Perform $k$ merges of two SPQR-tree node labels into one label in $\orderO(k\alpha(k,m+n))$ time.
\item
  Perform $k$ queries of the end-vertices of any edge $e$ in $\orderO(k\alpha(k,m+n))$time.
\item
  Add a vertex or edge to a skeleton in $\orderO(1)$.
\item
  Add a new node with an empty skeleton to $\T$ in $\orderO(1)$ time.
\item
  Given a virtual edge $e\in G_\mu$ that is paired with $f\in G_\nu$, find $\nu$ in $\orderO(k\alpha(k,m+n))$ time for $k$ queries.
\item
  Find the SPQR-tree $\T$ that contains an SPQR-forest node $\mu \in V(\mathcal{F})$ in $\orderO(k\alpha(k,m+n))$ time for $k$ queries.
\end{itemize}

Note that for many operations, we incur a $\alpha(k,m+n)$ overhead for $k$ queries, that depends on the size of the whole matrix $(m+n)$, even though the size of the SPQR-tree in which they occur might be smaller, if the matrix has multiple blocks. In particular, this is because the used union-find data structures represent the whole matrix, and not just a single SPQR-tree.

\section{High-level algorithm}
\label{sec_high_level}

In order to develop an algorithm that correctly determines graphicness of row augmentation, we need to combine the results of the previous sections.
First, recall that in \cref{sec_general_row_augmentation} we characterized graphic row augmentation by means of determining whether a given graph has a $Y$-splittable vertex.
Second, in \cref{sec_spqr_trees} we observed that the set of graphs represented by connected graphic matrices are exactly the set of graphs represented by a minimal SPQR-tree.
Combining these two ideas, the row augmentation problem then becomes to determine if the SPQR-tree represents a graph that has a $Y$-splittable vertex, and if so, to find the unique minimal SPQR-tree of the matrix after it has been updated by splitting a $Y$-splittable vertex in one of the represented graphs.

Note that, by \cref{thm_spqr_realizations}, it is sufficient to show that a single realization $G$ of the original SPQR-tree has a $Y$-splittable vertex: we could find all represented graphs of $M'$ by computing the unique minimal SPQR-tree of $G'$, where $G'$ is obtained by splitting said $Y$-splittable vertex in $G$.
In our algorithm we do not recompute the SPQR-tree of $G'$ from scratch, but instead construct it by modifying the SPQR-tree of $G$.
To establish the correctness of our algorithm, we additionally argue that the updated SPQR-tree is minimal.
In case of a non-graphic row augmentation we need to show that none of the graphs represented by the SPQR-tree contains a $Y$-splittable vertex, which then by \cref{thm_spqr_realizations} shows that $M'$ is not graphic.

We start with a few structural results relating the concepts of \cref{sec_general_row_augmentation} to $2$-connected graphs.
We first show that the bipartition of $H^v_Y$ is unique for any $Y$-splittable vertex $v$, which shows that the resulting neighborhood split is also unique.

\begin{lemma}
  \label{thm_2connected_aux_connected_unique}
  Let $G = (V, E)$ be a 2-connected multigraph with spanning tree $T \subseteq E$, let $Y \subseteq E \setminus T$ and $v \in V$.
  Then $H^v_Y$ is a connected graph.
  Moreover, if $H^v_Y$ is bipartite then it has a unique bipartition.
\end{lemma}

\begin{proof}
  Let $C$ be a connected component of $H^v_Y$ and let $U \coloneqq \bigcup_{h \in V(C)} V(h)$ be the vertices of $G^v_Y$ that belong to any vertex of $H^v_{Y}$ belonging to $C$.
  Since $G$ is $2$-connected, $G[V \setminus \{v\}]$ is connected.
  This implies that either $W \coloneqq V \setminus (U \cup \{v\})$ is empty or that, by connectivity, $E$ contains an edge $e = \{u,w\}$ with $u \in U$ and $w \in W$.
  By construction, we would have $e \in Y$, which would contradict the assumption that $C$ is a connected component of $H^v_Y$.
  We conclude that $W = \emptyset$ holds, which proves that $C$ is the only connected component of $H^v_Y$.

  The second statement clearly follows from bipartiteness and connectivity of $H^v_Y$.
\end{proof}

One of the main difficulties in finding a realization of the SPQR-tree with a $Y$-splittable vertex is that the SPQR-tree can have exponentially many realizations, as shown in \Cref{fig_example_many_graphs_few_realizations}.
Thus, we cannot hope to efficiently identify realizations that have $Y$-splittable vertices by enumerating all represented graphs and determining if they have $Y$-splittable vertices.
The central idea to our approach is that the identification of a $Y$-splittable vertex in a given realization can be decomposed into identifying $Y$-splittable vertices in each of its skeleton graphs.
A key part of this approach is to understand how $Y$-splittable vertices interact with 2-separations.

Consider a $2$-connected graph $G$ that has a 2-separation $(E_1,E_2)$ with separating vertices $\{u,w\}$.
Then, for a vertex $v\in V(G)$ either all its neighbors are contained in $E_1$, in $E_2$, or $v \in \{u,w\}$ holds.
In all cases, we aim to find a $Y$-splittable vertex in the realization in terms of $Y$-splittable vertices in the skeleton graphs defined by $E_1$ and $E_2$.
In \cref{thm_shared_splittable_combine} we show how the case $v\in \{u,w\}$ is reduced to two cases corresponding to the two sides of the 2-separation.

\begin{lemma}
  \label{thm_shared_splittable_combine}
  Let $G$ be a $2$-connected multigraph that has a 2-separation $(E_1,E_2)$ with separating vertices $\{u,w\}$.
  For $i=1,2$, let $G_i$ denote the graph with vertex set $V(E_i)$ and edge set $E_i \cup \{e_i\}$ for a new edge $e_i \coloneqq \{u,w\}$, and let $Y_i \coloneqq Y \cap E_i$.
  Then $v \in \{u,w\}$ is $Y$-splittable in $G$ if and only if $v$ is $Y_i$-splittable in $G_i$ for $i \in \{1,2\}$.
\end{lemma}

\begin{proof}
  Since the statement of the lemma is symmetric in $u$ and $w$, we only show it for $v = u$.
  First, observe that $G - v$ has the articulation vertex $w$ and that $G - v$ can be obtained from $G_1 - v$ and $G_2 - v$ by taking their union and identifying the vertex $w$ that belongs to both graphs.
  Let $C$, $C_1$ and $C_2$ denote the sets of connected components of $G^v_Y = (G \setminus Y) - v$, $(G_1 \setminus Y_1) - v$ and $(G_2 \setminus Y_2) - v$, respectively.
  Moreover, let $h_w$, $h_w^1$ and $h_w^2$ denote the component of $C$, $C_1$ and $C_2$, respectively, that contains the vertex $w$.
  From $V(G_1 - v) \cap V(G_2 - v) = \{w\}$ we obtain that $C \setminus \{h_w\}$ is the disjoint union of $C_1 \setminus \{h_w^1\}$ and $C_2 \setminus \{h_w^2\}$.
  Finally, each $y \in Y$ belongs to some $E_i$ for some $i \in \{1,2\}$ and thus connects only components that both belong to $C_i$.
  This shows that in the auxiliary graph $H^v_Y(G)$, the component $h_w$ is also an articulation vertex (or contains $V(G_i)$ for some $i \in \{1,2\}$).
  We conclude that $H^v_Y(G)$ is bipartite if and only if $H^v_{Y_i}(G_i)$ is bipartite for both $i=1,2$, which yields the claimed result.
\end{proof}

An important observation about \cref{thm_shared_splittable_combine} is that if $u$ and $w$ are both $Y$-splittable for $G_2$, then $Y$-splittability of a vertex $v\in \{u,w\}$ only depends on the $Y$-splittability of $v$ in $G_1$.
Thus, in this case the computation of the $Y$-splittable vertex can be \emph{reduced} to computing a $Y$-splittable vertex in $G_1$.
Moreover, in the case where $u$ or $w$ is $Y$-splittable in $G_1$,  we can see that both orientations of the 2-separation yield a realization that contains a $Y$-splittable vertex.

In fact, we will show that the graph $G'$, obtained from $G$ by splitting the $Y$-splittable vertex in $G$, can be obtained by considering $G_1$ and $G_2$ individually. If we split the $Y$-splittable vertex in $G_1$ to obtain $G'_1$, and subsequently reverse the reduction on $G'_1$ by replacing the inserted virtual edge with $G_2$ again, we will show that the resulting graph is $G'$.
Although the above discussion only dealt with the case where the $Y$-splittable vertex is on the 2-separation boundary, we will show that the application of the proposed reduction ensures that the existence of $Y$-splittable vertices is preserved more generally.

In \cref{sec_augmentation_reductions}, we discuss the proposed reduction in further detail.
We divide the reduction into three cases based on the tree structure and conditions on the $Y$-edges in $E_2$.
For each case we prove that:
\begin{enumerate}
\item
  The reduced graph has a $Y$-splittable vertex if and only if the original graph has a $Y$-splittable vertex.
\item
  The updated graphic realization can be found by splitting a $Y$-splittable vertex in the reduced graph and reversing the reduction.
\end{enumerate}
Since the reduction is only applicable if both $u$ and $w$ are $Y$-splittable in $G_2$, the reduction is valid for both graphic realizations represented by the two orientations of 2-separation.
This is very convenient, as it means that the reduction can be directly applied to the SPQR-tree, effectively performing the reduction in multiple graphs at once.
In the first step of our algorithm, we iteratively consider the leaf nodes of the SPQR-tree.
If a leaf node can be reduced, then we carry out the reduction and remove it from the SPQR-tree.
By iteratively considering the leaf nodes of the tree, we shrink the tree and exhaustively apply the reduction.
The remaining SPQR-tree obtained in this way is called the \emph{reduced SPQR-tree} and is denoted by $\T_R$.

In the remainder of the algorithm, our goal is to find the $Y$-splittable vertices in the reduced SPQR-tree.
In order to achieve this, we need algorithms that can efficiently find $Y$-splittable vertices in the skeletons of each type.
These algorithms are also important to find the reductions described above efficiently.
In \cref{sec_augmentation_singleskeleton}, we present algorithms that can identify the $Y$-splittable vertices for each skeleton graph.
For the skeletons of type~\eqref{node_S},~\eqref{node_P} or~\eqref{node_Q}, this is relatively simple.
Most of the results in this section consider nodes of type~\eqref{node_R}, which is the most complicated case.
In particular, we show that for each node of type~\eqref{node_R}, it is only necessary to construct the auxiliary graph for a constant number of candidate vertices.
Furthermore, in \cref{sec_augmentation_singleskeleton} we explain how to update the reduced SPQR-tree if it consists of only a single node.

In \cref{sec_augmentation_merging} we tackle the case where $\T_R$ contains at least two nodes.
Consider a 2-separation structure as in \cref{thm_shared_splittable_combine}.
Since we applied all the reductions exhaustively, at most one of $u$ and $w$ is $Y$-splittable in $G_1$ and in $G_2$.
In fact, we show that on both sides, exactly one of the vertices $u, w$ must be $Y$-splittable.
In particular, we show that a vertex $v$ in a graph $G_R$ realized by $\T_R$ is $Y$-splittable if and only if the following hold:
\begin{enumerate}
\item
  $v$ is one of the 2-separation vertices of \emph{every} 2-separation of $G_R$, and
\item
  $v$ is $Y$-splittable in the skeleton of every node of $\T_R$.
\end{enumerate}
The first condition requires that in $\T_R$, the vertices that map to $v$ must be adjacent to all virtual edges of the skeleton.
Throughout the proofs for these claims, we extensively use the fact that $\T_R$ is reduced.
Moreover, we show that the graph $G'_R$ obtained from splitting a unique $Y$-splittable vertex $v$ in $G_R$ is $3$-connected and simple, and thus of type~\eqref{node_R}.
Furthermore, we will see that the $3$-connected graph $G'_R$ can be obtained by splitting $v$ in each skeleton individually and merging the skeleton graphs together by choosing the correct orientation of $\T_R$.

In \cref{sec_algorithm} we review the overall algorithm, including the reversal of the reductions.
Additionally, we handle a few technical details regarding minimality of the computed SPQR-tree and discuss the overall running time of the algorithm.

To summarize, the goal of the results and proofs in \cref{sec_augmentation_singleskeleton} -- \ref{sec_algorithm} is primarily to answer the following questions, that are necessary for the correctness, exactness and efficiency of our algorithm:
\begin{enumerate}
\item
  Do the proposed algorithms correctly identify a realization with a $Y$-splittable vertex, or prove that no such realization exists?
\item
  Do the proposed algorithms label the nodes of the updated SPQR-tree correctly? For example, we must show that changed~\eqref{node_R}-nodes indeed have simple $3$-connected skeleton graphs.
\item
  Is the updated SPQR-tree computed by the proposed algorithm a minimal SPQR-tree?
\item
  What is the running time of the proposed algorithm?
\end{enumerate}

\section{Reductions for graphic row augmentation}
\label{sec_augmentation_reductions}

In this section we present three reductions that preserve the existence of $Y$-splittable vertices and shrink the represented graph $G$.
Since we are interested in reductions that work not just on $G$ but also on its SPQR-tree, it is natural to consider 2-separations of the represented $2$-connected graphs.
All three proposed reductions work in a similar fashion.
For each reduction, we consider a 2-separation with a certain structure, and then replace one of its sides by a single edge, whilst preserving the existence of a splittable vertex in the remaining graph. Note that it is sufficient to guarantee that a single $Y$-splittable vertex is preserved, rather than preserving all $Y$-splittable vertices, since we can split it to obtain a new graphic realization, and obtain the SPQR-tree from the updated realization.
Although we formulate the results in terms of general graphs, the relevance of the following results on the SPQR-tree are due to the fact that every 2-separation is encoded in the SPQR-tree. Moreover, the proposed reductions do not depend on the orientation of the 2-separation; they work for both the original graph and its reversal, so that we can apply them directly to an SPQR-tree $\T$, simultaneously performing the reduction on all represented graphs of $\T$.

First, let us introduce some notation to define the structure of a graph-tree pair $(G,T)$ when considering a 2-separation of $G$.

\begin{definition}[tree partition induced by a 2-separation]
  \label{def_propagationStructure}
  Let $G$ be a $2$-connected multigraph with spanning tree $T$ and let $(E_1,E_2)$ be a 2-separation of $G$ with separating vertices $u$ and $w$
  and the corresponding graphs $G_1$ and $G_2$ such that $P_{u,w}(T) \subseteq E(G_1)$ holds.
  The \emph{tree partition of $T$ induced by the 2-separation} is the partition $T = T_1 \cup T_u \cup T_w$ such that $T_1 \coloneqq T \cap E(G_1)$ and $T_v \coloneqq \{ e \in T \cap E(G_2) \mid \text{$e$ and $v$ are in the same connected component of $T \setminus T_1$} \}$ for $v \in \{u,w\}$.
  The \emph{corresponding vertex partition} is $V(G_2) = V_u \cup V_w$ with $V_v \coloneqq V(T_v) \cup \{v\}$ for $v \in \{u,w\}$.
  The edges whose fundamental path contains $P_{u,w}(T)$ are denoted by
  \[
    P^{-1}_{u,w}(G,T) \coloneqq \{e \in E(G) \setminus T \mid P_{u,w}(T) \subseteq P_e(T)\} .
  \]
\end{definition}

Note that the definition above ensures for $v \in \{u,w\}$ that $T_v = \emptyset$ implies $V_v = \{v\}$.
Moreover, note that $P^{-1}_{u,w}(G,T)\cap E_2 $ consists of exactly those edges that connect $V_u$ to $V_w$ in $E_2$.

\begin{definition}[Reductions]
  \label{def_reductions}
  Let $G$ be a $2$-connected multigraph with spanning tree $T$ and let $Y \subseteq E(G) \setminus T$ be non-empty.
  Let $(E_1,E_2)$ be a 2-separation of $G$ with separating vertices $u$ and $w$ and the corresponding graphs $G_1$ and $G_2$ such that $P_{u,w}(T) \subseteq E_1$.
  A \emph{reduction} $G'$ at this 2-separation is one of the following three subgraphs.
  \begin{enumerate}[label=(\Alph*)]
  \item
    \label{def_reductions_A}
    If $Y \cap E_2 = \emptyset$ then $G'$ is $G_1$ augmented by a new edge $e = \{u,w\}$, $T' \coloneqq T \cap E_1$, and $Y' \coloneqq Y$.
  \item
    \label{def_reductions_B}
    If $Y \cap E_1 = \emptyset$ then $G'$ is $G_2$ augmented by a new edge $e = \{u,w\}$, $T' \coloneqq (T \cap E_2) \cup \{ e \}$, and $Y' \coloneqq Y$.
  \item
    \label{def_reductions_C}
    If $Y \cap E_1 \neq \emptyset$ and $Y \cap E_2 = P^{-1}_{u,w}(G,T) \cap E_2$ then $G'$ is $G_1$ augmented by a new edge $e = \{u,w\}$, $T' \coloneqq T \cap E_1$, and $Y' \coloneqq (Y \cap E_1) \cup \{ e \}$.
  \end{enumerate}
  The $2$-separation is \emph{reducible} if the conditions of any reduction are satisfied, and \emph{irreducible} otherwise.
  Moreover, the graph $G$ is called \emph{reducible} if it has a reducible 2-separation, and \emph{irreducible} otherwise.
\end{definition}

An overview of all three reductions can be found in \Cref{fig_reductions_overview}.
In subsequent lemmas we will show that reductions \ref{def_reductions_A}, \ref{def_reductions_B} and \ref{def_reductions_C} of \cref{def_reductions} preserve splittability, i.e., that $G'$ has a $Y'$-splittable vertex if and only if $G$ has a $Y$-splittable vertex.
The respective results are \cref{thm_empty_propagation_cotree,thm_empty_propagation_tree1,thm_full_propagation_cotree}.
Before establishing them, we will provide a basic auxiliary result.

\begin{figure}[htpb]
  \begin{subfigure}{0.45\textwidth}
    \centering
    \begin{tikzpicture}[node distance = 1.8 cm]
      \node[main] (1) {};
      \node[main] (2) [right of=1] {};
      \node[main] (3) [below right of=2] {};
      \node[main] (4) [above right of=2] {};
      \node[main] (5) [above right of=3] {};
      \node[main] (6) [right of=5]{};
      \draw[tree] (1) -- (2);
      \draw[marked] (1) -- (3);
      \draw[marked] (1) -- (4);
      \draw[tree] (2) -- (4);
      \draw[tree] (3) -- (2);

      \draw[cotree] (3) -- (5);
      \draw[tree] (6) -- (3);
      \draw[tree] (5) -- (6);
      \draw[cotree] (4) -- (6);
      \draw[cotree] (5) -- (4);
    \end{tikzpicture}
    \caption{%
      A case in which $Y \cap E_2 = \emptyset$ holds.
    }%
    \label{fig_reductions_overview_a}
  \end{subfigure}
\hfill
  \begin{subfigure}{0.45\textwidth}
    \centering
    \begin{tikzpicture}[node distance = 1.8 cm]
      \node[main] (1) {};
      \node[main] (2) [right of=1] {};
      \node[main] (3) [below right of=2] {};
      \node[main] (4) [above right of=2] {};
      \node[main,fill=white,draw=white] (5) [above right of=3] {};
      \node[main,fill=white,draw=white] (6) [right of=5]{};
      \draw[tree] (1) -- (2);
      \draw[marked] (1) -- (3);
      \draw[marked] (1) -- (4);
      \draw[tree] (2) -- (4);
      \draw[tree] (3) -- (2);
      \draw[virtualcotree] (3) -- (4);
    \end{tikzpicture}
    \caption{%
      After applying Reduction~\ref{def_reductions_A} to \subref{fig_reductions_overview_a}.
    }%
    \label{fig_reductions_overview_b}
  \end{subfigure}

  \begin{subfigure}[t]{0.45\textwidth}
    \centering
    \begin{tikzpicture}[node distance = 1.8 cm]
      \node[main] (1) {};
      \node[main] (2) [right of=1] {};
      \node[main] (3) [below right of=2] {};
      \node[main] (4) [above right of=2] {};
      \node[main] (5) [above right of=3] {};
      \node[main] (6) [right of=5]{};
      \draw[tree] (1) -- (2);
      \draw[cotree] (1) -- (3);
      \draw[cotree] (1) -- (4);
      \draw[tree] (2) -- (4);
      \draw[tree] (3) -- (2);

      \draw[marked] (3) -- (5);
      \draw[tree] (6) -- (3);
      \draw[tree] (5) -- (6);
      \draw[cotree] (4) -- (6);
      \draw[marked] (5) -- (4);
    \end{tikzpicture}
    \caption{%
      A case in which $Y \cap E_1 = \emptyset$ holds.
    }%
    \label{fig_reductions_overview_c}
  \end{subfigure}
\hfill
  \begin{subfigure}[t]{0.45\textwidth}
    \centering
    \begin{tikzpicture}[node distance = 1.8 cm]
      \node[main,fill=white,draw=white] (1) {};
      \node[main,fill=white,draw=white] (2) [right of=1] {};
      \node[main] (3) [below right of=2] {};
      \node[main] (4) [above right of=2] {};
      \node[main] (5) [above right of=3] {};
      \node[main] (6) [right of=5]{};

      \draw[marked] (3) -- (5);
      \draw[tree] (6) -- (3);
      \draw[tree] (5) -- (6);
      \draw[cotree] (4) -- (6);
      \draw[marked] (5) -- (4);
      \draw[virtualtree] (3) -- (4);
    \end{tikzpicture}
    \caption{%
      After applying Reduction~\ref{def_reductions_B} to \subref{fig_reductions_overview_c}.
    }%
    \label{fig_reductions_overview_d}
  \end{subfigure}

  \begin{subfigure}[t]{0.45\textwidth}
    \centering
    \begin{tikzpicture}[node distance = 1.8 cm]
      \node[main] (1) {};
      \node[main] (2) [right of=1] {};
      \node[main] (3) [below right of=2] {};
      \node[main] (4) [above right of=2] {};
      \node[main] (5) [above right of=3] {};
      \node[main] (6) [right of=5]{};
      \draw[tree] (1) -- (2);
      \draw[marked] (1) -- (3);
      \draw[cotree] (1) -- (4);
      \draw[tree] (2) -- (4);
      \draw[tree] (3) -- (2);

      \draw[cotree] (3) -- (5);
      \draw[tree] (6) -- (3);
      \draw[tree] (5) -- (6);
      \draw[marked] (4) -- (6);
      \draw[marked] (5) -- (4);
    \end{tikzpicture}
    \caption{%
      A case in which $Y \cap E_2 = P^{-1}_{u,w}(G,T) \cap E_2$ \\
      and $Y \cap E_1 \neq \emptyset$ hold.
    }%
    \label{fig_reductions_overview_e}
  \end{subfigure}
\hfill
  \begin{subfigure}[t]{0.49\textwidth}
    \centering
    \begin{tikzpicture}[node distance = 1.8 cm]
      \node[main] (1) {};
      \node[main] (2) [right of=1] {};
      \node[main] (3) [below right of=2] {};
      \node[main] (4) [above right of=2] {};
      \node[main,fill=white,draw=white] (5) [above right of=3] {};
      \node[main,fill=white,draw=white] (6) [right of=5]{};
      \draw[tree] (1) -- (2);
      \draw[marked] (1) -- (3);
      \draw[cotree] (1) -- (4);
      \draw[tree] (2) -- (4);
      \draw[tree] (3) -- (2);
      \draw[virtualmarked] (3) -- (4);
    \end{tikzpicture}
    \caption{%
      After applying Reduction~\ref{def_reductions_C} to \subref{fig_reductions_overview_e}.
    }%
    \label{fig_reductions_overview_f}
  \end{subfigure}
  \caption{%
    Each subfigure on the left depicts a graph-tree pair $(G,T)$ with a 2-separation $(E_1,E_2)$ (edges left/right of the two center vertices) and a set $Y$ such that one of the conditions in \cref{def_reductions} is applicable.
    The corresponding subfigure on the right depicts the reduced graph-tree pair $(G',T')$ with $Y'$.
    The trees $T$ and $T'$ are depicted in bold and red, and sets $Y$ and $Y'$ with two stripes.
  }%
  \label{fig_reductions_overview}
\end{figure}

\begin{lemma}
  \label{thm_path_in_twosep_side}
  Let $G$ be a $2$-connected multigraph with spanning tree $T$ and let $(E_1,E_2)$ be a 2-separation of $G$ with separating vertices $u$ and $w$ and the corresponding graphs $G_1$ and $G_2$ such that $P_{u,w}(T) \subseteq E_1$ holds.
  Then any edge $e \in E_1 \setminus T$ has $P_e(T) \subseteq E_1$, and any edge $e\in E_2\setminus T$ has $P_e(T)\subseteq E_2 \cup P_{u,w}(T)$.
\end{lemma}

\begin{proof}
  Consider an edge $e \in E_1 \setminus T$.
  As $T_1$ is a spanning tree, there exists a path between every vertex pair $u',w'\in V(G_1)$, and $P_e(T) \subseteq T_1$ must hold.
  By definition of $T_1$ we have $P_e(T) \subseteq T_1\subseteq E_1$.

  Similarly, if both end-vertices of $e \in E_2 \setminus T$ lie in $T_u$ or $T_w$ then $P_e(T) \subseteq E_2$ holds as there exists a unique path between them in $T_u$ or $T_w$.
  In the other case, the end-vertices of $e$ lie on both $T_u$ and $T_w$.
  Then $P_e(T)$ must be a subset of $T_u\cup T_w\cup P_{u,w}(T)$, as $u$ and $w$ are connected by a unique path in $T$.
\end{proof}

\begin{lemma}
  \label{thm_empty_propagation_cotree}
  Let graph-tree pair $(G,T)$, edge subset $Y \subseteq T$, 2-separation $(E_1,E_2)$ with vertices $u,w$, and subgraphs $G_1,G_2$ be as in \cref{def_reductions} such that Reduction~\ref{def_reductions_A} is applicable.
  Define $G'$, $e$, $T'$ and $Y'$ as in the reduction.
  Then the following hold:
  \begin{enumerate}
  \item
    A vertex $v \in V(G)$ is $Y$-splittable with respect to $G$ if and only if $v \in V(G_1)$ and $v$ is $Y'$-splittable with respect to $G'$.
  \item
    For any $v \in V(G_1)$ that is $Y$-splittable in $G$, let $\widehat{G}$ be the graph obtained from $G$ by applying \cref{thm_SplitConstructionProof} to $v$.
    Let $\widehat{G}'$ be the graph obtained from $G'$ by applying \cref{thm_SplitConstructionProof} to $v$ with respect to $T'$ and $Y'$, and let $\widehat{G}''$ be the graph obtained from $\widehat{G}'$ by replacing $e$ with $G_2$.
    Then $\widehat{G}'' = \widehat{G}$.
  \end{enumerate}
\end{lemma}

\begin{proof}
  Recall that Reduction~\ref{def_reductions_A} is applicable if $Y \cap E_2 = \emptyset$ holds, that $G'$ is $G_1$ augmented by the new edge $e = \{u,w\}$, and that $T' = T \cap E_1$ and $Y' = Y$ hold.
  Let us prove the first statement.
  We consider an arbitrary vertex $v\in V(G)$ and create a case distinction based on the side of the 2-separation that includes $v$.

\medskip
\noindent
  \textbf{Case 1:} $v\in V(G_2) \setminus V(G_1)$.
  Let $y \in Y$, which exists since $Y\neq\emptyset$, and note that $y \in E_1$ since $Y \cap E_2 = \emptyset$ holds.
  By \cref{thm_path_in_twosep_side}, $P_y(T) \subseteq E_1$, and by \cref{thm_MustLieOnPath}, all $Y$-splittable vertices lie on $P_y(T)$.
  We conclude that $v$ is not $Y$-splittable.

\medskip
\noindent
  \textbf{Case 2:} $v\in V(G_1)$.
  Since $G$ is $2$-connected, $v$ is not an articulation vertex of $G$, and thus all vertices of $V(G_2) \setminus \{v\}$ are in the same connected component of $G$ after removal of $v$.
  Since $Y \cap E_2 = \emptyset$ holds, all these vertices are even in the same connected component $h$ of $G_Y^v$.
  Similarly, by definition of $G'$, all vertices in $\{u,w\} \setminus \{v\}$ are in the same connected component $h'$ of ${G'}_Y^v$.
  For any vertex $v'\in V(G_1)\setminus\{u,w\}$, the neighborhood of $v'$ is identical in $G$ and $G'$. Hence, the connected components of $G^v_Y$ and $G'^v_Y$ are identical, except that in one component $V(G_2)\setminus\{v\}$ is replaced by $\{u,w\}\setminus\{v\}$.
  This implies that there exists an isomorphism between $H_Y^v(G)$ and $H_Y^v(G')$ that identifies $h$ with $h'$ and maps all other vertices to themselves.
  This implies that $v \in V(G_1)$ is $Y$-splittable with respect to $G$ if and only if $v$ is $Y$-splittable with respect to $G'$.

  We proceed to the proof of the second statement.
  First, note that $\widehat{G}''$ and $\widehat{G}$ have the same edge set and the same number of vertices.
  To show their equality it thus suffices to show that the incident edges for each vertex are equal.
  Both $\widehat{G}''$ and $\widehat{G}$ were obtained by performing operations that only change the end-vertices of edges incident to $v$ and re-link these edges either to $v_1$ or to $v_2$.
  In particular, replacing $E_2$ by $e$ and replacing $e$ by $E_2$ again does not change the edges incident to vertices $V(G_2) \setminus\{u,w\}$.
  Thus, it suffices to show for the vertices $v_1$ and $v_2$ that were obtained from the neighborhood split of $v$ in $G$, and $v'_1$ and $v'_2$ that were obtained from the neighborhood split of $v$ in $G'$, that $\delta_{\widehat{G}}(v_1) = \delta_{\widehat{G}''}(v'_1)$ holds.

  Without loss of generality, we assume that $v_1$ was obtained from the $I$-part of the neighborhood split, such that $\delta_{\widehat{G}}(v_1) = \delta^I_G(v)$ holds.
  Since $E_1$ and $E_2$ partition the edges of the graph, $\delta^I_G(v)\cap E_1$ and $\delta^I_G(v)\cap E_2$ form a partition of $\delta^I_G(v)$.
  We consider both sets separately and show that $\delta^I_G(v)\cap E_i = \delta_{\widehat{G}''}(v'_1)\cap E_i$ holds for $i=1,2$.

  First, we consider $\delta^I_G(v) \cap E_1$. We know that $G$ is 2-connected, and it is easy to see that also $G'$ is 2-connected.
  Then, because $G$ and $G'$ are 2-connected, $H^v_Y(G)$ and $H^v_Y(G')$ both have a unique bipartition by \cref{thm_2connected_aux_connected_unique}, which we denote by $(I,J)$ and $(I',J')$ respectively.
  In particular, it must be the case that $I$ and $I'$ both contain the same connected components (after identification) of $G^v_Y$ and $G'^v_Y$.
  Without loss of generality we assume that $h\in I$ and $h'\in I'$ for the connected components $h$ and $h'$ that contain $\{u,w\}\setminus\{v\}$.
  Because all edges of $E_1$ are also contained in $G'$, we then have that $\delta^I_G(v) \cap E_1 = \delta^{I'}_{G'}(v) \cap E_1$.
  Then, by \cref{def_SplitConstruction}, we have that $\delta^{I'}_{G'}(v) \cap E_1 = \delta_{\widehat{G}'}(v'_1) \cap E_1$.
  Clearly, $\delta_{\widehat{G}'}(v'_1) \cap E_1 = \delta_{\widehat{G}''}(v'_1) \cap E_1$ holds, since $e\notin E_1$ and $e$ is replaced only by edges in $E_2$.
  Thus, we conclude that $\delta^I_G(v)\cap E_1 = \delta_{\widehat{G}''}(v'_1)\cap E_1$.

  Secondly, consider $\delta^I_G(v)\cap E_2$.
  In $G'$, we replace $E_2$ by $e$.
  If $v \notin \{u,w\}$, then $\delta_G(v) \cap E_2 = \emptyset$, and replacing $G_2$ by $e$ and subsequently replacing $e$ back by $G_2$ does not modify the neighborhood of any vertex of the graph.
  In particular, we then have that $\delta_{\widehat{G}''}(v_1) \cap E_2 = \emptyset$.
  If $v\in\{u,w\}$, then replacing $G_2$ by $e$, splitting the vertex, and then replacing $e$ by $G_2$, means that all edges of $\delta_{G'}(v)\cap E_2$ lie in the same partition of the neighborhood split of $G'$.
  However, we proved earlier that this must exactly be the case; since $h$ and $h'$ lie in $I$ and $I'$ respectively, we have $E_2 \subseteq \delta^I_G(v)$ and $e\in \delta^{I'}_{G'}(v)$, which implies that $e\in \delta_{\widehat{G'}}(v'_1)$.
  Then, by replacing $e$ with $E_2$ in $\widehat{G}''$, we have that $\delta^I_G(v) \cap E_2 = E_2 = \delta_{\widehat{G}''}(v'_1) \cap E_2$ holds.
  Thus, $\delta_{\widehat{G}''}(v'_1) \cap E_2 = \delta^I_G(v) \cap E_2$.

  We conclude that $\delta_{\widehat{G}}(v_1) = \delta^I_G(v) = (\delta^I_G(v) \cap E_1) \cup (\delta^I_G(v) \cap E_2) = (\delta_{\widehat{G}''}(v'_1) \cap E_1)\cup (\delta_{\widehat{G}''}(v'_1) \cap E_2) = \delta_{\widehat{G}''}(v'_1)$ holds, which establishes $\widehat{G} = \widehat{G}''$.
\end{proof}

\begin{lemma}
  \label{thm_empty_propagation_tree1}
  Let graph-tree pair $(G,T)$, edge subset $Y \subseteq T$, 2-separation $(E_1,E_2)$ with vertices $u,w$, and subgraphs $G_1,G_2$ be as in \cref{def_reductions} such that Reduction~\ref{def_reductions_B} is applicable.
  Define $G'$, $e$, $T'$ and $Y'$ as in the reduction.
  Then the following hold:
  \begin{enumerate}
  \item
    A vertex $v \in V(G_2)$ is $Y$-splittable with respect to $G$ if and only if $v$ is $Y'$-splittable with respect to $G'$.
  \item
    For any $v \in V(G_2)$ that is $Y$-splittable in $G$, let $\widehat{G}$ be the graph obtained from $G$ by applying \cref{thm_SplitConstructionProof} to $v$.
    Let $\widehat{G}'$ be the graph obtained from $G'$ by applying \cref{thm_SplitConstructionProof} to $v$ with respect to $T'$ and $Y'$, and let $\widehat{G}''$ be the graph obtained from $\widehat{G}'$ by replacing $e$ with $G_1$.
    Then $\widehat{G}'' = \widehat{G}$.
  \end{enumerate}
\end{lemma}

\begin{proof}
  Recall that Reduction~\ref{def_reductions_B} is applicable if $Y \cap E_1 = \emptyset$ holds, that $G'$ is $G_2$ augmented by the new edge $e = \{u,w\}$, and that $T' = (T \cap E_2) \cup \{e\}$ and $Y' = Y$ hold.
  Note that $T'$ is indeed a spanning-tree, so that $Y'$-splittability of a vertex in $G'$ is well-defined.
  First, let us prove the first claim.
  Consider a vertex $v \in V(G_2)$.
  Since $G$ is $2$-connected, $v$ is not an articulation vertex of $G$, and thus all vertices of $V(G_1) \setminus \{v\}$ are in the same connected component of $G$ after removal of $v$.
  Since $Y \cap E_1 = \emptyset$ holds, all these vertices are even in the same connected component $h$ of $G_Y^v$.
  Similarly, by definition of $G'$, all vertices in $\{u,w\} \setminus \{v\}$ are in the same connected component $h'$ of ${G'}_Y^v$.
  This implies that there exists an isomorphism between $H^v_Y(G)$ and $H^v_Y(G')$ that maps $h$ to $h'$ and all other vertices to themselves.
  We conclude that $v \in V(G_2)$ is $Y$-splittable with respect to $G$ if and only if $v$ is $Y$-splittable with respect to $G'$.

  The proof for the second statement is identical to the proof of the second statement of \cref{thm_empty_propagation_cotree}, except that $E_1$ and $E_2$ as well as $G_1$ and $G_2$ are swapped.
\end{proof}

Although \cref{thm_empty_propagation_tree1} provides a strong argument for performing its proposed reduction, it does not guarantee that $G'$ has a $Y$-splittable vertex if $G$ has a $Y$-splittable vertex in $V(G_1)\setminus \{u,w\}$.
However, \cref{thm_empty_propagation_tree2} shows that in this case the two vertices $u$ and $w$ of the 2-separation must also be $Y$-splittable and a very specific set of conditions holds. Crucially, this shows that applying the reduction proposed by \cref{thm_empty_propagation_tree1} still valid, as it maintains the existence of a $Y$-splittable vertex.

\begin{lemma}
  \label{thm_empty_propagation_tree2}
  Let graph-tree pair $(G,T)$, edge subset $Y \subseteq T$, 2-separation $(E_1,E_2)$ with vertices $u,w$, and subgraphs $G_1,G_2$ be as in \cref{def_reductions} such that Reduction~\ref{def_reductions_B} is applicable.
  Then for each vertex $v \in V(G_1) \setminus \{u,w\}$, $G^v_Y$ consists of one or two connected components.
  Moreover, let $h_u$ and $h_w$ be the components containing $u$ and $w$, respectively.
  Then $v$ is $Y$-splittable with respect to $G$ if and only if all of the following conditions hold:
  \begin{enumerate}
      \item $v$ is an internal vertex of $P_{u,w}(T)$,
      \item $Y= P^{-1}_{u,w}(G,T)$,
      \item $u$ and $w$ are $Y$-splittable, and
      \item $h_u \neq h_w$.
  \end{enumerate}
\end{lemma}

\begin{proof}
  Recall that Reduction~\ref{def_reductions_B} is applicable if $Y \cap E_1 = \emptyset$ holds.
  We start by showing that $G^v_Y$ has at most two connected components.
  To this end, we claim that each vertex $s \in V(G) \setminus \{v\}$ is connected to $u$ or to $w$ within $G^v_Y$.
  Since $(T_u \cup T_w) \cap E_2$ belongs to $G^v_Y$, the statement is obvious for $s \in V(G_2)$.
  Otherwise, in case $s \in V(G_1) \setminus \{u,w\}$, the $2$-connectivity of $G$ implies that there is a path $P \subseteq E(G^v)$ from $s$ to $\{u,w\}$.
  Let $P$ have minimum length among such paths, which implies $P \subseteq E_1$.
  Since $E_1 \cap Y = \emptyset$, even $P \subseteq E(G^v_Y)$ holds, and the claim follows.
  In particular, this implies that $H^v_Y(G)$ can have at most two components $h_u$ and $h_w$, which contain $u$ and $w$, respectively.

  We first assume that $v \in V(G_1) \setminus \{u,w\}$ is a $Y$-splittable vertex and show that each of the four statements holds.
  From $\delta(v) \cap Y \subseteq E_1 \cap Y = \emptyset$ and $Y \neq \emptyset$ it follows that $Y$ is not a star centered at $v$.
  By \cref{thm_splittableNotStarArticulationLemma}, $v$ is an articulation vertex of $G_Y$.
  By the above argument, this implies that $h_u \neq h_w$, as these are the only two possible components of $G^v_Y$, which shows the fourth statement.
  Next, we show that $v$ is an internal vertex of $P_{u,w}(T)$.
  \cref{thm_MustLieOnPath} implies $v \in P_y(T)$ for some $y \in Y$ and thus $y \in E_2$ holds.
  Since $v \in V(G_1) \setminus \{u,w\}$, we must have $P_{u,w}(T) \subseteq P_y(T)$.
  Then, since $V(P_{u,w}(T))$ are the only vertices of $P_y(T)$ that are in $V(G_1)$ by \cref{thm_path_in_twosep_side}, we have $v \in V(P_{u,w}(T)) \setminus \{u,w\}$, which shows the first statement. We proceed to prove the second statement. Note that $Y\subseteq P^{-1}_{u,w}(G,T)$ holds, since $P_{u,w}(T)\subseteq P_y(T)$ holds for any $y\in Y$ using \cref{thm_MustLieOnPath} and the fact that $y\in E_2$.
  Suppose, for the sake of contradiction, that an edge $e \in P^{-1}_{u,w}(G,T) \setminus Y$ exists.
  This implies that a path that connects $u$ and $w$ in $G^v_Y$ exists in $T_u \cup T_w \cup \{e\}$.
  However, this contradicts our observation that $h_u \neq h_w$ must hold.
  Consequently, $Y = P^{-1}_{u,w}(G,T)$ holds, which proves the second statement. Furthermore, note that we have also shown that each edge $y\in Y$ connects $V_u$ to $V_w$ since we have $E_1 \cap Y = \emptyset$.

  It remains to show the third statement, which claims that $u$ and $w$ are $Y$-splittable. We only show that $u$ is $Y$-splittable; the argument for $w$ is symmetric, with $u$ and $w$ swapped.
  Consider $G^u_Y$ and let $h'_w$ be the component of $G^u_Y$ that contains $w$.
  Since $T_w$ is a tree, it follows that $V_w$ is also contained in $h'_w$.
  Additionally, note that $E_1 \setminus \delta(u)$ is connected since $G$ is $2$-connected.
  Since $E_1 \cap Y = \emptyset$ and $w$ is incident to some edges from $E_1$, also $E_1 \setminus \delta(u)$ is contained in $h'_w$.

  Due to $Y = P^{-1}_{u,w}(G,T)$, the edges connecting $V_u$ and $V_w$ in $E_2$ all belong to $Y$.
  Then, since we remove $u$ in $G^u_Y$ and $\{u,w\}$ are separating vertices, there are no paths in $G^u_Y$ connecting any vertex $s \in V_u \setminus \{u\}$ to any vertex $t \in V_w$, which implies that each vertex from $V_u \setminus \{u\}$ lies in some connected component different from $h'_w$.
  This, together with the observation that each $Y$-edge is incident to a vertex from $h'_w$, implies that $H^u_Y$ is a star centered at $h'_w$, and thus bipartite.
  We conclude that $u$ (and thus $w$) is $Y$-splittable, and hence the third statement follows.

  To show the reverse direction we assume that $v$ is an internal vertex of $P_{u,w}(T)$, that $Y = P^{-1}_{u,w}(G,T)$ and $h_u \neq h_w$ hold and that $u$ and $w$ are $Y$-splittable vertices and show that this implies that $v$ is $Y$-splittable.
  Note that $V_u$ is contained in $h_u$ and $V_w$ is contained in $h_w$.
  Due to $Y = P^{-1}_{u,w}(G,T)$, every edge in $Y$ connects $V_u$ to $V_w$, which implies that every edge $y \in Y$ connects $h_u$ to $h_w$.
  Since $V(H^v_Y) = \{h_u,h_w\}$ holds, $\{h_u\}$ and $\{h_w\}$ are a bipartition of $H^v_Y$.
  Thus, $v$ is $Y$-splittable.
\end{proof}

Note that since $\{u,w\}\subseteq V(G_2)$, Reduction~\ref{def_reductions_B} is also valid for $u$ and $w$ and thus $u$ and $w$ are $Y$-splittable in $G$ if and only if $u$ and $w$ are $Y'$-splittable in $G'$ (as defined in \cref{def_reductions}).
Since \cref{thm_empty_propagation_tree2} shows that a vertex $V(G_1)\setminus\{u,w\}$ is $Y$-splittable only if $u$ and $w$ are $Y$-splittable, this shows that Reduction~\ref{def_reductions_B} is indeed safe and valid, as we then always guarantee that $G'$ has a $Y'$-splittable vertex if $G$ has a $Y$-splittable vertex.

\Cref{thm_empty_propagation_cotree,thm_empty_propagation_tree1} show that if one side of a 2-separation contains no $Y$-edges, then we can replace it by a single edge that is not in $Y$.
\cref{thm_full_propagation_cotree} does something similar in the setting where the $Y$-edges on one side constitute a particular cut of that side; in this case, one can replace this side of the 2-separation by a single edge that is then added to $Y$, which intuitively represents this cut.

\begin{lemma}
  \label{thm_full_propagation_cotree}
  Let graph-tree pair $(G,T)$, edge subset $Y \subseteq T$, 2-separation $(E_1,E_2)$ with vertices $u,w$, and subgraphs $G_1,G_2$ be as in \cref{def_reductions} such that Reduction~\ref{def_reductions_C} is applicable.
  Define $G'$, $e$, $T'$ and $Y'$ as in the reduction.
  Then the following hold:
  \begin{enumerate}
  \item
    A vertex $v \in V(G)$ is $Y$-splittable with respect to $G$ if and only if $v \in V(G_1)$ and $v$ is $Y'$-splittable with respect to $G'$.
  \item
    For any $v \in V(G_1)$ that is $Y$-splittable in $G$, let $\widehat{G}$ be the graph obtained from $G$ by applying \cref{thm_SplitConstructionProof} to $v$.
    Let $\widehat{G}'$ be the graph obtained from $G'$ by applying \cref{thm_SplitConstructionProof} to $v$ with respect to $T'$ and $Y'$, and let $\widehat{G}''$ be the graph obtained from $\widehat{G}'$ by replacing $e$ with $G_2$.
    Then $\widehat{G}'' = \widehat{G}$.
  \end{enumerate}
\end{lemma}

\begin{proof}
  Recall that Reduction~\ref{def_reductions_C} is applicable if $Y \cap E_1 \neq \emptyset$ and $Y \cap E_2 = P^{-1}_{u,w}(G,T) \cap E_2$ hold, that $G'$ is $G_1$ augmented by the new edge $e = \{u,w\}$, and that $T' = T \cap E_1$ and $Y' = (Y \cap E_1) \cup \{ e \}$ hold.
  We start by proving the first statement, which we treat by a case analysis for a vertex $v\in V(G)$.

\medskip
\noindent
  \textbf{Case 1:} $v \in V(G_2) \setminus V(G_1)$. Additionally, we consider some $y\in Y\cap E_1$, which exists by our assumptions.
  By \cref{thm_path_in_twosep_side}, $P_y(T) \subseteq E_1$ holds, and by \cref{thm_MustLieOnPath}, all $Y$-splittable vertices lie on $P_y(T)$.
  We conclude that $v$ is not $Y$-splittable.

\medskip
\noindent
  \textbf{Case 2: $v\in V(G_1)\setminus V(G_2)$.}
  Since both trees $T_u$ and $T_w$ are part of $G_Y^v$, the connected components $h_u$ and $h_w$ that contain $u$ and $w$, respectively, cover all vertices in $V(G_2)$.
  Additionally, observe that $P_{u,w}^{-1}(G,T) \cap E_2$ is the set of edges from $G_2$ that connect $V_u$ with $V_w$.
  Also observe that the set is non-empty since otherwise $u$ or $w$ would be an articulation vertex of $G$, contradicting $2$-connectivity of $G$.
  Note that by the assumption from the lemma, this set is equal to $Y \cap E_2$. Thus, all edges in $Y\cap E_2$ connect $h_u$ with $h_w$.
  Similarly, the connected components $h'_u$ and $h'_w$ of ${G'}^v_{Y'}$ that contain $u$ and $w$, respectively, are connected by the new edge $e \in Y'$.
  All other connected components of $G_Y^v$ and ${G'}^v_{Y'}$ are identical since they are disjoint from $V(G_2)$.
  This shows that $H_Y^v(G)$ and $H_Y^v(G')$ are isomorphic by identifying $h_u$ with $h_u'$ and $h_w$ with $h_w'$.
  We conclude that $v$ is $Y$-splittable with respect to $G$ if and only if $v$ is $Y'$-splittable with respect to $G'$.

\medskip
\noindent
  \textbf{Case 3: $v\in \{u,w\}$.} We only prove the case $v = u$; the proof for $v = w$ has a similar proof.
  Again, the tree $T_w$ is part of $G_Y^v$ and hence the connected component $h_w$ that contains $w$ covers $V_w$.
  Let $H_u$ denote the set of all connected components of $G_Y^v$ that contain vertices from $V_u \setminus \{u\}$.
  Each such component $h_u \in H_u$ contains only vertices from $V_u \setminus \{u\}$ and is not disconnected just by removal of $v = u$ since otherwise the latter would be an articulation vertex.
  Hence, there exists an edge $y' \in Y$ that connects $h_u$ to some other component $h$.
  Since $y'$ must lie in $P^{-1}_{u,w}(G,T)\cap E_2$, we must have $h = h_w$.
  Similarly, let $h'_w$ denote the connected component of ${G'}^v_{Y'}$ that contains $w$.
  All other connected components of $G_Y^v$ and ${G'}^v_{Y'}$ are identical since they are disjoint from $V(G_2)$.
  This shows that $H_{Y'}^v(G')$ and the graph obtained from $H_{Y}^v(G)$ by removing vertices $H_u$ are isomorphic by identifying $h_w$ with $h'_w$.
  Since all vertices $h_u \in H_u$ of $H_Y^v(G)$ have degree~1, this implies that $H_Y^v(G)$ is bipartite if and only if $H_{Y'}^v(G')$ is bipartite.
  We conclude that $v$ is $Y$-splittable with respect to $G$ if and only if $v$ is $Y'$-splittable with respect to $G'$.

  We proceed with the proof of the second statement.
  Note that $\widehat{G}''$ and $\widehat{G}$ have the same edge set and the same number of vertices.
  Thus, it suffices to show that the incident edges for each vertex are equal.
  For both $\widehat{G}''$ and $\widehat{G}$ we obtained the graph from $G$ by performing operations that only change the end-vertices of edges incident to $v$ and distribute these edges over $v_1$ and $v_2$.
  In particular, replacing $E_2$ by $e$ and replacing $e$ by $E_2$ again does not change the edges incident to vertices in $V(G_2) \setminus\{u,w\}$.
  Thus, it suffices to check for the vertices $v_1$ and $v_2$ obtained from the neighborhood split of $v$ in $G$, and $v'_1$ and $v'_2$ obtained from the neighborhood split of $v$ in $G'$ that $\delta_{\widehat{G}}(v_1) = \delta_{\widehat{G}''}(v'_1)$ holds.

  Because $G$ and $G'$ are 2-connected, $H^v_Y(G)$ and $H^v_Y(G')$ both have a unique bipartition by \cref{thm_2connected_aux_connected_unique}, which we denote by $(I,J)$ and $(I',J')$ respectively.
  In particular, $I$ and $I'$ both contain the same connected components (after identification) of $G^v_Y$ and $G'^v_Y$.
  Moreover, we have shown already that $I$ and $I'$ contain exactly the same vertices of $V(G_1)$, such that $\delta_{\widehat{G}}(v_1) = \delta^I_G(v)$ and $\delta_{\widehat{G}'}(v'_1) = \delta^{I'}_{G'}(v)$ hold.
  Since $E_1$ and $E_2$ partition the edges of the graph, $\delta^I_G(v)\cap E_1$ and $\delta^I_G(v)\cap E_2$ form a partition of $\delta^I_G(v)$.
  We consider both sets separately and show that $\delta^I_G(v)\cap E_i = \delta_{\widehat{G}''}(v'_1)\cap E_i$ holds for $i=1,2$.

  Consider $\delta^I_G(v) \cap E_1$.
  Then, $\delta^I_G(v) \cap E_1 = \delta^{I'}_{G'}(v) \cap E_1 $ holds, because $I$ and $I'$ contain the same vertices of $V(G_1)$.
  By applying \cref{def_SplitConstruction}, we have that $\delta^{I'}_{G'}(v) \cap E_1 = \delta_{\widehat{G}'}(v'_1) \cap E_1$, which is equal to $\delta_{\widehat{G}''}(v'_1)\cap E_1$ because replacing $e$ by $E_2$ does not affect the edges of $E_1$.
  Thus, we have $\delta^I_G(v)\cap E_1 = \delta_{\widehat{G}''}(v'_1)\cap E_1$.

  Secondly, we consider $\delta^I_G(v) \cap E_2$.
  First, note that if $v\notin\{u,w\}$, then $\delta^I_G(v) \cap E_2 = \emptyset = \delta_{\widehat{G}''}(v'_1)$ holds, since then $v$ is not incident to $E_2$ or to $e$ in any of the graphs.
  Next, consider the case $v \in \{u,w\}$, and let $v' \in \{u,w\} \setminus \{v\}$.
  First, since $Y \cap E_2 = P^{-1}_{u,w}(G,T) \cap E_2$, any edge connecting $v$ to $V_{v'}$ must be in $Y$.
  Then, $V_{v'}$ is contained in one connected component $h$, since $T_{v'}$ connects all these vertices.
  Because $Y \cap E_2 = P^{-1}_{u,w}(G,T) \cap E_2$, $h$ must be connected in $H^v_Y(G)$ to any component $h_v\in H_v$ where $h_v$ contains vertices in $V_v$.

  Since $H^v_Y$ is bipartite, this shows that these edges must be on opposite sides.
  By, $Y\cap E_2 = P^{-1}_{u,w}(G,T) \cap E_2$, there can exist no edges in $Y$ connecting $v$ to any vertex in $V_{v}$.
  Thus, in $E_2$, $v$ can only connect to vertices in $V_{v}$ using edges that are not in $Y$ and $v$ can only connect to vertices in $V_{v'}$ using edges in $Y$, and all vertices in $V_{v}$ and $V_{v'}$ are on opposite sides of the bipartition $(I,J)$.
  Thus, we either have $\delta_G(v)\cap E_2 \subseteq \delta^I_G(v)$ or $\delta_G(v)\cap E_2 \subseteq \delta^J_G(v)$.
  From this, it follows that either $\delta^I_G(v)\cap E_2 = \delta_G(v) \cap E_2$ or $\delta^I_G(v)\cap E_2 = \emptyset$ holds.

  Because we defined $I$ and $I'$ to have similar vertices, we have $e\in \delta^{I'}_{G'}(v)$ if and only if $\delta_G(v)\cap E_2 = \delta^I_G(v) \cap E_2$.
  We observe that if $e\in\delta^{I'}_{G'}(v)$ holds, then $e\in \delta_{\widehat{G}'}(v'_1)$ holds too.
  Then, in $\widehat{G}''$, all edges from $\delta_G(v)\cap E_2$ are exactly placed incident to $v'_1$, i.e., $\delta_G(v)\cap E_2 = \delta_{\widehat{G}''}(v'_1)$ holds. Then, by the above, it follows that $\delta^I_G(v)\cap E_2 = \delta_{\widehat{G}''}(v'_1)$.
  Similarly, if $e \notin \delta^{I'}_{G'}(v)$ then $\delta^I_G(v)\cap E_2 = \emptyset = \delta_{\widehat{G}''}(v'_1) \cap E_2$ holds because all edges are placed incident to $v'_2$, instead.

  We obtain $\delta_{\widehat{G}}(v_1) = \delta^I_G(v) = (\delta^I_G(v) \cap E_1) \cup (\delta^I_G(v) \cap E_2) = (\delta_{\widehat{G}''}(v'_1) \cap E_1)\cup (\delta_{\widehat{G}''}(v'_1) \cap E_2) = \delta_{\widehat{G}''}(v'_1)$, which shows that $\widehat{G} = \widehat{G}''$ holds.
\end{proof}

Although we argued that all three reductions are valid on a single graph, it is not immediately clear that they are also applicable to the SPQR-tree that represents all graphs for the matrix. In particular, in order to create an algorithm that always finds a realization with a $Y$-splittable vertex if one exists, every realization of the matrix must be considered.
Luckily, the important conditions for the reductions from \cref{def_reductions} that require either $E_2 \cap Y = \emptyset$ or $E_1 \cap Y = \emptyset$ (Reductions~\ref{def_reductions_C} and~\ref{def_reductions_B}) or require $Y \cap E_2 = P^{-1}_{u,w}(G,T) \cap E_2$ (Reduction~\ref{def_reductions_C}), are invariant under performing reversals at the 2-separations given by $u$ and $w$.
Thus, we can apply them directly to the 2-separations of the SPQR-tree to reduce all represented graphs at the same time.

We have not yet mentioned how to check the condition $Y \cap E_2 = P^{-1}_{u,w}(G,T) \cap E_2$ used in \cref{thm_full_propagation_cotree}.
However, in the case where $G_2$ is a skeleton of type \eqref{node_S}, \eqref{node_P} or \eqref{node_Q} this is not a problem, as then it simply amounts to checking whether $Y = E \setminus T$.
For nodes of type \eqref{node_R} we show that this can be done by testing for splittability.

\begin{lemma}
  \label{thm_full_propagation_3connected}
  Let $G$ be a simple $3$-connected graph with spanning tree $T$, let $Y \subseteq E(G) \setminus T$ be non-empty, and let $\{u,w\} \in T$ be a tree edge.
  Then $Y = P^{-1}_{u,w}(G,T)$ holds if and only if $u$ and $w$ are both $Y$-splittable.
\end{lemma}

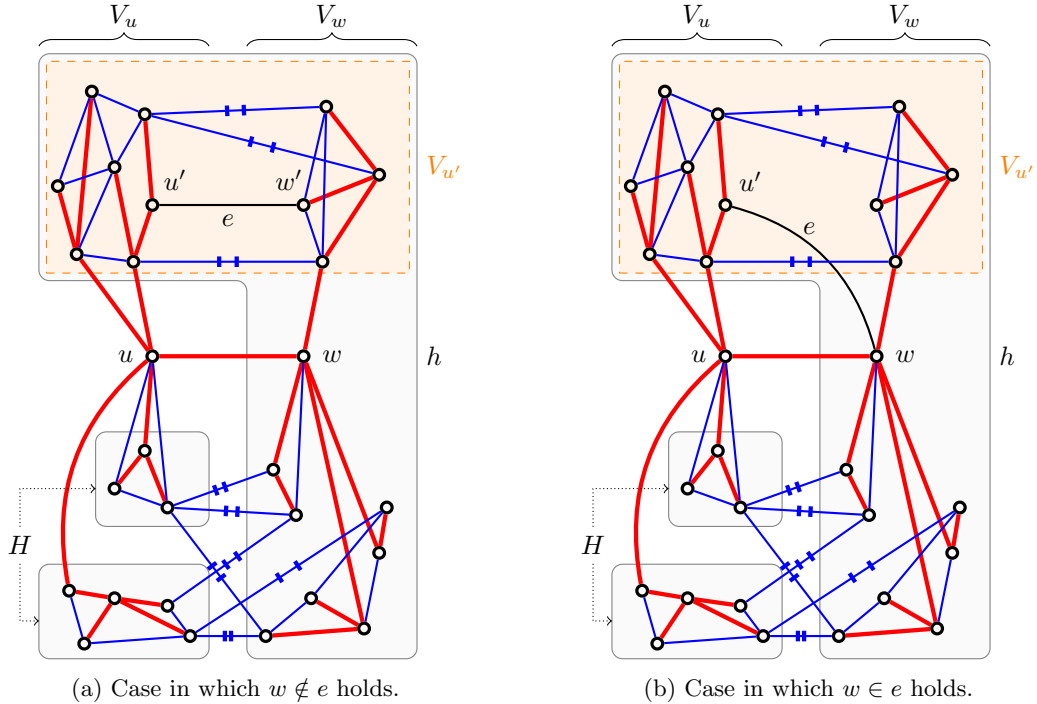
\begin{figure}[htpb]
  \begin{subfigure}{0.5\textwidth}
  \centering
  \begin{tikzpicture}
    \draw[draw=black!50, rounded corners, fill=black!2!white] (-2.5,1) -- (0.25,1) -- (0.25,-4) -- (2.5,-4) -- (2.5,4) -- (-2.5,4) -- cycle;
    \draw[draw=black!50, rounded corners, fill=black!2!white] (-1.75,-1) -- (-0.25,-1) -- (-0.25,-2.25) -- (-1.75,-2.25) -- cycle;
    \draw[draw=black!50, rounded corners, fill=black!2!white] (-2.5,-2.75) -- (-0.25,-2.75) -- (-0.25,-4) -- (-2.5,-4) -- cycle;

    \draw[dashed, draw=orange, fill=orange!10!white] (-2.4,1.1) -- (2.4,1.1) -- (2.4,3.9) -- (-2.4,3.9) -- cycle;

    \node[main] (u) at (-1,0) {};
    \node[main] (w) at (1,0) {};
    \node[main] (uprime) at (-1,2) {};
    \node[main] (wprime) at (1,2) {};

    \node[main] (ua1) at (-1.25,1.25) {};
    \node[main] (ua2) at (-2.00,1.35) {};
    \node[main] (ua3) at (-2.25,2.25) {};
    \node[main] (ua4) at (-1.50,2.50) {};
    \node[main] (ua5) at (-1.80,3.50) {};
    \node[main] (ua6) at (-1.10,3.20) {};

    \node[main] (ub1) at (-1.1,-1.25) {};
    \node[main] (ub2) at (-1.5,-1.75) {};
    \node[main] (ub3) at (-0.8,-2.00) {};

    \node[main] (uc1) at (-2.10,-3.10) {};
    \node[main] (uc2) at (-1.50,-3.20) {};
    \node[main] (uc3) at (-1.90,-3.80) {};
    \node[main] (uc4) at (-0.8,-3.30) {};
    \node[main] (uc5) at (-0.5,-3.70) {};

    \node[main] (wa1) at (1.25,1.25) {};
    \node[main] (wa2) at (2.0,2.4) {};
    \node[main] (wa3) at (1.3,3.3) {};

    \node[main] (wb1) at (0.6,-1.5) {};
    \node[main] (wb2) at (0.9,-2.1) {};

    \node[main] (wc1) at (1.80,-3.60) {};
    \node[main] (wc2) at (1.1,-3.20) {};
    \node[main] (wc3) at (0.5,-3.70) {};
    \node[main] (wc4) at (2.10,-2.00) {};
    \node[main] (wc5) at (2.00,-2.60) {};

    \draw[thick] (uprime) to node[below] {$e$} (wprime);

    \foreach \i/\j in {%
      u/ua1, ua1/uprime, uprime/ua6, ua1/ua4, u/ua2, ua2/ua3, ua2/ua5,%
      u/ub1, ub1/ub2, ub1/ub3,%
      uc1/uc2, uc2/uc3, uc2/uc4, uc2/uc5,%
      w/wa1, wa1/wa2, wa2/wprime, wa2/wa3,%
      w/wb1, wb1/wb2,%
      w/wc1, wc1/wc2, wc1/wc3, w/wc5, wc4/wc5%
    }{
      \draw[tree] (\i) -- (\j);
    }
    \draw[tree, ultra thick] (u) -- (w);
    \draw[tree] (u) to[bend right] (uc1);

    \foreach \i/\j in {%
      ua5/ua6, ua4/ua6, ua3/ua4, ua3/ua5, ua4/ua5, ua2/ua4, ua1/ua2,%
      ub2/ub3, u/ub2, u/ub3,%
      uc1/uc3, uc3/uc5, uc4/uc5,%
      wa1/wprime, wa1/wa3, wa3/wprime,%
      w/wb2,%
      wc4/wc2, wc5/wc1, wc2/wc3%
    }{
      \draw[cotree] (\i) -- (\j);
    }

    \foreach \i/\j in {%
      ua6/wa3, ua6/wa2, ua1/wa1,%
      wb1/ub3, wc3/ub3, wc3/uc5, wb2/ub3, wb2/uc4, wc4/uc5%
    }{
      \draw[marked] (\i) -- (\j);
    }

    \node[left=1pt of u] {$u$};
    \node[right=1pt of w] {$w$};
    \node[above right=0.3pt and -1pt of uprime] {$u'$};
    \node[above left=0.3pt and -5pt of wprime] {$w'$};

    \node[anchor=west] (h) at (2.5,0) {$h$};
    \node (H) at (-2.75,-2.5) {$H$};
    \draw[densely dotted, ->] (H) |- +(1.00,+0.75);
    \draw[densely dotted, ->] (H) |- +(0.25,-1.0);

    \node[anchor=west, orange] (Vuprime) at (2.5,2.5) {$V_{u'}$};

    \draw[decorate, decoration={brace,amplitude=5pt}]
  (-2.5,4.1) -- (-0.25,4.1) node[midway,yshift=4mm]{$V_u$};
    \draw[decorate, decoration={brace,amplitude=5pt}]
  (0.25,4.1) -- (2.5,4.1) node[midway,yshift=4mm]{$V_w$};
    \end{tikzpicture}
    \caption{Case in which $w \notin e$ holds.}
    \label{fig_full_propagation_3connected_generic}
  \end{subfigure}
  \hfill
  \begin{subfigure}{0.5\textwidth}
  \centering
  \begin{tikzpicture}
    \draw[draw=black!50, rounded corners, fill=black!2!white] (-2.5,1) -- (0.25,1) -- (0.25,-4) -- (2.5,-4) -- (2.5,4) -- (-2.5,4) -- cycle;
    \draw[draw=black!50, rounded corners, fill=black!2!white] (-1.75,-1) -- (-0.25,-1) -- (-0.25,-2.25) -- (-1.75,-2.25) -- cycle;
    \draw[draw=black!50, rounded corners, fill=black!2!white] (-2.5,-2.75) -- (-0.25,-2.75) -- (-0.25,-4) -- (-2.5,-4) -- cycle;

    \draw[dashed, draw=orange, fill=orange!10!white] (-2.4,1.1) -- (2.4,1.1) -- (2.4,3.9) -- (-2.4,3.9) -- cycle;

    \node[main] (u) at (-1,0) {};
    \node[main] (w) at (1,0) {};
    \node[main] (uprime) at (-1,2) {};
    \node[main] (wprime) at (1,2) {};

    \node[main] (ua1) at (-1.25,1.25) {};
    \node[main] (ua2) at (-2.00,1.35) {};
    \node[main] (ua3) at (-2.25,2.25) {};
    \node[main] (ua4) at (-1.50,2.50) {};
    \node[main] (ua5) at (-1.80,3.50) {};
    \node[main] (ua6) at (-1.10,3.20) {};

    \node[main] (ub1) at (-1.1,-1.25) {};
    \node[main] (ub2) at (-1.5,-1.75) {};
    \node[main] (ub3) at (-0.8,-2.00) {};

    \node[main] (uc1) at (-2.10,-3.10) {};
    \node[main] (uc2) at (-1.50,-3.20) {};
    \node[main] (uc3) at (-1.90,-3.80) {};
    \node[main] (uc4) at (-0.8,-3.30) {};
    \node[main] (uc5) at (-0.5,-3.70) {};

    \node[main] (wa1) at (1.25,1.25) {};
    \node[main] (wa2) at (2.0,2.4) {};
    \node[main] (wa3) at (1.3,3.3) {};

    \node[main] (wb1) at (0.6,-1.5) {};
    \node[main] (wb2) at (0.9,-2.1) {};

    \node[main] (wc1) at (1.80,-3.60) {};
    \node[main] (wc2) at (1.1,-3.20) {};
    \node[main] (wc3) at (0.5,-3.70) {};
    \node[main] (wc4) at (2.10,-2.00) {};
    \node[main] (wc5) at (2.00,-2.60) {};

    \draw[thick] (uprime) to[bend left] node[above,pos=0.4] {$e$} (w);

    \foreach \i/\j in {%
      u/ua1, ua1/uprime, uprime/ua6, ua1/ua4, u/ua2, ua2/ua3, ua2/ua5,%
      u/ub1, ub1/ub2, ub1/ub3,%
      uc1/uc2, uc2/uc3, uc2/uc4, uc2/uc5,%
      w/wa1, wa1/wa2, wa2/wprime, wa2/wa3,%
      w/wb1, wb1/wb2,%
      w/wc1, wc1/wc2, wc1/wc3, w/wc5, wc4/wc5%
    }{
      \draw[tree] (\i) -- (\j);
    }
    \draw[tree, ultra thick] (u) -- (w);
    \draw[tree] (u) to[bend right] (uc1);

    \foreach \i/\j in {%
      ua5/ua6, ua4/ua6, ua3/ua4, ua3/ua5, ua4/ua5, ua2/ua4, ua1/ua2,%
      ub2/ub3, u/ub2, u/ub3,%
      uc1/uc3, uc3/uc5, uc4/uc5,%
      wa1/wprime, wa1/wa3, wa3/wprime,%
      w/wb2,%
      wc4/wc2, wc5/wc1, wc2/wc3%
    }{
      \draw[cotree] (\i) -- (\j);
    }

    \foreach \i/\j in {%
      ua6/wa3, ua6/wa2, ua1/wa1,%
      wb1/ub3, wc3/ub3, wc3/uc5, wb2/ub3, wb2/uc4, wc4/uc5%
    }{
      \draw[marked] (\i) -- (\j);
    }

    \node[left=1pt of u] {$u$};
    \node[right=1pt of w] {$w$};
    \node[above right=-0.2pt and -0.2pt of uprime] {$u'$};

    \node[anchor=west] (h) at (2.5,0) {$h$};
    \node (H) at (-2.75,-2.5) {$H$};
    \draw[densely dotted, ->] (H) |- +(1.00,+0.75);
    \draw[densely dotted, ->] (H) |- +(0.25,-1.0);

    \node[anchor=west, orange] (Vuprime) at (2.5,2.5) {$V_{u'}$};

    \draw[decorate, decoration={brace,amplitude=5pt}]
  (-2.5,4.1) -- (-0.25,4.1) node[midway,yshift=4mm]{$V_u$};
    \draw[decorate, decoration={brace,amplitude=5pt}]
  (0.25,4.1) -- (2.5,4.1) node[midway,yshift=4mm]{$V_w$};
    \end{tikzpicture}
    \caption{Case in which $w \in e$ holds.}
    \label{fig_full_propagation_3connected_degenerate}
  \end{subfigure}
  \caption{%
    Construction of a 2-separation in the proof of \cref{thm_full_propagation_3connected}.
    In both pictures, the trees $T_u$ and $T_w$ are at the left and right, respectively, and the gray components are those of $G^u_Y$ with $h$ being the one that contains $u'$.
    The dashed box indicates $V_{u'}$, which is the component of $G_Y^{u,w}$ that lies in $h$.
    In the proof it is shown that all edges with at least one vertex in $V_{u'}$ constitute one part of a 2-separation with $u$ and $w$ as separating vertices.
  }
  \label{fig_full_propagation_3connected}
\end{figure}

\begin{proof}
  We first show that $Y=P^{-1}_{u,w}(G,T)$ implies $Y$-splittability of both $u$ and $w$. Since the statement of the lemma is symmetric in $u$ and $w$, it suffices to show that $u$ is $Y$-splittable.
  Let $T_u$ and $T_w$ be the subtrees of $T \setminus \{\{u,w\}\}$ containing $u$ and $w$, respectively, and let $V_u$ and $V_w$ be their respective sets of vertices.
  Due to $Y \neq \emptyset$ there is an edge $y \in Y$.
  Since $G$ is simple, $y \neq \{u,w\}$ holds, so there exists a vertex $v \in V(G)\setminus\{u,w\}$ with $y\in\delta(v)$.
  Our assumption $Y = P^{-1}_{u,w}(G,T)$ implies that the only edge from $E \setminus Y$ that connects $V_u$ with $V_w$ is $\{u,w\}$.
  Let $h_w$ denote the connected component of $G^u_Y$ consisting of the vertices $V_w$ and let $H$ denote the set of all other components, whose vertices form a partition of $V_u \setminus\{u\}$.
  Now, consider any edge $y'\in Y \setminus \delta(u)$.
  Since $\{u,w\}\in P_{y'}(T)$ holds, the two end-vertices of $y'$ belong to both $V_u$ and $V_w$, which means that they belong to $h_w$ and to one of the components in $H$.
  This proves that $H$ and $\{h_w\}$ form a bipartition of $H^u_Y$, so we conclude that $u$ (and thus $w$ as well) is $Y$-splittable.

  To show the reverse direction, we assume that $u$ and $w$ are $Y$-splittable and aim to show that $Y=P^{-1}_{u,w}(G,T)$ holds.
  By \cref{thm_MustLieOnPath}, both $u$ and $w$ must lie on the intersection of the paths $P_y$ for all $y \in Y$ (which is again a path due to $Y \neq \emptyset$).
  This implies $Y \subseteq P^{-1}_{u,w}(G,T)$.
  We assume, for the sake of contradiction, that there exists an edge $e \in P^{-1}_{u,w}(G,T) \setminus Y$, such that $\{u,w\} \in P_e(T)$ holds.
  Since $G$ is simple, $e \neq \{u,w\}$ holds, which implies that at least one of $G^u_Y$ or $G^w_Y$ must contain $e$.
  Without loss of generality we assume that this is the case for $G^u_Y$ since the argument for $G^w_Y$ is similar.
  This implies $e \notin \delta(u)$.
  Furthermore, note that $e$ connects $T_u$ with $T_w$ since $e\in P^{-1}_{u,w}(G,T)$ holds.
  Our goal is to construct a 2-separation of $G$, which contradicts $3$-connectivity of $G$, and exploit splittability of $u$ and $w$ to achieve this.

  Consider the connected components of $G^u_Y$ and let $h_e$ be the connected component containing edge $e$ (see \cref{fig_full_propagation_3connected} for an example).
  Note that since $h_e$ is connected to $T_w$, $T_w$ is also contained in $h_e$.
  Since $T$ is connected and disjoint from $Y$, every connected component of $G^u_Y$ must be connected to $u$ by some tree edge in $T$.
  Denote the set of connected components other than $h_e$ by $H \coloneqq V(H^u_Y)\setminus\{h_e\}$.
  In particular, this implies that $V(h) \subseteq V_u$ holds for each $h \in H$.
  Since $e$ is not incident to $u$, there must exist some $u'\in V_{u}\setminus\{u\}$ to which $e$ is incident.
  Let $G^{u,w}_Y$ be the graph $G$ with vertex set $V(G) \setminus \{u,w\}$ and edge set $E(G) \setminus (\delta(u) \cup \delta(w) \cup Y)$, and denote by $V_{u'}$ and $E_{u'}$ the vertices and edges of its connected component containing $u'$ (in \cref{fig_full_propagation_3connected} the set $V_{u'}$ is depicted by a dashed box).

  Then  $|\delta(u) \cap \delta(V_{u'})| \geq 1$ holds since the set must contain a tree edge connecting to $V_{u'}$ by the above argumentation.
  Similarly, $|\delta(w) \cap \delta(V_{u'})| \geq 1$ holds: if there is some vertex $w' \in V_w \setminus \{w\}$ contained in $V_{u'}$, then there must exist a tree edge by the above reasoning (see \cref{fig_full_propagation_3connected_generic}).
  Otherwise, $e = \{u',w\}$ must hold since $w$ is the only vertex in $T_w$ within $E_{u'}$, which shows that $e \in \delta(w) \cap \delta(V_{u'})$ (see \cref{fig_full_propagation_3connected_degenerate}).
  In the following, we will argue that $E_1 \coloneqq E_{u'} \cup (\delta(u) \cap \delta(V_{u'})) \cup (\delta(w) \cap \delta(V_{u'}))$ and $E_2 \coloneqq E \setminus E_1$ form a 2-separation of $G$ with separating vertices $u$ and $w$, which contradicts $3$-connectivity of $G$.
  Note that $|\delta(u) \cap \delta(V_{u'})| \geq 1$ and $|\delta(w) \cap \delta(V_{u'})| \geq 1$ together imply $|E_1| \geq 2$.

  Next, observe that $E_{u'} \subseteq E[h_e]$ holds since $G^{u,w}_Y$ is a subgraph of $G^u_{Y}$.
  Since $u$ is $Y$-splittable, $h_e$ has no self-loops.
  In particular, this implies there is no edge in $Y$ that connects $V_{u'}$ with $T_w$ (and thus with $V_w$).
  Since $w$ is also $Y$-splittable, the component containing $T_u$ in $H^w_Y$ has no self-loops and hence there are also no edges in $Y$ that connect $V_{u'}$ to $V_{u}$ (since $V_{u'} \subseteq V_{u}$).
  Note that $\delta(u)\cap \delta(V_{u'})$ and $\delta(w)\cap \delta(V_{u'})$ connect $E_{u'}$ to $u$ and $w$, and that by the above argumentation there are no edges in $Y$ incident to any vertex from $V_{u'}$ since $V_{u}\cup V_w = V$ spans all possible vertices. Additionally, since we defined $E_{u'}$ as a connected component of $G^{u,w}_Y$, there can be no other edges connecting $V_{u'}$ to vertices in $V\setminus (V_{u'}\cup\{u,w\}). $
  Thus, we can then conclude that $E_1$ and $E_2$ are only adjacent in $u$ and $w$.
  Next, we make a case distinction on the number of connected components of $G^{u,w}_Y$.

  In the case where $E_{u'}$ is the only connected component of $G^{u,w}_Y$, $h_e$ is the only connected component of $G^{u}_Y$.
  Similarly, $G^{w}_Y$ also consists of a single connected component.
  Since $u$ and $w$ are both $Y$-splittable, these components cannot have self-loops.
  This implies that $Y \setminus\delta(u) = \emptyset$ and $Y \setminus\delta(w) = \emptyset$. As $G$ is simple, there is no edge in $y\in Y$, with $y = \{u,w\}$, which contradicts the assumption $Y \neq \emptyset$.

  In the alternative case, $G^{u,w}_Y$ has more than one connected component, which implies that there exists some vertex $q \in V \setminus (V_{u'} \cup \{u,w\})$.
  Now $3$-connectivity of $G$ implies $|\delta(q)| \geq 2$, and $\delta(q) \subseteq E_2$ implies $|E_2| \geq 2$, which in turn implies that $E_1$ and $E_2$ form a 2-separation of $G$.
  This contradicts $3$-connectivity of $G$.
  We conclude that there exists no edge $e \in P^{-1}_{u,w}(G,T) \setminus Y$, and consequently that $Y = P^{-1}_{u,w}(G,T)$ holds.
  This completes the proof.
\end{proof}

The application of \cref{thm_full_propagation_3connected} is primarily that we can use it to efficiently check the condition of \cref{thm_full_propagation_cotree} for SPQR-tree nodes of type \eqref{node_R} (where $G = G_2$ as stated in \cref{thm_full_propagation_cotree}).
\cref{thm_full_propagation_3connected} hints at an important fact; rather than figuring out the intersection of all paths over $G$, it is sufficient to check splittability of the vertices of the smaller graph $G_2$ augmented by a single edge.
In fact, we will later see that this intuition holds more generally, and that we only need to test for splittability of vertices in the skeletons of the SPQR-tree, rather than on one of the represented graphs.

Although we have argued that the reduction is valid, it is still difficult to apply to 2-separations that are not leaves of the SPQR-tree.
The following lemma shows that if the reduction from \cref{thm_full_propagation_cotree} is applicable, then every other 2-separation contained in it is also reducible.
This shows that it is sufficient to only consider leaf nodes of the reduced SPQR-tree to find all reductions.

\begin{lemma}
  \label{thm_nested_reductions}
  Let $(G,T)$ be a graph-tree pair and $Y \subseteq T$ with a reducible 2-separation $(E_1,E_2)$.
  Consider another 2-separation $(E_1',E_2')$.
  If
  \begin{enumerate}
  \item
    Reduction~\ref{def_reductions_A} or~\ref{def_reductions_C} is applicable to $(E_1,E_2)$, and $E_2' \subsetneqq E_2$ holds, or
  \item
    Reduction~\ref{def_reductions_B} is applicable to $(E_1,E_2)$ and $E_1' \subsetneqq E_1$ holds,
  \end{enumerate}
  then $(E_1',E_2')$ or $(E_2',E_1')$ is also reducible.
\end{lemma}

\begin{proof}
  Let $u,w$ and $u',w'$ be the separating vertices of $(E_1,E_2)$ and of $(E_1',E_2')$, respectively.

  If Reduction~\ref{def_reductions_A} is applicable to $(E_1,E_2)$ then $\emptyset = E_2 \cap Y \supseteq E_2' \cap Y$, that is, $Y \cap E_2' = \emptyset$.
  Hence, if $P_{u',w'}(T) \subseteq E_1'$ then $(E_1',E_2')$ admits Reduction~\ref{def_reductions_A}, and otherwise, $(E_2',E_1')$ admits Reduction~\ref{def_reductions_B}.

  Similarly, if Reduction~\ref{def_reductions_B} is applicable to $(E_1,E_2)$ and $E_1' \subsetneqq E_1$ holds then $\emptyset = E_1 \cap Y \supseteq E_1' \cap Y$, that is, $Y \cap E_1' = \emptyset$.
  Hence, if $P_{u',w'}(T) \subseteq E_1'$ then $(E_1',E_2')$ admits Reduction~\ref{def_reductions_B}, and otherwise, $(E_2',E_1')$ admits Reduction~\ref{def_reductions_A}.

  Finally, assume that Reduction~\ref{def_reductions_C} is applicable to $(E_1,E_2)$, that is, $Y \cap E_1 \neq \emptyset$ and $Y \cap E_2 = P^{-1}_{u,w}(G,T) \cap E_2$ holds.
  Suppose $P_{u',w'}(T) \supseteq P_{u,w}(T)$ holds.
  For an edge $e \in E_2'$ we obtain $P_e(T) \supseteq P_{u',w'}(T)$ if and only if $P_e(T) \supseteq P_{u,w}(T)$, and hence $P^{-1}_{u',w'}(G,T) \cap E_2' = P^{-1}_{u,w}(G,T) \cap E_2'$.
  We conclude that $(E_1',E_2')$ admits Reduction~\ref{def_reductions_C}.

  Otherwise, if $P_{u',w'}(T) \not\supseteq P_{u,w}(T)$ holds (and Reduction~\ref{def_reductions_C} is applicable to $(E_1,E_2)$), then $P_{u',w'}(T) \subseteq E_2$.
  For an edge $e \in E_2'$ we obtain $P_e(T) \supseteq E_2' \cup P_{u',w'}(T) \subseteq E_2$ by \cref{thm_path_in_twosep_side}.
  Hence, $Y \cap E_2' = P^{-1}_{u,w}(G,T) \cap E_2' = \emptyset$ holds.
  We conclude that, depending on whether $P_{u',w'}(T) \subseteq E_2'$ holds, either Reduction~\ref{def_reductions_B} is applicable to $(E_2',E_1')$ or Reduction~\ref{def_reductions_A} is applicable to $(E_1',E_2')$.
\end{proof}

SPQR-trees naturally encode some reductions using the virtual edge pairs.
Consider a node $\mu$ of an SPQR-tree $\T$ and let $e$ be a virtual edge of $\mu$ that connects $\mu$ to some subtree $\T_S$ of $\T$.
In the reductions we replace one side of the 2-separation with a single edge; in the SPQR-tree this corresponds to replacing any realization of $\T_S$ by $e$, which is equivalent to setting $\T_R = \T \setminus \T_S$ and making $e$ non-virtual. Additionally, in case the reduction from \cref{thm_full_propagation_cotree} is used, we add $e$ to $Y$ and remove any edges in $\T_S$ from $Y$. Then $e$ is virtual in $\T$, but has become a regular edge in $\T_R$.
In our algorithm, we iteratively consider the leaves of $\T_R$ since \cref{thm_nested_reductions} shows that larger subtrees can only be reduced if all 2-separations in the subtree can be reduced as well.
Thus, if a leaf node cannot be reduced, we cannot reduce a larger subtree containing it, and thus it suffices to focus on the leaf nodes.

Because local 2-separations do not have an associated virtual edge pair, we instead replace 2 or more regular edges that define a reducible local 2-separation by a single edge. For a local 2-separation of a node $\mu\in\V(\T)$, we can also view this as splitting $\mu$ into two members (of the same type as $\mu$), dividing the edges based on the 2-separation, adding a virtual edge pair, and then reducing the 2-separation using one of the appropriate results.

First, let us define a $Y$-reduced SPQR-tree, which is simply an SPQR-tree that has been fully reduced. We show a few properties that must hold, regardless of the algorithm used to derive the reduced SPQR-tree.
\begin{definition}[$Y$-reduced SPQR-tree]
  \label{def_structureReduced}
  Let $\T$ be an SPQR-tree.
  We say that $\T$ is $Y$-reduced if $\emptyset \neq Y \subseteq \spqrNonvirtual(\T)$ holds and if either $\T$ consists of a a single cycle skeleton or all of the represented graphs are irreducible.
\end{definition}

Note that \cref{def_structureReduced} has one notable exception, which is that if $\T$ consists of a single cycle skeleton, that not necessarily all 2-separations are irreducible.
We now present elementary properties of $Y$-reduced SQPR trees.
\begin{proposition}
  \label{thm_structureReducedSimple}
  Every $Y$-reduced SPQR-tree $\T$ satisfies these properties:
  \begin{enumerate}
      \item Each leaf $\mu$ of $\T$ satisfies $E_\mu\cap Y \neq \emptyset$.
      \item If $|\V(\T)| \geq 2$, no leaf $\mu$ of $\T$ is of type~\eqref{node_S}.
      \item If $|\V(\T)| \geq 2$, then each $\mu \in \V(\T)$ of type~\eqref{node_S} satisfies $|\spqrNonvirtual_\mu(\T)| \leq 1$.
  \end{enumerate}
\end{proposition}

\begin{proof}
  The first statement follows from $Y \neq \emptyset$ in case $|\V(\T)| = 1$ holds.
  Otherwise, a leaf $\mu \in \V(\T)$ with $E_\mu \cap Y = \emptyset$ cannot exist since then Reduction~\ref{def_reductions_A} or Reduction~\ref{def_reductions_B} would be applicable, in contradiction to \cref{def_structureReduced}.

  For the second statement, consider a leaf $\mu$ of type \eqref{node_S} and assume $|\V(\T)| \geq 2$.
  By the first statement, we have $E_\mu \cap Y \neq \emptyset$.
  Since $|T_\mu| = |E_\mu| - 1$ holds for nodes of type \eqref{node_S}, the single edge $e \in E_\mu \setminus T_\mu$ lies in $Y$.
  Since $\mu$ is a leaf node, there must exist some virtual tree edge $e'\in T_\mu$.
  Denoting the vertices of $e'$ by $\{u,w\}$, we observe that Reduction~\ref{def_reductions_C} is applicable since the edge $e$ is the only co-tree edge in $\mu$, whose path contains $e'$.
  Again this contradicts \cref{def_structureReduced}.

  For the third statement, assume for the sake of contradiction, that there exist at least two regular edges $e_1,e_2 \in \spqrNonvirtual_\mu(\T)$, where we assume without loss of generality that $e_1$ and $e_2$ are adjacent.
  Clearly, $E_1 \coloneqq \{e_1,e_2\}$ and $E_2 \coloneqq \spqrNonvirtual_\mu(\T) \setminus \{e_1,e_2\}$ form a local 2-separation (of $\mu$), where $|E_2| \geq |\spqrVirtual_\mu(\T)| \geq 2$ holds because $\mu$ is, by the second statement, not a leaf.
  If $e_1,e_2 \notin Y$ then either Reduction~\ref{def_reductions_B} is applicable to $(E_1,E_2)$ or Reduction~\ref{def_reductions_A} is applicable to $(E_2,E_1)$.
  Otherwise, since all edges but one belong to $T_\mu$, we can assume without loss of generality that $e_1 \in T_\mu$ and $e_2 \in Y$ hold, which means that Reduction~\ref{def_reductions_C} is applicable to $(E_2,E_1)$.
  In all cases, applicability of the reductions contradicts \cref{def_structureReduced}.
\end{proof}

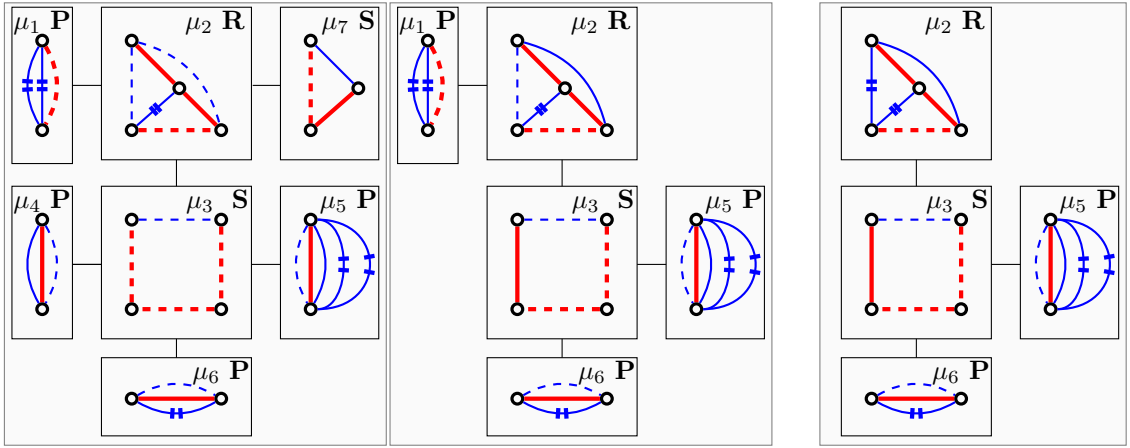
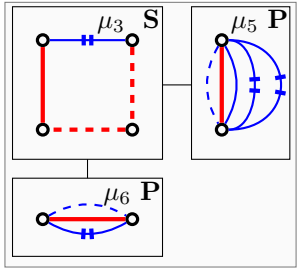
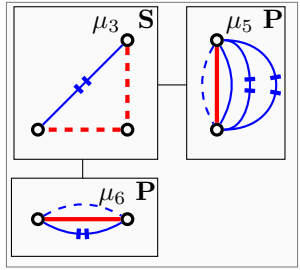
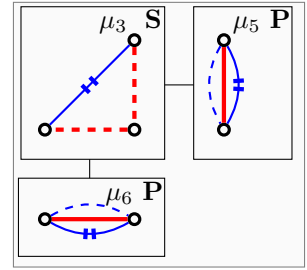
\begin{figure}[htpb]
    \usetikzlibrary{positioning}

    \begin{subfigure}[t]{0.33\textwidth}
       \usetikzlibrary{positioning}
        \begin{tikzpicture}
        \draw[behindrect] (-0.5,0.5) rectangle (4.55, -5.35);
    \node[main] (A) {};
    \node[main] (B) [below = of A]{};
    \draw[marked] (A)  to [bend right] (B) ;
    \draw[marked] (A) -- (B);
    \draw[virtualtree] (A) to [bend left] (B);

    \node[main] (3) [right = of A]{};
    \node[main] (4) [below = of 3]{};
    \node[main] (5) [below right = 0.707cm of 3]{};
    \node[main] (6) [right = of 4]{};

    \draw[virtualcotree] (3) -- (4);
    \draw[virtualtree] (4) -- (6);
    \draw[marked] (4) -- (5);
    \draw[tree] (5) -- (6);
    \draw[tree] (3) -- (5);
    \draw[virtualcotree] (3) to [bend left] (6);

    \node[main] (17) [right = of 6]{};
    \node[main] (18) [above = of 17]{};
    \node[main] (19) [below right = 0.707cm of 18]{};
    \draw[virtualtree] (17) to (18);
    \draw[tree] (17) to (19);
    \draw[cotree] (18) to (19);

    \node[main] (7) [below = of 4]{};
    \node[main] (8) [right = of 7]{};
    \node[main] (9) [below = of 8]{};
    \node[main] (10) [left = of 9]{};
    \draw[virtualcotree] (7) -- (8);
    \draw[virtualtree] (8) -- (9);
    \draw[virtualtree] (9) -- (10);
    \draw[virtualtree] (10) -- (7);

    \node[main] (11) [left = of 7]{};
    \node[main] (12) [below = of 11]{};
    \draw[cotree] (11)  to [bend right] (12) ;
    \draw[tree] (11) -- (12);
    \draw[virtualcotree] (11) to [bend left] (12);

    \node[main] (13) [right = of 8]{};
    \node[main] (14) [below = of 13]{};
    \draw[cotree] (13)  to [bend left] (14) ;
    \draw[marked] (13)  to [in=0,out=0,looseness = 1.0] (14) ;
    \draw[marked] (13)  to [in=0,out=0,looseness = 2.0] (14) ;

    \draw[tree] (13) -- (14);

    \draw[virtualcotree] (13) to [bend right] (14);

    \node[main] (15) [below = of 10]{};
    \node[main] (16) [right = of 15]{};
    \draw[marked] (15) to [bend right] (16);
    \draw[tree] (15) -- (16) ;
    \draw[virtualcotree] (15) to [bend left] (16);
    \draw[black] ($(A.north)+(-0.4,0.35)$) rectangle ($(B.south)+(0.4,-0.35)$);
    \draw[black] ($0.5*(A.north)+0.5*(B.south) + (0.4,0.0)$) -- ($0.5*(3.north)+0.5*(4.south)+ (-0.4,0.0)$) ;
    \draw[black] ($(3.north)+(-0.4,0.35)$) rectangle ($(6.south)+(0.4,-0.3)$);
    \draw[black] ($0.5*(4.south)+0.5*(6.south) + (0.0,-0.3)$) -- ($0.5*(7.north)+0.5*(8.north)+ (0.0,0.35)$) ;

    \draw[black] ($0.5*(9.south)+0.5*(10.south) + (0.0,-0.3)$) -- ($0.5*(15.north)+0.5*(16.north)+ (0.0,0.45)$) ;

    \draw[black] ($0.5*(17.south)+0.5*(18.north) -(0.4,0.0)$) -- ($0.5*(3.north)+0.5*(4.south)+ (1.6,0.0)$) ;

    \node[black] (Box1) at ($(A.north) + 0.5*(0.0,0.3)$) {$\mu_1$~\bf{P}};
    \node[black] (Box2) at ($(3.north) + 0.5*(2.2,0.3)$) {$\mu_2$~\bf{R}};
    \node[black] (Box3) at ($(8.north) + 0.5*(-0.1,0.3)$) {$\mu_3$ ~\bf{S}};
    \node[black] (Box4) at ($(11.north) + 0.5*(0.0,0.3)$) {$\mu_4$~\bf{P}};
    \node[black] (Box5) at ($(13.north) + (0.5,0.15)$) {$\mu_5$~\bf{P}};
    \node[black] (Box6) at ($(16.north) + 0.5*(0.0,0.5)$) {$\mu_6$~\bf{P}};
    \node[black] (Box7) at ($(18.north) + (0.5,0.15)$) {$\mu_7$~\bf{S}};

    \draw[black] ($(7.north)+(-0.4,0.35)$) rectangle ($(9.south)+(0.4,-0.3)$);
    \draw[black] ($0.5*(11.north)+0.5*(12.south) + (0.4,0.0)$) -- ($0.5*(7.north)+0.5*(10.south)+ (-0.4,0.0)$) ;
    \draw[black] ($0.5*(11.north)+0.5*(12.south) + (0.4,0.0)$) -- ($0.5*(7.north)+0.5*(10.south)+ (-0.4,0.0)$) ;
    \draw[black] ($0.5*(8.north)+0.5*(9.south) + (0.4,0.0)$) -- ($0.5*(13.north)+0.5*(14.south)+ (-0.4,0.0)$) ;
    \draw[black] ($(11.north)+(-0.4,0.35)$) rectangle ($(12.south)+(0.4,-0.3)$);
    \draw[black] ($(18.north)+(-0.4,0.35)$) rectangle ($(17.south)+(0.9,-0.3)$);

    \draw[black] ($(13.north)+(-0.4,0.35)$) rectangle ($(14.south)+(0.9,-0.3)$);

    \draw[black] ($(15.north)+(-0.4,0.45)$) rectangle ($(16.south)+(0.4,-0.4)$);

    \end{tikzpicture}
        \subcaption{Original SPQR-tree $\T$ with\\ edge set $Y$.}
    \end{subfigure}\hfill
    \begin{subfigure}[t]{0.33\textwidth}
       \usetikzlibrary{positioning}
        \begin{tikzpicture}
        \draw[behindrect] (-0.5,0.5) rectangle (4.55, -5.35);
    \node[main] (A) {};
    \node[main] (B) [below = of A]{};
    \draw[marked] (A)  to [bend right] (B) ;
    \draw[marked] (A) -- (B);
    \draw[virtualtree] (A) to [bend left] (B);

    \node[main] (3) [right = of A]{};
    \node[main] (4) [below = of 3]{};
    \node[main] (5) [below right = 0.707cm of 3]{};
    \node[main] (6) [right = of 4]{};

    \draw[virtualcotree] (3) -- (4);
    \draw[virtualtree] (4) -- (6);
    \draw[marked] (4) -- (5);
    \draw[tree] (5) -- (6);
    \draw[tree] (3) -- (5);
    \draw[cotree] (3) to [bend left] (6);

    \node[main] (7) [below = of 4]{};
    \node[main] (8) [right = of 7]{};
    \node[main] (9) [below = of 8]{};
    \node[main] (10) [left = of 9]{};
    \draw[virtualcotree] (7) -- (8);
    \draw[virtualtree] (8) -- (9);
    \draw[virtualtree] (9) -- (10);
    \draw[tree] (10) -- (7);

    \node[main] (13) [right = of 8]{};
    \node[main] (14) [below = of 13]{};
    \draw[cotree] (13)  to [bend left] (14) ;
    \draw[marked] (13)  to [in=0,out=0,looseness = 1.0] (14) ;
    \draw[marked] (13)  to [in=0,out=0,looseness = 2.0] (14) ;

    \draw[tree] (13) -- (14);

    \draw[virtualcotree] (13) to [bend right] (14);

    \node[main] (15) [below = of 10]{};
    \node[main] (16) [right = of 15]{};
    \draw[marked] (15) to [bend right] (16);
    \draw[tree] (15) -- (16) ;
    \draw[virtualcotree] (15) to [bend left] (16);
    \draw[black] ($(A.north)+(-0.4,0.35)$) rectangle ($(B.south)+(0.4,-0.35)$);
    \draw[black] ($0.5*(A.north)+0.5*(B.south) + (0.4,0.0)$) -- ($0.5*(3.north)+0.5*(4.south)+ (-0.4,0.0)$) ;
    \draw[black] ($(3.north)+(-0.4,0.35)$) rectangle ($(6.south)+(0.4,-0.3)$);
    \draw[black] ($0.5*(4.south)+0.5*(6.south) + (0.0,-0.3)$) -- ($0.5*(7.north)+0.5*(8.north)+ (0.0,0.35)$) ;

    \draw[black] ($0.5*(9.south)+0.5*(10.south) + (0.0,-0.3)$) -- ($0.5*(15.north)+0.5*(16.north)+ (0.0,0.45)$) ;

    \node[black] (Box1) at ($(A.north) + 0.5*(0.0,0.3)$) {$\mu_1$~\bf{P}};
    \node[black] (Box2) at ($(3.north) + 0.5*(2.2,0.3)$) {$\mu_2$~\bf{R}};
    \node[black] (Box3) at ($(8.north) + 0.5*(-0.1,0.3)$) {$\mu_3$ ~\bf{S}};
    \node[black] (Box5) at ($(13.north) + (0.5,0.15)$) {$\mu_5$~\bf{P}};
    \node[black] (Box6) at ($(16.north) + 0.5*(0.0,0.5)$) {$\mu_6$~\bf{P}};

    \draw[black] ($(7.north)+(-0.4,0.35)$) rectangle ($(9.south)+(0.4,-0.3)$);
    \draw[black] ($0.5*(8.north)+0.5*(9.south) + (0.4,0.0)$) -- ($0.5*(13.north)+0.5*(14.south)+ (-0.4,0.0)$) ;

    \draw[black] ($(13.north)+(-0.4,0.35)$) rectangle ($(14.south)+(0.9,-0.3)$);

    \draw[black] ($(15.north)+(-0.4,0.45)$) rectangle ($(16.south)+(0.4,-0.4)$);

    \end{tikzpicture}
    \subcaption{%
      $\T_1$ with edge set $Y_1$ is obtained from $\T$ by removing $\mu_4$ using \\ Reduction~\ref{def_reductions_B} and removing $\mu_7$ \\ using Reduction~\ref{def_reductions_A}.}

    \end{subfigure}\hfill
    \begin{subfigure}[t]{0.32\textwidth}
       \usetikzlibrary{positioning}
        \begin{tikzpicture}
    \node[main,draw=white,fill=white] (A) {};
    \node[main,draw=white,fill=white] (B) [below = of A]{};
        \draw[behindrect] (0.5,0.5) rectangle (4.55, -5.35);
    \node[main] (3) [right = of A] {};
    \node[main] (4) [below = of 3]{};
    \node[main] (5) [below right = 0.707cm of 3]{};
    \node[main] (6) [right = of 4]{};

    \draw[marked] (3) -- (4);
    \draw[virtualtree] (4) -- (6);
    \draw[marked] (4) -- (5);
    \draw[tree] (5) -- (6);
    \draw[tree] (3) -- (5);
    \draw[cotree] (3) to [bend left] (6);

    \node[main] (7) [below = of 4]{};
    \node[main] (8) [right = of 7]{};
    \node[main] (9) [below = of 8]{};
    \node[main] (10) [left = of 9]{};
    \draw[virtualcotree] (7) -- (8);
    \draw[virtualtree] (8) -- (9);
    \draw[virtualtree] (9) -- (10);
    \draw[tree] (10) -- (7);

    \node[main] (13) [right = of 8]{};
    \node[main] (14) [below = of 13]{};
    \draw[cotree] (13)  to [bend left] (14) ;
    \draw[marked] (13)  to [in=0,out=0,looseness = 1.0] (14) ;
    \draw[marked] (13)  to [in=0,out=0,looseness = 2.0] (14) ;

    \draw[tree] (13) -- (14);

    \draw[virtualcotree] (13) to [bend right] (14);

    \node[main] (15) [below = of 10]{};
    \node[main] (16) [right = of 15]{};
    \draw[marked] (15) to [bend right] (16);
    \draw[tree] (15) -- (16) ;
    \draw[virtualcotree] (15) to [bend left] (16);

    \draw[black] ($(3.north)+(-0.4,0.35)$) rectangle ($(6.south)+(0.4,-0.3)$);
    \draw[black] ($0.5*(4.south)+0.5*(6.south) + (0.0,-0.3)$) -- ($0.5*(7.north)+0.5*(8.north)+ (0.0,0.35)$) ;

    \draw[black] ($0.5*(9.south)+0.5*(10.south) + (0.0,-0.3)$) -- ($0.5*(15.north)+0.5*(16.north)+ (0.0,0.45)$) ;

    \node[black] (Box2) at ($(3.north) + 0.5*(2.2,0.3)$) {$\mu_2$~\bf{R}};
    \node[black] (Box3) at ($(8.north) + 0.5*(-0.1,0.3)$) {$\mu_3$ ~\bf{S}};
    \node[black] (Box5) at ($(13.north) + (0.5,0.15)$) {$\mu_5$~\bf{P}};
    \node[black] (Box6) at ($(16.north) + 0.5*(0.0,0.5)$) {$\mu_6$~\bf{P}};

    \draw[black] ($(7.north)+(-0.4,0.35)$) rectangle ($(9.south)+(0.4,-0.3)$);
    \draw[black] ($0.5*(8.north)+0.5*(9.south) + (0.4,0.0)$) -- ($0.5*(13.north)+0.5*(14.south)+ (-0.4,0.0)$) ;

    \draw[black] ($(13.north)+(-0.4,0.35)$) rectangle ($(14.south)+(0.9,-0.3)$);

    \draw[black] ($(15.north)+(-0.4,0.45)$) rectangle ($(16.south)+(0.4,-0.4)$);

    \end{tikzpicture}
    \subcaption{$\T_2$ with edge set $Y_2$ is obtained from $\T_1$ and $Y_1$ by removing $\mu_1$ using Reduction~\ref{def_reductions_C}.}
    \end{subfigure}

    \begin{subfigure}[t]{0.3\textwidth}
        \centering
       \usetikzlibrary{positioning}
        \begin{tikzpicture}
                \draw[behindrect] (-0.5,0.5) rectangle (3.35, -3.0);
    \node[main] (7) {};
    \node[main] (8) [right = of 7]{};
    \node[main] (9) [below = of 8]{};
    \node[main] (10) [left = of 9]{};
    \draw[marked] (7) -- (8);
    \draw[virtualtree] (8) -- (9);
    \draw[virtualtree] (9) -- (10);
    \draw[tree] (10) -- (7);

    \node[main] (13) [right = of 8]{};
    \node[main] (14) [below = of 13]{};
    \draw[cotree] (13)  to [bend left] (14) ;
    \draw[marked] (13)  to [in=0,out=0,looseness = 1.0] (14) ;
    \draw[marked] (13)  to [in=0,out=0,looseness = 2.0] (14) ;

    \draw[tree] (13) -- (14);

    \draw[virtualcotree] (13) to [bend right] (14);

    \node[main] (15) [below = of 10]{};
    \node[main] (16) [right = of 15]{};
    \draw[marked] (15) to [bend right] (16);
    \draw[tree] (15) -- (16) ;
    \draw[virtualcotree] (15) to [bend left] (16);

    \draw[black] ($0.5*(9.south)+0.5*(10.south) + (0.0,-0.3)$) -- ($0.5*(15.north)+0.5*(16.north)+ (0.0,0.45)$) ;

    \node[black] (Box3) at ($(8.north) + 0.5*(-0.1,0.3)$) {$\mu_3$ ~\bf{S}};
    \node[black] (Box5) at ($(13.north) + (0.5,0.15)$) {$\mu_5$~\bf{P}};
    \node[black] (Box6) at ($(16.north) + 0.5*(0.0,0.5)$) {$\mu_6$~\bf{P}};

    \draw[black] ($(7.north)+(-0.4,0.35)$) rectangle ($(9.south)+(0.4,-0.3)$);
    \draw[black] ($0.5*(8.north)+0.5*(9.south) + (0.4,0.0)$) -- ($0.5*(13.north)+0.5*(14.south)+ (-0.4,0.0)$) ;

    \draw[black] ($(13.north)+(-0.4,0.35)$) rectangle ($(14.south)+(0.9,-0.3)$);

    \draw[black] ($(15.north)+(-0.4,0.45)$) rectangle ($(16.south)+(0.4,-0.4)$);

    \end{tikzpicture}
    \subcaption{%
      $\T_3$ with edge set $Y_3$ is \\
      obtained from $\T_2$ and $Y_2$ by \\
      removing $\mu_2$ using \\
      Reduction~\ref{def_reductions_C}.
    }
    \end{subfigure}\hfill
    \begin{subfigure}[t]{0.3\textwidth}
        \centering
       \usetikzlibrary{positioning}
        \begin{tikzpicture}
                    \draw[behindrect] (-1.6,0.5) rectangle (2.25, -3.0);
    \node[main] (8) {};
    \node[main] (9) [below = of 8]{};
    \node[main] (10) [left = of 9]{};
    \draw[virtualtree] (8) -- (9);
    \draw[virtualtree] (9) -- (10);
    \draw[marked] (10) -- (8);

    \node[main] (13) [right = of 8]{};
    \node[main] (14) [below = of 13]{};
    \draw[cotree] (13)  to [bend left] (14) ;
    \draw[marked] (13)  to [in=0,out=0,looseness = 1.0] (14) ;
    \draw[marked] (13)  to [in=0,out=0,looseness = 2.0] (14) ;

    \draw[tree] (13) -- (14);

    \draw[virtualcotree] (13) to [bend right] (14);

    \node[main] (15) [below = of 10]{};
    \node[main] (16) [right = of 15]{};
    \draw[marked] (15) to [bend right] (16);
    \draw[tree] (15) -- (16) ;
    \draw[virtualcotree] (15) to [bend left] (16);

    \draw[black] ($0.5*(9.south)+0.5*(10.south) + (0.0,-0.3)$) -- ($0.5*(15.north)+0.5*(16.north)+ (0.0,0.45)$) ;

    \node[black] (Box3) at ($(8.north) + 0.5*(-0.1,0.3)$) {$\mu_3$ ~\bf{S}};
    \node[black] (Box5) at ($(13.north) + (0.5,0.15)$) {$\mu_5$~\bf{P}};
    \node[black] (Box6) at ($(16.north) + 0.5*(0.0,0.5)$) {$\mu_6$~\bf{P}};

    \draw[black] ($(8.north)+(-1.5,0.35)$) rectangle ($(9.south)+(0.4,-0.3)$);
    \draw[black] ($0.5*(8.north)+0.5*(9.south) + (0.4,0.0)$) -- ($0.5*(13.north)+0.5*(14.south)+ (-0.4,0.0)$) ;

    \draw[black] ($(13.north)+(-0.4,0.35)$) rectangle ($(14.south)+(0.9,-0.3)$);

    \draw[black] ($(15.north)+(-0.35,0.45)$) rectangle ($(16.south)+(0.4,-0.4)$);

    \end{tikzpicture}
    \subcaption{%
      $\T_4,Y_4$ is obtained from $\T_3,Y_3$ by replacing two edges in $\mu_3$ with one edge by using \\
      Reduction~\ref{def_reductions_C}, as in \\
      \algoReduceSeries.
    }
    \end{subfigure}\hfill
    \begin{subfigure}[t]{0.3\textwidth}
        \centering
       \usetikzlibrary{positioning}
        \begin{tikzpicture}
                    \draw[behindrect] (-1.6,0.5) rectangle (2.25, -3.0);
    \node[main] (8) {};
    \node[main] (9) [below = of 8]{};
    \node[main] (10) [left = of 9]{};
    \draw[virtualtree] (8) -- (9);
    \draw[virtualtree] (9) -- (10);
    \draw[marked] (10) -- (8);

    \node[main] (13) [right = of 8]{};
    \node[main] (14) [below = of 13]{};
    \draw[marked] (13)  to [bend left] (14) ;
    \draw[tree] (13) -- (14);

    \draw[virtualcotree] (13) to [bend right] (14);

    \node[main] (15) [below = of 10]{};
    \node[main] (16) [right = of 15]{};
    \draw[marked] (15) to [bend right] (16);
    \draw[tree] (15) -- (16) ;
    \draw[virtualcotree] (15) to [bend left] (16);

       \draw[black] ($0.5*(9.south)+0.5*(10.south) + (0.0,-0.3)$) -- ($0.5*(15.north)+0.5*(16.north)+ (0.0,0.45)$) ;

    \node[black] (Box3) at ($(8.north) + 0.5*(-0.1,0.3)$) {$\mu_3$ ~\bf{S}};
    \node[black] (Box5) at ($(13.north) + (0.5,0.15)$) {$\mu_5$~\bf{P}};
    \node[black] (Box6) at ($(16.north) + 0.5*(0.0,0.5)$) {$\mu_6$~\bf{P}};

    \draw[black] ($(8.north)+(-1.5,0.35)$) rectangle ($(9.south)+(0.4,-0.3)$);
    \draw[black] ($0.5*(8.north)+0.5*(9.south) + (0.4,0.0)$) -- ($0.5*(13.north)+0.5*(14.south)+ (-0.4,0.0)$) ;

    \draw[black] ($(13.north)+(-0.4,0.35)$) rectangle ($(14.south)+(0.9,-0.3)$);

    \draw[black] ($(15.north)+(-0.35,0.45)$) rectangle ($(16.south)+(0.4,-0.4)$);

    \end{tikzpicture}
    \subcaption{%
      $\T_5,Y_5$ is obtained from $\T_4,Y_4$ by replacing two edges in $\mu_5$ by one edge using Reduction~\ref{def_reductions_C} and by replacing two more edges in $\mu_5$ with one edge using Reduction~\ref{def_reductions_B}, as in \algoReduceParallel.
    }
    \label{fig_reducetreeexample_final}
    \end{subfigure}

    \caption{%
      A sample run of \algoReduceTree (\Cref{algo_reducetree}), applying the reductions to the SPQR-tree $\T$ from \cref{SPQRTreeExampleFigure} for a particular set $Y$.
      SPQR-trees $\T_1,T_2,\dotsc,\T_5$, along with updated edge sets $Y_1,\dots Y_5$ are derived from $\T$ by repeatedly applying reductions to $\T$.
      Virtual edges in the SPQR-tree are given by dashed edges, tree edges are marked in bold and red and all other edges are marked in blue.
      Edges in $Y,Y_1,\dotsc,Y_5$ are indicated by two stripes.
    }
    \label{fig:reducetreeexample}
\end{figure}

Let us now present the complete reduction algorithm \algoReduceTree (\Cref{algo_reducetree}), which performs all possible reductions on $\T$.
It first tries to apply the reductions from \cref{def_reductions} to leaf skeletons of the SPQR-tree.
Then it performs the reductions given by \algoReduceSeries (\Cref{algo_reduceseries}) and \algoReduceParallel (\Cref{algo_reduceParallel}), which remove the local 2-separations contained in~\eqref{node_S} and~\eqref{node_P} nodes. \Cref{fig:reducetreeexample} shows an example of a run of \algoReduceTree.

The procedures for removing local 2-separations are given in \algoReduceSeries and \algoReduceParallel.
In particular, if $|Y_\mu| \geq 2$ for an SPQR-tree node $\mu$ of type~\eqref{node_S} or~\eqref{node_P}, then $Y_\mu$ can be replaced by a single edge using Reduction~\ref{def_reductions_C} due to \cref{thm_full_propagation_cotree}.
Similarly, Reductions~\ref{def_reductions_A} or~\ref{def_reductions_B} can be applied if $|E_\mu\setminus(\spqrNonvirtual(\T_R)\cup Y)| \geq 2$ holds.

\begin{algorithm}[hb]
  \label{algo_reduceParallel}
    \LinesNumbered
    \footnotesize
    \caption{Removing local 2-separations within \eqref{node_P}-nodes}
    \TitleOfAlgo{ReduceParallel$(\mu, \mathcal{T}_R, Y_R)$}
    \KwIn{SPQR-tree node $\mu$ of type~\eqref{node_P}, partially reduced SPQR-tree $\T_R$, marked edges $Y_R$}
    \KwOut{further reduced SPQR-tree $\T_R$, marked edges $Y_R$}
    Let $Y_\mu \coloneqq Y_R \cap E_\mu$. \;
    \lIf{$|Y_\mu| > 1$}{%
        replace $Y_\mu$ by a new edge $e$ in $\T_R$ and update $Y_R \coloneqq (Y_R \cup \{e\})\setminus Y_\mu$. \label{algo_reduceparallel_reduction_1}
    }
    Let $Z \coloneqq E_\mu \setminus (\spqrVirtual(\T_R)\cup Y_R)$. \label{algo_reduceparallel_Z} \;
    \If{$|Z| > 1$}{%
        replace $Z$ by $f$ in $\T_R$.\label{algo_reduceparallel_reduction_2}\;
        \lIf{$T_\mu\cap Z\neq\emptyset$}{
        Update $T_\mu\coloneqq (T_\mu \cup\{f\}) \setminus Z$.
        }
    }
    \lIf{$\spqrVirtual(\T_R) = \emptyset$}{change $\mu$ to type \eqref{node_Q}.}

    \Return $(\T_R,Y_R)$
\end{algorithm}
\begin{algorithm}[ht]
  \label{algo_reduceseries}
    \LinesNumbered
    \footnotesize
    \caption{Removing local 2-separations within \eqref{node_S}-nodes}
    \TitleOfAlgo{ReduceSeries$(\mu,\T_R,Y_R)$}
    \KwIn{SPQR-tree node $\mu$ of type~\eqref{node_S}, partially reduced SPQR-tree $\T_R$, marked edges $Y_R$}
    \KwOut{further reduced SPQR-tree $\T_R$, marked edges $Y_R$}
    Let $Z \coloneqq E_\mu \setminus \spqrVirtual(\T_R)$. \label{algo_reduceseries_Z} \;
    \If{$|Z| \neq |E_\mu|$ and $|Z| > 1$ }{
        Replace $Z$ by a new edge $e$ in $\T_R$. \;
        Let $Y_\mu \coloneqq Z\cap Y_R$. \;
        \lIf{$ Y_\mu \neq \emptyset$}{%
          update $Y_R\coloneqq Y_R\cup \{e\} \setminus Y_\mu $.
        }
        \lIf{$Z \subseteq T_\mu$}{update $T_\mu\coloneqq (T_\mu \cup\{e\})\setminus Z$.
        }\lElse{%
        update $T_\mu\coloneqq T_\mu \setminus Z$.
        }
    }
    \Return $(\T_R,Y_R)$
\end{algorithm}
\newpage
\begin{algorithm}[!hpt]
  \label{algo_reducetree}
  \LinesNumbered
  \small
  \TitleOfAlgo{ReduceTree$(\T, Y)$}
  \caption{Compute the reduced minimal SPQR-tree $\T_R$ of a minimal SPQR-tree $\T$.}
  \KwIn{Minimal SPQR-tree $\T$, edges $Y$}
  \KwOut{Reduced SPQR-tree $\T_R$, edges $Y_R$}
  Let $\T_R \coloneqq \T$ and $Y_R\coloneqq Y$. \;
  Let $\mathcal{L} \subseteq \mathcal{V}$ be a list of all leaves of $\T_R$. \;
  Let $\mathcal{L}_R \coloneqq \emptyset$. \;
  \While(\tcp*[f]{Reductions~\ref{def_reductions_A} and~\ref{def_reductions_B}}){$\mathcal{L}\neq\emptyset$ and $|\V(\T_R)| \geq 2$\label{algo_emptyLoopStart}}{ Let $\mu \in \mathcal{L}$ be an arbitrary leaf of $\T_R$. \;
    Let $\nu$ be the unique neighbor of $\mu$ in $\T_R$. \;
    Update $\mathcal{L} \coloneqq \mathcal{L} \setminus \{\mu\}$. \;
    Let $e \in E(G_\mu)$ and $f \in E(G_\nu)$ be the virtual edge pair connecting $\mu$ and $\nu$. \;
    \uIf{$E_\mu \cap Y_R = \emptyset$}{
      Remove $\mu$ from $\T_R$. \;
      Mark $f$ as a non-virtual edge.\;
      \lIf{$\nu$ is a leaf of $\T_R$}{%
        update $\mathcal{L} \coloneqq \mathcal{L} \cup \{\nu\}$.
      }
    }\Else{
        Update $\mathcal{L}_R \coloneqq \mathcal{L}_R \cup \{\mu\}$.
    }
  } \label{algo_emptyLoopEnd}
  \While(\tcp*[f]{Reductions~\ref{def_reductions_C}}){$\mathcal{L}_R \neq \emptyset$ and $|\V(\T_R)| \geq 2$\label{algo_fullLoopBegin}}{
    Let $\mu \in \mathcal{L}_R$ be an arbitrary leaf of $\T_R$. \;
    Let $\nu$ be the unique neighbor of $\mu$ in $\T_R$. \;
    Update $\mathcal{L}_R \coloneqq \mathcal{L}_R \setminus \{\mu\}$. \;
    Let $e \in E(G_\mu)$ and $f \in E(G_\nu)$ be the virtual edge pair connecting $\mu$ and $\nu$. \;
    \If{$e\in T_\mu$ and $E_\mu \cap Y_R = P^{-1}_{e}(G_\mu,T_\mu)$\label{algo_reducetree_test_full}}{
      Remove $\mu$ from $\T_R$. \;
      Mark $f$ as a non-virtual edge. \;
      \lIf{$\nu$ is a leaf of $\T_R$}{%
        update $\mathcal{L}_R \coloneqq \mathcal{L}_R \cup \{\nu\}$.
      }
      Update $Y_R \coloneqq (Y_R\cup \{f\}) \setminus E_\mu$. \;
    }
  } \label{algo_fullLoopEnd}
  \For(\tcp*[f]{Reductions~\ref{def_reductions_A}, \ref{def_reductions_B} and~\ref{def_reductions_C} within skeletons}){$\mu\in \V(\T_R)$\label{algo_splitLoopStart}}{
    \uIf{$\mu$ is of type \eqref{node_S}}{
      Update $(\T_R,Y_R) \coloneqq\algoReduceSeries(\mu,\T_R,Y_R)$. \;
    }\ElseIf{$\mu$ is of type \eqref{node_P}}{
      Update $(\T_R,Y_R) \coloneqq \algoReduceParallel(\mu, \T_R, Y_R)$. \;
    }
  }
  \label{algo_splitLoopEnd}
  \Return $(\T_R,Y_R)$
\end{algorithm}
\newpage
In order to show that $\algoReduceTree$ outputs an SPQR-tree, it is crucial that we show that the node labels of $\T_R$ are correct.
Thus, we need to consider \algoReduceSeries and \algoReduceParallel, which can modify the skeletons and node labels.

\begin{lemma}
  \label{thm_nodelabelsvalid_reduction}
  Let $\T$ be an SPQR-tree, let $Y \subseteq \spqrNonvirtual(\T)$ be non-empty and let $(\T_R,Y_R)$ be obtained from $\algoReduceTree(\T,Y)$ (\Cref{algo_reducetree}).
  Then each of the following holds:
  \begin{enumerate}
  \item Each $\mu \in \V(\T_R)$ of type~\eqref{node_S} has $|E_\mu| \geq 3$.
  \item Each $\mu\in \V(\T_R)$ of type~\eqref{node_P} has $|E_\mu| \geq 3$.
  \item If $\V(\T_R) = \{\mu\}$, then $\mu$ is not of type~\eqref{node_P}.
  \item If $(\T_R) = \{\mu\}$ of type~\eqref{node_Q}, then $|E_\mu| = 2$.
  \end{enumerate}
\end{lemma}

\begin{proof}
  The only part of \algoReduceTree (\Cref{algo_reducetree}) where skeleton graphs are modified is the loop \ref{algo_splitLoopStart}--\ref{algo_splitLoopEnd}.
  Let $\T'$ and $Y'$ be the tree and the reduced edges before this loop.
  Note that $\T'$ is a minimal SPQR-tree since it is a subtree of $\T$.
  We denote by $E^{\T'}_\mu$ the set of edges of the skeleton $G_\mu$ before execution of the loop, and by $E^{\T_R}_\mu$ the set of edges after execution of the loop.
  Since $\T$ is a valid SPQR-tree, we initially have $|E^{\T'}_\mu| \geq 3$ for each $\mu \in \V(\T')$ of type~\eqref{node_S} or~\eqref{node_P}.

  First, consider the case in which $\V(\T') = \{\mu\}$ holds.
  If $\mu$ is of type \eqref{node_S}, then \algoReduceSeries (\Cref{algo_reduceseries}) does not modify $E^{\T'}_\mu$, which implies $|E^{\T_R}_\mu| = |E^{\T'}_\mu| \geq 3$.
  In case $\mu$ is of type~\eqref{node_P}, then \algoReduceParallel (\Cref{algo_reduceParallel}) changes it to a node of type~\eqref{node_Q}. Because this is the only place in \algoReduceTree where a node's type is changed, it thus cannot occur that $\V(\T_R)=\{\mu\}$ where $\mu$ is of type~\eqref{node_P}, which proves the third statement.

  To show the fourth statement, first note that if $\T$ is a single node of type~\eqref{node_Q} then \algoReduceTree does not change it.
  In the other case where $\T_R$ is a single node that is changed to type~\eqref{node_Q} by \algoReduceParallel, note that we must have $\spqrVirtual_\mu(\T_R) = \emptyset$, which implies that $Y' \cap E^{\T'}_\mu$ and $E^{\T'}_\mu \setminus Y'$ partition the edges of $E_\mu$.
  Since $T_\mu \neq \emptyset$, we have $E^{\T'}_\mu \setminus Y' \neq \emptyset$ is non-empty, and since the reductions preserve existence of edges in $Y$, it follows that $Y' \cap E^{\T'}_\mu \neq \emptyset$.
  Since both of these two sets are replaced by exactly one edge, we have $|E^{\T_R}_\mu| = 2$, showing the fourth statement.

  Second, consider the case in which $|\V(\T')| \geq 2$ holds and $\mu$ is of type~\eqref{node_S}.
  Because \cref{thm_full_propagation_cotree} is applicable if $\mu$ is a leaf of type \eqref{node_S}, $\mu$ must have degree at least $2$, since otherwise $\mu$ would have been removed before reaching line~\ref{algo_splitLoopStart}.
  In particular, this implies $|\spqrVirtual(\T') \cap E_\mu^{\T'}|\geq 2$.
  Since $\T'$ is a valid SPQR-tree,
  $|E_\mu^{\T'}| \geq 3$ holds.
  Moreover, \algoReduceSeries does not remove edges in $\spqrVirtual(\T') \cap E_\mu^{\T'}$ and retains at least one edge $\spqrNonvirtual(\T') \cap E_\mu^{\T'}$ (if present), which shows $|E_\mu^{\T_R}| \geq 3$.

  Third, consider the case in which $|\V(\T')| \geq 2$ holds, and where $\mu$ is a leaf of type \eqref{node_P}, which implies $|\spqrVirtual_\mu(\T')| = 1$.
  Since $\mu$ is a leaf of $\T'$, the set $Y'_\mu = Y' \cap E_\mu$ must be non-empty.
  Additionally, we have that $Z = E^{\T'}_\mu \setminus (\spqrVirtual(T') \cup Y')$, as defined in line~\ref{algo_reduceparallel_Z} of \algoReduceParallel,
  is non-empty, as otherwise \cref{thm_full_propagation_cotree} could be applied to $\mu$ since $E_\mu$ contains exactly one virtual tree edge with all other edges in $Y$.
  Note that each reduction in \algoReduceParallel preserves exactly one edge from $Y_\mu$ and from $Z$.
  Then, $E^{\T_R}_\mu$ contains exactly one edge from each set and the single edge in $\spqrVirtual(\T') \cap E^{\T'}_\mu$, which shows $|E^{\T_R}_\mu| = 3$.

  Finally, consider the case where $|\V(\T')| \geq 2$ holds, and where $\mu$ is of type \eqref{node_P}, but not a leaf, which implies $|\spqrVirtual(\T') \cap E^{\T'}_\mu| \geq 2$.
  Since $E^{\T'}_\mu$ has at least 3 edges, there must exist a third edge $e'$ which lies in either $Y'_\mu$, $Z$ (as defined in line~\ref{algo_reduceparallel_Z} of \algoReduceParallel) or $\spqrVirtual(\T')\cap E^{\T}_\mu$.
  Since $\algoReduceParallel$ preserves at least one edge from $Y'_\mu$ or $Z$ and keeps the edges from $\spqrVirtual(\T) \cap E^{\T}_\mu$, this shows that $|E^{\T_R}_\mu|\geq 3$ holds.
\end{proof}

\begin{theorem}
  \label{thm_reductions_overview}
  Let $\T$ be a minimal SPQR-tree, $Y \subseteq \spqrNonvirtual(\T)$ be non-empty and let $(\T_R,Y_R)$ be obtained from $\algoReduceTree(\T, Y)$ (\Cref{algo_reducetree}).
  Then $\T_R$ is a $Y_R$-reduced and minimal SPQR-tree.
  Moreover, there exists a graph that is represented by $\T$ that has a $Y$-splittable vertex if and only if there exists a graph that is represented by $\T_R$ that has a $Y_R$-splittable vertex.
\end{theorem}

\begin{proof}
  We start by showing that there exists a graph represented by $\T_R$ that has a $Y_R$-splittable vertex.
  First, observe that the loop in lines \ref{algo_emptyLoopStart}--\ref{algo_emptyLoopEnd} of $\algoReduceTree$ (\Cref{algo_reducetree}) performs Reductions~\ref{def_reductions_A} and~\ref{def_reductions_B}, where, for the latter, \cref{thm_empty_propagation_tree2} shows that the possible existence of a $Y_R$-splittable vertex in $\T_R$ is preserved.
  Additionally, after line \ref{algo_emptyLoopEnd} we have that every leaf $\mu\in\V(\T_R)$ has $E_\mu\cap Y_R\neq \emptyset$.

  In particular, this also shows that after the loop exits, no 2-separations that are given by the edges of the SPQR-tree can admit a Reduction~\ref{def_reductions_A} or~\ref{def_reductions_B} (see \cref{thm_nested_reductions}).

  Lines \ref{algo_fullLoopBegin}--\ref{algo_fullLoopEnd} perform Reduction~\ref{def_reductions_C} on the remaining tree $\T_R$.
  In particular, this implies that after exiting the loop, each leaf $\mu$ of $\T_R$ cannot be reduced using Reduction~\ref{def_reductions_C}.
  Note that Reduction~\ref{def_reductions_C} preserves the existence of a $Y_R$-edge in every leaf, which implies that we do not need to check for Reductions~\ref{def_reductions_A} and~\ref{def_reductions_B} again.
  \Cref{thm_nested_reductions} shows that it is sufficient to only consider reductions at edges incident to leaf nodes of the SPQR-tree.
  In particular, it shows that if a 2-separation given by an edge $\{\mu,\nu\} \in \T$ is reducible then all leaves on one side of the subtree formed by removing $\{\mu,\nu\}$ from $\T$ must be reducible.
  Thus, once the loop terminates, no edges of the SPQR-tree can be further reduced.

  Finally, in lines \ref{algo_splitLoopStart}--\ref{algo_splitLoopEnd} we ensure that none of the skeletons $G_\mu$ has local 2-separations that can be reduced.
  Note that we do not check nodes of type~\eqref{node_R} since these are $3$-connected, and nodes of type~\eqref{node_Q} can occur only when the SPQR-tree consists of a single skeleton $G_\mu$ with $|E_\mu| \leq 2$, and thus has no 2-separations at all.

  There is one case, where $\algoReduceParallel$ (\Cref{algo_reduceParallel}) can perform a reduction that we have not explicitly shown to be valid.
  In particular, if $\mu$ is a node of type~\eqref{node_P} and $\spqrVirtual_\mu(\T) = \emptyset$, i.e. $\mu$ is the only node in $\T$, and  $|E_\mu|= 3$ holds, then we still apply Reduction~\ref{def_reductions_B} or~\ref{def_reductions_C} to it, even though we do not have a 2-separation.
  However, it can easily be seen that both nodes in $V_\mu$ are star-nodes, and that their $Y$-splittability is preserved by replacing two edges by a single edge as done in $\algoReduceParallel$.

  For technical reasons, there is one notable exception, which is that \algoReduceSeries (\Cref{algo_reduceseries}) does not perform local reductions on nodes $\mu$ of type~\eqref{node_S} when $E_\mu \cap \spqrVirtual(\T_R) = \emptyset$.
  In this case, $\T_R$ must consist of a single cycle skeleton $\mu$, which implies that $\T_R$ is still $Y_R$-reduced.

  To show that $\T_R$ is an SPQR-tree, we show that each node $\mu\in\V(\T_R)$ is correctly labeled. Since $\T$ is an SPQR-tree, initially all nodes $\mu\in\V(\T)$ are correctly labeled.
  Hence, we only need to check those nodes for which the skeleton $G_\mu$ or the label changes. Then, \cref{thm_nodelabelsvalid_reduction} proves that the node labels of $\T_R$ are still correct.

  To observe minimality of $\T_R$, note that the node labels of $\T$ are not modified, unless we have $\V(\T_R) = \{\mu\}$ where $\mu$ is of type \eqref{node_P} in \algoReduceParallel.
  In this case $\T_R$ consists of a single \eqref{node_Q}-node by \cref{thm_nodelabelsvalid_reduction}, which is clearly minimal.
  Thus, since $\T$ is minimal, $\T_R$ must be minimal, too.

  By \cref{thm_spqr_tree_2sep_local_edge}, all 2-separations of any graph $G$ represented by $\T$ are either local 2-separations or given by an SPQR-tree edge.
  Consequently, after line~\ref{algo_splitLoopEnd} the algorithm has applied reductions from \cref{def_reductions} to all applicable 2-separations of $\T$.
  Since each of the reductions preserves existence of a $Y_R$-splittable vertex in any represented graph $G$, it follows that $\T_R$ is $Y_R$-reduced.
\end{proof}

We analyze the running time of \algoReduceTree in \cref{thm_reducetree_complexity}, except for the step of determining the $Y$-splittable vertices of a single skeleton, which we describe and analyze later.

\begin{lemma}
  \label{thm_reducetree_complexity}
  For a matrix $M \in \{0,1\}^{m \times n}$ and SPQR-tree $\T$ of the graph $G = (V,E)$ with the column edges $Y \subseteq E$,
  $\algoReduceTree(\T, Y)$ (\Cref{algo_reducetree}) runs in $\orderO( \alpha(|E|, m + n) \cdot |E|)$ time plus the running time of line~\ref{algo_reducetree_test_full} for determining the $Y$-splittable vertices of each single skeleton of type~\eqref{node_R}.
\end{lemma}

\begin{proof}
  First, we find the leaves of $\T$, which can be done in $\orderO(|\V(\T)|) = \orderO(|E|)$ time by \cref{thm_SPQRTreeNodeAndSkeletonBound}.
  Then we consider the direct requirements for the reduction of each node $\mu\in \V(\T)$.
  First, $Y_\mu = \emptyset$ can be checked in constant time.
  Second, for the more complicated requirement
  $Y_\mu = P_{e}^{-1}(G_\mu,T_\mu)$, we distinguish several cases.
  The case in which $\mu$ is of type~\eqref{node_R} is left out in the statement of the lemma.

  If $\mu$ is of type~\eqref{node_P} then we simply need to check if $|Y_\mu| = |E_\mu| - 1$ holds, i.e., that all non-tree edges are in $Y_\mu$, which takes $\orderO(|E_\mu|)$ time.
  Nodes of type~\eqref{node_S} are always propagated by \cref{thm_structureReducedSimple}, so deciding this takes $\orderO(1)$ time.
  The reductions performed by \algoReduceParallel and \algoReduceSeries run in $\orderO(|E_\mu|)$ time since they only iterate over the edges and perform set operations.
  All other (sub-)steps are set operations that can be done in $\orderO(1)$ time using the proposed data structures, and happen at most once for each node.
  Thus, the total time for processing each node $\mu$ is bounded by $\orderO(\alpha(|E_\mu|,m+n) \cdot |E_\mu|)$.
  Summing up over all $\mu \in \V(\T)$, we find that the total run time is bounded by $\orderO(\alpha(|E|, m+n) \cdot |E|)$ using \cref{thm_SPQRTreeEdgeBound}.
\end{proof}

\algoReduceTree does not present the most efficient way to perform the reductions. In particular, they can be performed more efficiently by starting with the smallest SPQR-subtree that contains the edges $Y$.
Doing so, one implicitly performs the reductions where $E_i\cap Y = \emptyset$ for $i=1,2$.
Thereafter, such reductions cannot happen because each leaf will have an edge from $Y$, and this property is preserved when applying \cref{thm_full_propagation_cotree}.

Note that we have not yet discussed how to test in line~\ref{algo_reducetree_test_full} of \algoReduceTree whether \cref{thm_full_propagation_cotree} is applicable.
\Cref{thm_full_propagation_3connected} shows that it suffices to determine all $Y$-splittable vertices of the 3-connected skeleton. In the following sections, we will discuss how to (efficiently) find the $Y$-splittable vertices in further detail. More precisely, we will characterize when the reduced SPQR-tree $\T_R$ represents a graph that contains a $Y_R$-splittable vertex, and describe the structure of such a graph.

In order to effectively use the $Y$-reduced SPQR-tree, we must use some of its structural properties. Recall that in \cref{thm_shared_splittable_combine}, we observed that $Y$-splittable vertices that lie on the boundary of a 2-separation can be decomposed into two $Y$-splittability vertices that lie in two subgraphs augmented by an edge. In \cref{thm_shared_splittable}, we show that for graphs realized by the $Y$-reduced SPQR-tree that \emph{every} 2-separation must contain the $Y$-splittable vertex.

\begin{restatable}{lemma}{sharedsplit}
  \label{thm_shared_splittable}
  Let $G$ be a $2$-connected multigraph with spanning tree $T$ and $Y \subseteq E(G) \setminus T$.
  Let $(E_1,E_2)$ be an irreducible 2-separation of $G$ with separating vertices $u$ and $w$ such that $P_{u,w}(T) \subseteq E_1$.
    Let $(G_1,T_1)$ be the graph-tree pair given by  $E_1$ augmented by an edge $e = \{u,w\}$ with $T_1 = E_1 \cap T$ and let $(G_2,T_2)$ be the graph-tree pair formed by $E_2$ augmented by an edge $e_2 = \{u,w\}$ with $T_2 = (E_2 \cap T)\cup\{ e \}$.
    Then no vertex $V(G) \setminus \{u,w\}$ is $Y$-splittable, and $v \in \{u,w\}$ is $Y$-splittable if and only if $v$ is $(Y \cap E_i)$-splittable in $(G_i,T_i)$ for $i=1,2$.
\end{restatable}

\begin{proof}
  First, let us show that no vertex in $V(G) \setminus\{u,w\}$ can be $Y$-splittable.
  Without loss of generality, assume that $P_{u,w}(T)\subseteq E_1$.
  Since Reductions~\ref{def_reductions_A} and~\ref{def_reductions_B} are not applicable, there exists $y_i\in Y\cap E_i$ for $i=1,2$.
  Then by \cref{thm_path_in_twosep_side}, $P_{y_1}(T)\subseteq E_1$ implies that no vertex in $V(G_2)\setminus\{u,w\}$ can be $Y$-splittable.
  Next, consider the case $v \in V(G_1) \setminus\{u,w\}$.
  If there exists a $y \in Y \cap E_2$ such that $v \notin P_y(T)$, then by \cref{thm_MustLieOnPath} $v$ is not $Y$-splittable.
  Otherwise, we must have that $v \in P_{u,w}(T)$ and $Y \cap E_2 \subseteq P^{-1}_{u,w}(G,T)\cap E_2$, which shows that $y_2\in  P^{-1}_{u,w}(G,T)\cap E_2$.
  Since Reduction~\ref{def_reductions_C} is not applicable, equality in the containment above does not hold.
  Hence, there exists an edge $c \in (P^{-1}_{u,w}(G,T)\cap E_2) \setminus Y$.
  However, since $y_2$ and $c$ both connect $T_u$ to $T_w$, they induce a self-loop in $H^v_Y$, which implies that $v$ is not $Y$-splittable.

  Finally, application of \cref{thm_shared_splittable_combine} to $v \in \{u,w\}$ shows the desired result.
\end{proof}

\Cref{thm_shared_splittable} presents the main structural property of $Y$-reduced SPQR-trees that we will exploit to efficiently detect $Y$-splittable vertices in the graphs that are represented by these trees.
However, note that it is only applicable if a 2-separation exists in one of the represented graphs.
The reduced tree $\T_R$ only contains a 2-separation if $|\V(\T_R)|\geq 2$ or if it consists of a single \eqref{node_S}-node with at least four edges.
From this, we can derive a natural case distinction.
First, we will show in the next section how to update the reduced SPQR-tree when $\T_R$ consists of a single node.
In \cref{sec_augmentation_merging}, we consider the case where $\T_R$ contains a 2-separation and has multiple nodes.
Throughout the following sections, we consider the reduced tree $\T_R$ and its marked edges $Y_R$.
For ease of notation, we use $\T$ and $Y$ rather than $\T_R$ and $Y_R$ to indicate the reduced tree and its marked edges.

\section{Updating a single skeleton}
\label{sec_augmentation_singleskeleton}

First, we consider the case where $\V(\T) = \{\mu\}$, so that the reduced tree $\T$ consists of a single skeleton graph.
An important part of this algorithm is to efficiently find $Y$-splittable vertices of the skeletons graphs.
For skeletons of type \eqref{node_S},\eqref{node_P} or \eqref{node_Q}, we will see that finding such a $Y$-splittable vertex is rather easy.
In fact, all vertices are $Y$-splittable for skeletons of this type.
Finding the $Y$-splittable vertex for~\eqref{node_R} nodes whose skeletons are 3-connected is more challenging.
First, we will show a few results that help us to formulate an almost linear-time algorithm to find the $Y$-splittable vertices in $G$.

Two results from \cref{sec_general_row_augmentation} are very relevant in the context of detecting $Y$-splittable vertices.
In \cref{thm_splittableNotStarArticulationLemma}, we observed that a $Y$-splittable vertex $v$ can occur if $Y$ is a star centered at $v$, or if $v$ is an articulation node of $G\setminus Y$.
The former case is somewhat easy to detect and handle, and our focus will be on the second case.
Moreover, in \cref{thm_MustLieOnPath}, we observed that $v$ must lie on $P_y(T)$ for all $y\in Y$; this excludes many possibilities for $Y$-splittable vertices.
In fact, we will use these observations to show that a $3$-connected graph can have at most two $Y$-splittable vertices (if $Y$ is non-empty), and that there are at most four vertices for which one needs to construct the auxiliary graph to check if it is indeed $Y$-splittable.

\begin{lemma}
  \label{thm_4_splittable_candidates}
  Let $G = (V,E)$ be a $3$-connected graph with spanning tree $T\subseteq E$.
  Consider a non-empty set $Y \subseteq E \setminus T$ and let $Q \coloneqq \bigcap_{y\in Y} P_y(T)$ be the intersection of the fundamental paths of all $Y$-edges.
  Then, the following hold.
  \begin{enumerate}
      \item All inner vertices of $Q$ that are articulation vertices of $G \setminus Y$ must be adjacent on $Q$. In particular, there are at most two such vertices.
      \item There are at most four vertices on $Q$ that are articulation vertices of $G\setminus Y$ or the center of a $Y$-star. Moreover, this upper bound is tight.
  \end{enumerate}
\end{lemma}

\begin{proof}
  First, we consider the proof of the first point.
  Note that, due to $Y \neq \emptyset$, $Q$ is a path.
  Let $a_1,a_2 \in V(Q)$ be two inner vertices of $Q$ that are both articulation vertices of $G \setminus Y$, and let $u_1$ and $u_2$ be the end-vertices of $Q$ such that $a_1$ comes first when traversing $Q$ from $u_1$ to $u_2$.
  Suppose that $a_1$ and $a_2$ are not adjacent on $Q$, which means that the sub-path of $Q$ from $a_1$ to $a_2$ has an inner vertex $m$.

  Since $G$ is $3$-connected, $G - \{a_1,a_2\}$ is connected, which implies that each connected component of $(G \setminus Y) - \{a_1,a_2\}$ is connected to some other component via an edge $y \in Y$.
  However, $Q \subseteq P_y(T)$ and the fact that $a_1,a_2$ lie on $Q$ imply that each of these components contains one end-vertex of $Q$.
  It follows that $(G \setminus Y) - \{a_1,a_2\}$ has at most two connected components.
  For $i=1,2$, let $C_i$ be the component that contains $u_i$.
  Assume, without loss of generality, that $m \in C_1$ holds.
  When adding back $a_2$, that is, considering $(G \setminus Y) - \{a_1\}$, the components $C_1$ and $C_2$ are merged into one.
  This contradicts the assumption that $a_1$ is an articulation vertex of $G \setminus Y$, which concludes the proof of the first statement.
  For the second statement, note that by their definition only end-vertices of $Q$ can be the center vertex of a $Y$-star.
  Then, considering the two end-vertices of $Q$ and the two internal vertices of $Q$ as shown by the first statement, the result follows. An example with exactly four vertices is shown in \Cref{fig_3connected_candidatebound_tight}, which proves tightness.
\end{proof}

\tikzset{pathmarked/.style={red, ultra thick, postaction={decorate, decoration={markings, mark=at position 0.5 with {\draw [-,red,ultra thick,solid] (0,-0.7mm) -- (0,0.7mm);}
}}}}

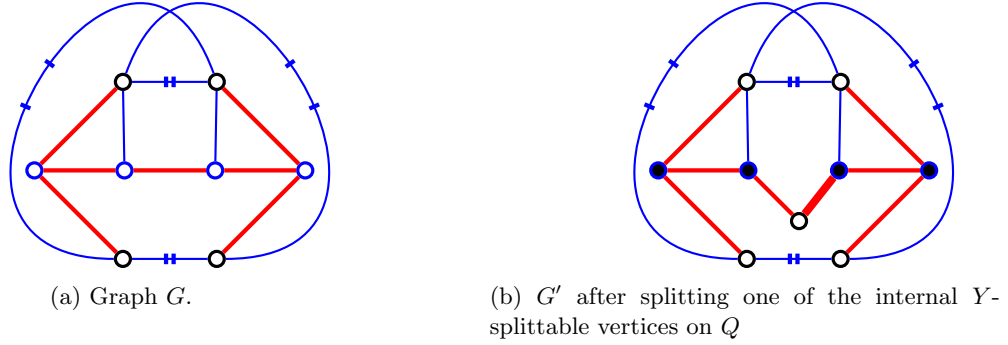
\begin{figure}[htpb]
  \vspace{-2.0cm}
  \begin{subfigure}[t]{0.45\textwidth}
    \centering
    \noindent\hspace{-1.0cm}
    \begin{tikzpicture}
      \node[main,inner sep= 2pt] (a) {};
      \node[main,inner sep= 2pt] (f) [right = of a]{};
      \node[main,fill=black,inner sep= 2pt] (b) [below left = of a]{};
      \node[main,fill=black,inner sep= 2pt] (c) [right = 0.95 cm of b]{};
      \node[main,fill=black,inner sep= 2pt] (e) [below right = of f]{};
      \node[main,fill=black,inner sep= 2pt] (d) [left = 0.95 cm of e]{};

      \node[main,inner sep= 2pt] (g) [below left = of e]{};
      \node[main,inner sep= 2pt] (h) [below right = of b]{};

      \draw[tree] (a) -- (b);
      \draw[pathmarked] (b) -- (c);
      \draw[pathmarked] (c) -- (d);
      \draw[pathmarked] (d) -- (e);
      \draw[tree] (e) -- (f);
      \draw[tree] (e) -- (g);
      \draw[tree] (b) -- (h);

      \draw[cotree] (a) -- (c);
      \draw[cotree] (d) -- (f);

      \draw[marked] (a) -- (f);
      \draw[marked] (a) to [in=0, out=70, looseness =3.0] (g);
      \draw[marked] (h) to [in=110, out=180, looseness =3.0] (f);
      \draw[marked] (h) -- (g);

    \end{tikzpicture}
    \caption{%
      Graph $G$.
    }
      \label{fig_3connected_candidatebound_tight_a}
  \end{subfigure}%
  \hfill
  \begin{subfigure}[t]{0.45\textwidth}
    \centering
    \noindent\hspace{-1.0cm}
    \begin{tikzpicture}
      \node[main,inner sep= 2pt] (a) {};
      \node[main,inner sep= 2pt] (f) [right = of a]{};
      \node[main,fill=black,inner sep= 2pt] (b) [below left = of a]{};
      \node[main,fill=black,inner sep= 2pt] (c) [right = 0.95 cm of b]{};
      \node[main,fill=black,inner sep= 2pt] (e) [below right = of f]{};
      \node[main,fill=black,inner sep= 2pt] (d) [left = 0.95 cm of e]{};

      \node[main,inner sep= 2pt] (g) [below left = of e]{};
      \node[main,inner sep= 2pt] (h) [below right = of b]{};
      \node[main,inner sep= 2pt] (i) [below right = 0.707cm of c]{};

      \draw[tree] (a) -- (b);
      \draw[tree] (b) -- (c);
      \draw[tree] (c) -- (i);
      \draw[tree,line width=3pt] (i) -- (d);
      \draw[tree] (d) -- (e);
      \draw[tree] (e) -- (f);
      \draw[tree] (e) -- (g);
      \draw[tree] (b) -- (h);

      \draw[cotree] (a) -- (c);
      \draw[cotree] (d) -- (f);

      \draw[marked] (a) -- (f);
      \draw[marked] (a) to [in=0, out=70, looseness =3.0] (g);
      \draw[marked] (h) to [in=110, out=180, looseness =3.0] (f);
      \draw[marked] (h) -- (g);

    \end{tikzpicture}
    \caption{$G'$ after splitting one of the internal $Y$-splittable vertices on $Q$.\label{fig_3connected_candidatebound_tight_b}}
  \end{subfigure}%
  \caption{%
    A graph-tree pair where $G\setminus Y$ has four articulation vertices on $Q$.
    Edges in $T$ are marked red and bold, all other edges are marked blue.
    Edges in $Y$ are blue and marked by two stripes. In \subref{fig_3connected_candidatebound_tight_a} edges in $Q$ are marked by one stripe.
    The articulation vertices are marked in black.
    The newly added row edge is marked extra bold.
  }
  \label{fig_3connected_candidatebound_tight}
\end{figure}

In~\cite{Truemper1990}, subroutine TEST-C, Truemper claims that a $3$-connected graph has a graphic row update if and only if it has a single articulation vertex or a star vertex.
However, \Cref{fig_3connected_candidatebound_tight} is a direct contradiction to this claim, since it contains four articulation vertices, of which the two internal to $Q$ are both $Y$-splittable.
In \cref{thm_4_splittable_candidates}, we observed that two articulation vertices of $G \setminus Y$ that are candidates for being $Y$-splittable vertices are always adjacent on $Q$.
In \cref{thm_splittable_adjacent_articulation} and \cref{thm_atmost2_splittable}, we show that more generally, there exist at most two $Y$-splittable vertices, and that if two $Y$-splittable vertices exist that they must be adjacent in $G$.
The structure of two $Y$-splittable vertices is interesting, since it needs to be treated differently during the update step.

\begin{lemma}
  \label{thm_splittable_adjacent_articulation}
  Let $G$ be a $3$-connected graph with spanning tree $T$ and a non-empty set $Y\subseteq E(G)\setminus T$.
  Let $v_1$ and $v_2$ be distinct $Y$-splittable vertices, where $v_1$ is an articulation vertex of $G\setminus Y$.
  Let $Q\coloneqq \bigcap_{y\in Y} P_y(T)$ be the intersection of the fundamental paths of all $Y$-edges.
  Then $v_1$ and $v_2$ must be adjacent on $Q$.
\end{lemma}

\begin{proof}
  First, note that by \cref{thm_MustLieOnPath}, both $v_1$ and $v_2$ must lie on $Q$.
  We distinguish two cases, depending on the structure of $Y$.

\medskip
\noindent
  \textbf{Case 1: $Y \subseteq \delta(v_2)$.} Since $G$ is 3-connected, $G-\{v_1,v_2\} = (G\setminus Y) - \{v_1,v_2\}$ must be connected.
  Since $v_1$ is an articulation vertex of $G \setminus Y$, $G^{v_1}_Y=(G\setminus Y)-v_1$ is disconnected.
  The connectivity of $(G \setminus Y) - \{v_1,v_2\}$ implies that $v_2$ must be disconnected from all other vertices in $G^{v_1}_Y$.
  Thus, we must have in particular that $\delta(v_2)\subseteq\delta(v_1)\cup Y$.
  However, since $T \cap Y = \emptyset$ and $T$ is a spanning tree, there must exist a spanning tree edge in $\delta(v_1) \cap \delta(v_2)$, which clearly belongs to $Q$ since $v_1$ and $v_2$ are on $Q$. Thus, $v_1$ and $v_2$ are adjacent on $Q$.

\medskip
\noindent
  \textbf{Case 2: $Y \not\subseteq \delta(v_2)$.}
  For the sake of contradiction, assume that $v_1$ and $v_2$ are not adjacent on $Q$.
  Then there exists a vertex $m$ on $Q$ that lies between $v_1$ and $v_2$.
  Since $v_2$ is $Y$-splittable, $v_2$ must be an articulation vertex of $G \setminus Y$ by \cref{thm_splittableNotStarArticulationLemma}.
  Since $G$ is 3-connected, $G - \{v_1,v_2\}$ is connected, which implies that each connected component of $(G \setminus Y) - \{v_1,v_2\}$ is connected to some other component via an edge $y \in Y$.
  Let $y^\star \in Y$ be an edge that connects the component $C_1$ containing $m$ to another component $C_2$.
  Furthermore, there can be no edge in $Y$ that connects to the vertices of the component of $T \setminus (\delta(v_1) \cup \delta(v_2))$ containing $m$, since such an edge would imply that either $v_1$ or $v_2$ does not lie on $Q$ because $m$ lies in between them on $T$.

  Then, by symmetry of $v_1$ and $v_2$, we can assume without loss of generality that the unique vertex $u \in C_1 \cap y^\star$ has the property that the $u$-$m$-path within $T$ traverses through $v_1$.
  The other end vertex $w \in C_2 \cap y^\star$ must have a unique $w$-$m$ path in $T$.
  Because we defined $u$ is in the same component as $M$,
  there exists a $u$-$m$ path $P$ in $(G \setminus Y) - \{v_1,v_2\}$.
  Clearly, $P$ also exists in $G \setminus Y - \{v_1\}$.
  When adding back $v_2$ along with its incident edges, that is, considering $G^{v_1}_Y = (G \setminus Y) - \{v_1\}$ , the $w$-$m$ path in $T$ connects $w$ with $m$ again.
  However, then $y^\star$ induces a loop in $H^{v_1}_Y$ that is given by the $w$-$m$ path, $P$ and $y^\star$ itself, contradicting that $v_1$ was $Y$-splittable.
  This concludes the proof.
\end{proof}

\begin{lemma}
  \label{thm_atmost2_splittable}
  Let $G$ be a 3-connected graph with at least four edges, a spanning tree $T$ and a non-empty set $Y \subseteq E(G) \setminus T$.
  Then $G$ has at most two $Y$-splittable vertices. Additionally, if $G$ has two $Y$-splittable vertices then these must be adjacent.
\end{lemma}

\begin{proof}
  Consider two $Y$-splittable vertices $v_1$ and $v_2$.
  If $Y \subseteq \delta(v_1) \cap \delta(v_2)$ then $v_1$ and $v_2$ are connected by some edge from $Y \neq \emptyset$, showing that they are adjacent.
  Otherwise, \cref{thm_splittableNotStarArticulationLemma} shows that one of them is an articulation vertex of $G_Y$ in which case \cref{thm_splittable_adjacent_articulation} shows that $v_1$ and $v_2$ are connected by an edge from $T$.
  This shows that any two $Y$-splittable vertices must be adjacent, which proves the second statement.

  Assume, for the sake of contradiction, that there exist three $Y$-splittable vertices $v_1$, $v_2$ and $v_3$.
  Since $T$ is a tree, they cannot be pair-wise connected via edges from $T$.
  Without loss of generality, assume that $v_1$ and $v_2$ are not adjacent on $T$. Then it follows from $Y$-splittability of $v_1$ and $v_2$ and the first part of the proof that $Y\subseteq \delta(v_1)\cap \delta(v_2)$ holds, since $v_1$ and $v_2$ are not adjacent on $T$.
Note that this implies $Y \not\subseteq \delta(v_3)$, and hence $v_3$ is connected to $v_1$ and $v_2$ by an edge from $T$ since it is a $Y$-splittable vertex.
  Since $G$ has at least four edges, none of which are parallel due to its $3$-connectivity, there must exist a fourth vertex $v_4 \in V(G) \setminus \{v_1,v_2,v_3\}$.
  Since $G$ is $3$-connected, $G - \{v_1,v_3\}$ and $G - \{v_2,v_3\}$ are both connected, which implies that there exist paths $P_1$ and $P_2$ from $v_4$ to $v_1$ and to $v_2$ that do not traverse any of the other two vertices $\{v_2,v_3\}$ or $\{v_1,v_3\}$, respectively.
  Hence, the path $P_1 \cup P_2$ connects $v_1$ with $v_2$ (via $v_4$) within $G_Y^{v_3}$, which induces a self-loop in $H_Y^{v_3}$, which in turn contradicts our assumption that $v_3$ is $Y$-splittable.
\end{proof}

In \cref{thm_adjacent_splittable} we highlight that if $v$ is $Y$-splittable then one can immediately determine the adjacent $Y$-splittable vertex if the neighborhood split has a certain structure.
In particular, the case where the neighborhood split has exactly one edge in one side of its split turns out to be interesting.
We show that for this case, the other vertex incident to this edge is also $Y$-splittable.

\begin{lemma}
  \label{thm_adjacent_splittable}
  Let $G$ be a connected multigraph with spanning tree $T$ and let $Y \subseteq E(G) \setminus T$ be such that $u$ is $Y$-splittable with respect to $G$.
  Let $v \in \delta(u)$ be an adjacent vertex.
  If the neighborhood split of $u$ is such that one part of it is non-empty and contains only edges between $u$ and $v$, then $v$ is $Y$-splittable with respect to $G$.
\end{lemma}

\begin{proof}
  Let $I$ and $J$ denote the bipartition of $H_Y^v$ such that $\delta^I(u)$ is non-empty and only consists of edges between $u$ and $v$.
  We apply the construction of \cref{thm_SplitConstructionProof} to $u$ and obtain the graph $G'$ by splitting $u$ into $i$ and $j$ with new edge $r^\star = \{i,j\}$, updated tree $T' \coloneqq T \cup \{r^\star\}$ and reassigned edges $\delta(u)$.
  Note that all edges from $\delta(u)$ are reassigned to $j$, except for those in $\delta^I(u)$.
  By the assumptions of the lemma, vertex $i$ only has the two neighbors $j$ and $v$.
  We now create the graph $\overline{G}'$ from $G'$ by reassigning the end-vertex of each edge incident to $i$ to the respective other vertex from $\{j,v\}$.
  Thereby, edge $r^\star = \{i,j\}$ is turned into $\overline{r}^\star = \{i,v\}$ and every edge $e \in \{i,v\}$ is turned into $\overline{e} = \{i,j\}$.
  Let $\overline{T}' \coloneqq T \cup \{\overline{r}^\star\}$.
  First, note that the sets of fundamental cycles of $G'$ with respect to $T'$ and of $\overline{G}'$ with respect to $\overline{T}'$ are identical.
  Second, contracting $\overline{r}^\star$ of $\overline{G}'$ also yields $G$.
  This implies that also $v$ must be $Y$-splittable, where application of \cref{thm_SplitConstructionProof} yields the graph $\overline{G}'$.
\end{proof}

Note that the statement of \cref{thm_adjacent_splittable} only shows that if the neighbourhood split has a certain structure, then an adjacent vertex is $Y$-splittable.
In \cref{thm_triconnected_proof_adjacent_split} we show the reverse direction under the additional assumption that $G$ is $3$-connected.
This particular case is relevant because the unique edge in the neighboorhood split must be placed in series with the newly added row in the update step (see \cref{fig_3connected_candidatebound_tight} for an example).
To argue correctness of the series label of the SPQR-tree node that is created, we need to ensure that two adjacent $Y$-splittable vertices can only occur if the neighbourhood split is given by a single edge.

\begin{lemma}
\label{thm_triconnected_proof_adjacent_split}
  Let $G$ be a $3$-connected multigraph with spanning tree $T$ and let $Y \subseteq E(G)\setminus T$ be non-empty such that two adjacent vertices $u,v \in V(G)$ are both $Y$-splittable.
  Then one side of the neighborhood split of $u$ is given by a set of edges that connects $u$ to $v$.
\end{lemma}

\begin{proof}
Since $G$ is 3-connected, $G-\{u,v\}$ is connected.
In particular, this implies that the connected components of $G\setminus Y - \{u,v\}$ are connected to each other by edges in $Y$.

Let $(I,J)$ denote the neighborhood split of $H^u_Y$.
If $u$ and $v$ are both $Y$-stars then we have $Y \subseteq \delta(u) \cap \delta(v)$, and $H^u_Y$ consists of only a single vertex.
Without loss of generality, let this be in $I$.
Then $\delta^J(u) = Y$ holds, which together with $Y \subseteq \delta(u) \cap \delta(v)$ shows that $\delta^J(u)$ consists only of edges connecting $u$ to $v$.

Otherwise, if not both $u$ and $v$ are $Y$-stars then \cref{thm_splittableNotStarArticulationLemma} implies that at least one of them must be an articulation vertex of $G \setminus Y$.
By \cref{thm_splittable_adjacent_articulation}, $u$ and $v$ are adjacent on $Q$, and are thus connected by a spanning tree edge $t\in T$. Since $T$ is a spanning tree and $u$ and $v$ are connected by $t\in T$ and $T\cap Y = \emptyset$, each connected component of $G\setminus Y - \{u,v\}$ must connect to either $u$ or $v$ using a spanning tree edge. Let $\mathcal{H}$ be the connected components of $G\setminus Y - \{u,v\}$. We partition $\mathcal{H}$ into $\mathcal{H}_u$ and $\mathcal{H}_v$, based on whether the component connects to $u$ or $v$.  Let $\mathcal{H}^{u,v}_Y$ be the auxilliary graph where each vertex represents a component of $G\setminus Y - \{u,v\}$. Two vertices $u,v\in \mathcal{H}^{u,v}_Y$ are connected by an edge if the components of $G\setminus Y - \{u,v\}$ are connected by a $Y$-edge.

Now consider $H^v_Y$.
Since $G\setminus Y - \{v\}$ can be obtained by adding back $u$ to $G\setminus Y - \{u,v\}$, we observe that the spanning tree edges connecting to $u$ imply that $\mathcal{H}_{u}$ is merged into one component $h_u \in V(H^v_Y)$.
Since $H^v_Y$ is bipartite, $h_u$ has no self-loops, and hence $\mathcal{H}_u$ must form an independent set in $\mathcal{H}^{u,v}_Y$. Similarly, $\mathcal{H}_v$ is also an independent set of $\mathcal{H}^{u,v}_Y$.

We argued above that $\mathcal{H}^{u,v}_Y$ must be connected, and that $\mathcal{H}_{u}$ and $\mathcal{H}_{v}$ are independent sets in $\mathcal{H}^{u,v}_Y$.
Since $\mathcal{H}_{u}$ and $\mathcal{H}_{v}$ partition the vertex set of $\mathcal{H}^{u,v}_Y$, this shows that $\mathcal{H}^{u,v}_Y$ is bipartite with a unique bipartition that is given by $\mathcal{H}_{u}$ and $\mathcal{H}_{v}$.
In particular, $H^u_Y$ is a star centered at $h_u$ (the component formed by merging all $\mathcal{H}_u$), and the bipartition of $H^u_Y$ is given by $I=\{h_u\}$ and $J = V(H^u_Y)\setminus \{h_u\}$.

Consider $\delta^J(u)$, which consists of edges $e=\{u,w\}$ for which either $e\in Y$ holds or there exists $h\in J$ with $w\in h$.
We show that for $w\neq v$, that existence of a such an edge leads to a contradiction.
In particular, note that $w\neq v$ implies that $w$ must lie in one of the vertices of the components of $\mathcal{H}_u\cup\mathcal{H}_v$.

First, consider the case where $w\in h$ with $h\in I$ and $e\in Y$.
Since $h \in I$ holds, we have $w\in h_u$.
Then, considering $H^v_Y$, existence of edge $e$ implies that $h_u$ has a self-loop, since there must also be a spanning tree edge connecting to $h_u$ from $u$.
This contradicts that $v$ is $Y$-splittable.

Second, consider the case where $w\in h$ with $h\in J$ and $e\notin Y$. Then, $w$ must lie in one of the components $\mathcal{H}_v$. We make a case distinction, based on the cardinality of $\mathcal{H}_u$.

\medskip
\noindent
  \textbf{Case 1: $\mathcal{H}_u= \emptyset$}. Since $\mathcal{H}^{u,v}_Y$ is bipartite and connected, we can conclude that $\mathcal{H}_v$ must consist of a single component.
In particular, this implies that $u$ is a leaf vertex of the spanning tree $T$.
Because we have $T \cap Y = \emptyset$, $u$ cannot be an articulation vertex of $G \setminus Y$, as $T$ spans the other vertices because $u$ is a leaf.
Thus, we must have $Y \subseteq \delta(u)$.
Since we assumed that not both $u$ and $v$ were articulation vertices of $G \setminus Y$, vertex $v$ must be an articulation vertex of $G \setminus Y$.
Since $\mathcal{H}_v$ consists of
a single connected component $h$, this implies that $G \setminus Y - \{v\}$ can only have two components if $h$ and $u$ are not connected using some non-$Y$ edge.
However, then the existence of
$e$ contradicts that $v$ is an articulation vertex of $G \setminus Y$ since it connects $u$ to $w$, implying that we only have a single connected component.

\medskip
\noindent
  \textbf{Case 2: $\mathcal{H}_u\neq \emptyset$.}
Let $h_2$ be the component of $\mathcal{H}_v$ in which $w$ lies.
Since $\mathcal{H}^{u,v}_Y$ is bipartite and connected and $\mathcal{H}_u$ is non-empty, there must exist an member $h'\in\mathcal{H}_u$ that is adjacent to $h_2$ in $\mathcal{H}^{u,v}_Y$.
Then, considering $H^v_Y$, we observe that $h'$ and $h_2$ both connect to $u$ using a edge not in $Y$, and $H^v_Y$ contains a self-loop for this component given by the $Y$ edge connecting $h'$ and $h_2$, which contradicts that $v$ is $Y$-splittable.

For $w\neq v$, we obtained a contradiction in all cases of the case distinctions that we considered. Hence, we conclude that $\delta^J(u)$ can only contain  edges $\{u,v\}$, which completes our proof.
\end{proof}

Next, we use the above insights to formulate algorithms that efficiently find $Y$-splittable vertices within skeletons graphs.
The algorithms that we present in this section do not immediately add a new edge, but rather modify the skeleton $G_\mu$ that will receive the new edge, and return a set of two vertices between which the new tree edge can be added.
We will say that an SPQR-tree that represents such a modified $G_\mu$ is a \emph{$Y$-processed} SPQR-tree.
In \Cref{def_Y_processed_tree}, we formally define $Y$-processed SPQR-trees.

\begin{definition}[$Y$-processed tree]
  \label{def_Y_processed_tree}
  Let $\T = (\V, \E)$ be an SPQR-tree and let $Y \subseteq \spqrNonvirtual(\T)$.
  We call another tree $\T' = (\V', \E')$ with vertices $v_1,v_2 \in V_\mu$ of a skeleton $\mu \in \V'$ the \emph{$Y$-processed tree of $\T$} if
  \begin{enumerate}[label=(\alph*)]
  \item
    \label{def_Y_processed_tree_spqr}
    the tree $\overline{\T}'$ that arises from $\T'$ by adding an edge $e'$ between $v_1$ and $v_2$ (in skeleton $\mu$) is an SPQR-tree,
  \item
    \label{def_Y_processed_tree_split}
    $\overline{\T}'$ represents the set of graph-tree pairs $(G',T')$ for which there exists a graph-tree pair $(G,T)$ represented by $\T$ such that applying the construction of \cref{thm_SplitConstructionProof} to any $Y$-splittable vertex
    yields $G'$ and $T'$, and
  \item
    \label{def_Y_processed_tree_minimal}
    $\overline{\T}'$ is a minimal SPQR-tree.
  \end{enumerate}
\end{definition}

The following lemma highlights that we need to show less than property~\ref{def_Y_processed_tree_split}.

\begin{lemma}
  \label{thm_Y_processed_single_sufficient}
  In order to establish \cref{def_Y_processed_tree}~\ref{def_Y_processed_tree_split} it suffices to consider only one graph-tree pair $(G,T)$ that is represented by $\T$ and only one $Y$-splittable vertex $v$ of $G$ such that the resulting graph-tree pair $(G',T')$ is represented by $\overline{\T}'$.
\end{lemma}

\begin{proof}
  By \cref{thm_spqr_realizations}, the graphs $G$ with spanning trees $T$ represented by $\T$ are those with the same matrix $M \coloneqq M(G,T)$.
  By \cref{thm_graphic_splittable}, applying the construction of \cref{thm_SplitConstructionProof}  to each such $G$ yields a graph $G'$ with spanning tree $T'$, which corresponds to the augmentation of $M$ with the binary vector $\transpose{b}$ with $\supp(b) = Y$ as in~\eqref{eq_augmented_matrix}.
  By construction, the unique resulting matrix $M'$ represents each of the graph-tree pairs $(G',T')$.
  Again by \cref{thm_spqr_realizations}, these are represented by the unique minimal SPQR-tree $\overline{\T}'$.
\end{proof}

 In a few algorithms, the virtual edges $\spqrVirtual_\mu(\T)$ connecting to skeletons of the reduced tree, are mentioned.
While in our case $\spqrVirtual_\mu(\T) = \emptyset$ holds, this will turn out to be useful later because we will re-use the algorithms in the setting where the reduced tree has more than one node.

First, we treat the case where $\mu$ is of type \eqref{node_Q} or \eqref{node_S} after applying \algoReduceTree (\Cref{algo_reducetree}).
It is not difficult to see that in this case, we can extend $\mu$ by elongating the existing cycle with the new edge to create a longer cycle.

\begin{proposition}
  \label{thm_singlecycle}
  Let $\T$ be a $Y$-reduced SPQR-tree with $\V(\T) = \{ \mu \}$ and let $\mu$ be of type \eqref{node_Q} or \eqref{node_S}.
  Let $G'_{\mu,v}$ be the graph obtained by applying the construction of \cref{thm_SplitConstructionProof} to $G_\mu$, $T_\mu$ at vertex $v\in V(G)$. For all $v\in V(G)$, $v$ is $Y$-splittable and $G'_{\mu,v}$ is a cycle of length $|E(G_\mu)| + 1$.
\end{proposition}

\begin{proof}
  Since $Y$ is nonempty and $\mu$ has a unique edge $\{e\} = E_\mu \setminus T_\mu$, we must have $Y = \{e\}$.
  Because $G_\mu$ is of type~\eqref{node_Q} or~\eqref{node_S}, $T_\mu$ forms a path.
  Applying the construction of \cref{thm_SplitConstructionProof} to any internal vertex of the path $T_\mu$, which has degree~2 and has two edges from $T_\mu$ incident to it, we observe that the resulting graph always places exactly one edge in $\delta^I(v)$ and one edge in $\delta^J(v)$.
  This implies that $G'_{\mu,v}$ is indeed a cycle of length $|V(G'_\mu)| + 1$.
  The reasoning for the end-vertices of $T_\mu$ is similar, where a single tree edge and $e$ must always be placed in different neighborhoods of $\delta^I(v)$ and $\delta^J(v)$.
\end{proof}

Note that \cref{thm_singlecycle} shows that \algoExtendSeries (\Cref{algo_extendseries}) is simply a specialized version of the construction of \cref{thm_SplitConstructionProof} for cycles.
In particular, we could also simply use \algoBipartiteSplit (\Cref{algo_bipartitesplit}).
\algoSplitSkeleton (\Cref{algo_splitskeleton}) contains the complete algorithm for splitting a single skeleton.

\newpage

\begin{algorithm}[!ht]
  \label{algo_findsplittablevertices}
  \footnotesize

  \LinesNumbered
  \caption{Find all $Y$-splittable vertices in a skeleton}
  \TitleOfAlgo{FindSplittableVertices$(\T, \mu, Y)$}
  \KwIn{SPQR-tree $\T$, a skeleton $\mu$, edges $Y \subseteq \spqrNonvirtual_\mu(\T) \setminus T_\mu$}
  \KwOut{Set $X \subseteq V(G_\mu)$ of $Y$-splittable vertices of $\mu$}
  \lIf{$Y =\emptyset$ or $\mu$ is of type \eqref{node_Q}, \eqref{node_S} or \eqref{node_P}}{%
    \Return $V(G_\mu)$
  } \label{algo_findsplittablevertices_trivial}
  Let $X \subseteq V(G_\mu)$ be the set of vertices $v$ to which all $y \in Y$ are incident. \label{algo_findsplittablevertices_star} \;
  \lIf{$|X| = 2$}{%
    \Return $X$
  }
  Let $Q \coloneqq V(G_\mu) \cap \bigcap\limits_{y \in Y} V(P_y(T_\mu))$ be the intersection of the vertices of $P_y(T_\mu)$ for all $y \in Y$.\;\label{algo_findsplittablevertices_intersect_path}
  Let $A \subseteq V(G_\mu)$ be the articulation vertices $a$ of $E(G_\mu) \setminus Y$ that also satisfy $a \in Q$. \;
  \For{$a \in A$}{
    Construct $H^a_{Y}$. \;
    \lIf{$H^a_Y$ is bipartite}{%
      update $X \coloneqq X \cup \{a\}$.
    }
  }
  \Return $X$
\end{algorithm}

\begin{algorithm}[!ht]
  \label{algo_findtreesplittablevertices}
    \footnotesize
    \caption{Find all $Y$-splittable vertices in a skeleton that are incident to all virtual edges}
    \LinesNumbered
    \TitleOfAlgo{FindTreeSplittableVertices$(\T, \mu, Y)$}
    \KwIn{SPQR-tree $\T$, a skeleton $\mu$, edges $Y \subseteq \spqrNonvirtual_\mu(\T)$}
  \KwOut{Set $X \subseteq V(G_\mu)$ of $Y$-splittable vertices of $\mu$ that are incident to all virtual edges}
    Let $X \coloneqq $\algoFindSplittableVertices$(\T, \mu, Y)$.\;
    \lFor{$e \in \spqrVirtual_\mu(\T)$}{%
       update $X \coloneqq X \cap e$.
    }
    \Return $X$
\end{algorithm}

\begin{algorithm}[!ht]
  \label{algo_extendseries}
  \footnotesize
  \caption{Extend a series skeleton with a new edge}
  \LinesNumbered
  \TitleOfAlgo{ExtendSeries$(\T, \mu)$}
  \KwIn{SPQR-tree $\T$ with skeleton $\mu$ of type~\eqref{node_Q} or type~\eqref{node_S}}
  \KwOut{$Y$-processed tree $\T'$ with vertices $\{v_1,v_2\}$}
  Obtain $\T'$ from $\T$ by splitting an arbitrary vertex $v \in V(G_\mu)$ into two vertices $v_1$ and $v_2$ such that both have degree~1 in $G_\mu$. \;
  \Return $(\T', \{v_1,v_2\})$
\end{algorithm}

\begin{algorithm}[!ht]
  \label{algo_bipartitesplit}
  \footnotesize
  \LinesNumbered
  \caption{Split a vertex as in \cref{thm_SplitConstructionProof}}
  \TitleOfAlgo{BipartiteSplit$(\T, \mu, Y, v)$}
  \KwIn{SPQR-tree $\T$, a skeleton $\mu$, edges $Y \subseteq \spqrNonvirtual_\mu(\T)$ and a vertex $v \in V_\mu$}
  \KwOut{$Y$-processed tree $\T'$ with vertices $\{v_1,v_2\}$}
  Construct the bipartite graph $H^v_{Y_\mu}$ with bipartition $I,J \subseteq V(H^v_Y)$. \;
  Obtain $\T'$ from $\T$ by the following modifications. \;
  Add the new vertices $v_1$ and $v_2$ to $G_\mu$. \;
  \For{$e = \{u,v\} \in \delta(v)$}{
    \uIf{\underline{either} $e \in Y_\mu$ \underline{or} there is a component $h \in I \text{ such that }u \in h$}{%
      Replace $e$'s end-vertex $v$ by $v_1$.
    }%
    \Else{%
      Replace $e$'s end-vertex $v$ by $v_2$.
    }
  }
  Remove $v$ from $G_\mu$. \;
  \Return $(\T', \{v_1,v_2\})$
\end{algorithm}

\begin{algorithm}[!ht]
  \label{algo_splitskeleton}
  \footnotesize
  \LinesNumbered
  \caption{Construct $Y$-processed tree using a $Y$-splittable vertex that belongs to a given skeleton and is incident to all virtual edges}
  \TitleOfAlgo{SplitSkeleton$(\T, \mu, Y)$}
  \KwIn{SPQR-tree $\T$, a skeleton $\mu$, edges $Y \subseteq \spqrNonvirtual_\mu(\T) \setminus T_\mu$}
  \KwOut{$Y$-processed tree $\T'$ with vertices $\{v_1,v_2\}$ or $(\T, \emptyset)$ if splitting is impossible}
  Let $A \coloneqq \algoFindTreeSplittableVertices(\T, \mu, Y)$. \;
  \uIf{$\mu$ is of type~\eqref{node_Q}}{
      \lIf{$|E_\mu| = 2$}{change $\mu$ to type~\eqref{node_S}.}
      \Return $ \algoExtendSeries(\mu)$
  }
  \uElseIf{$\mu$ is of type~\eqref{node_S}}{
    \lIf{$A = V(G_\mu)$}{%
      \Return $\algoExtendSeries(\mu)$
    }
    \uElseIf{$|A| =1$}{%
      Let $a \in A$ be the unique $Y$-splittable vertex. \;
      \Return $\algoBipartiteSplit(\T, \mu, Y, a)$
    }
    \lElse{\Return $(\T,\emptyset)$}
  }
  \uElseIf{$\mu$ is of type~\eqref{node_P}}{
    Pick an arbitrary $a\in A$. \;
    \Return $\algoBipartiteSplit(\T, \mu, Y, a)$
  }
  \ElseIf{$\mu$ is of type~\eqref{node_R}}{
    \lIf{$|A| = 0$}{\Return $(\T, \emptyset)$}
    \uElseIf{$|A| = 1$}{
      Let $a \in V(G_\mu)$ be such that $A = \{a\}$. \;
      \Return $\algoBipartiteSplit(\T, \mu, Y \cap E_\mu, a)$
    }
    \ElseIf{$|A| = 2$}{
      Let $a_1,a_2 \in V(G_\mu)$ be such that $A = \{a_1,a_2\}$. \;
      Let $e = \{a_1,a_2\}$ be the edge connecting $a_1$ and $a_2$. \;
      Create a new series node $\omega$ and move $e$ from $G_\mu$ to $G_\omega$. \;
      Create a virtual edge pair $(f,g)$ between $\mu$ and $\omega$ such that $f$ has $a_1$ and $a_2$ as end-vertices. \;
      \lIf{$e\in T_\mu$}{update $T_\mu\coloneqq (T_\mu\cup\{e\})\setminus\{f\}$.}
      \Return $\algoExtendSeries(\omega)$
    }
  }
\end{algorithm}

\begin{lemma}
  \label{thm_update_single_cycle}
  Let $\T$ be a $Y$-reduced SPQR-tree with $\V(\T) = \{ \mu \}$ and let $\mu$ be of type \eqref{node_Q} or \eqref{node_S}.
  Then the tree resulting from $\algoSplitSkeleton(\T, \mu, Y)$ (\Cref{algo_splitskeleton}) is $Y$-processed with respect to the vertices returned by the algorithm.
\end{lemma}

\begin{proof}
  \cref{thm_singlecycle} shows that \algoExtendSeries (\Cref{algo_extendseries}) indeed correctly updates $\T$. Note that if $\mu$ is of type~\eqref{node_Q} and has $|E_\mu|=2$ adding an edge to $E_\mu$ creates a cycle of length 3, which shows that $\mu$ must indeed become an~\eqref{node_S} node.
  Clearly, the returned SPQR-tree is minimal since it consists of only $\mu$.
\end{proof}

In \cref{thm_atmost2_splittable}, we fully explored the cases where a 3-connected graph $G$ can have multiple $Y$-splittable points. Now, we are ready to prove correctness of the proposed algorithms.
First, we consider the correctness and running time of $\algoFindSplittableVertices$.

\begin{lemma}
\label{thm_findsplittable_correct}
  Let $\T$ be a $Y$-reduced tree that contains a node $\mu$ of type \eqref{node_R}.
  Then \Cref{algo_findsplittablevertices}, \linebreak $\algoFindSplittableVertices(\T,\mu,Y)$, correctly identifies all $Y$-splittable vertices of $G_\mu$.
\end{lemma}
\begin{proof}
  For the first step which checks if $Y=\emptyset$, \cref{thm_yEmptyLemma} shows that if $Y=\emptyset$ holds, that the algorithm functions correctly.
  In the next steps, we check if $Y$ is a star for \cref{thm_yStarLemma}, and we find all articulation vertices of $G_\mu\setminus Y$ that lie on $Q$ (to check \cref{thm_MustLieOnPath}).
  Then we simply check if the auxiliary graph $H^a_Y$ is bipartite for each articulation vertex $a$ of $G \setminus Y$ that lies on $Q$.
  By \cref{thm_splittableNotStarArticulationLemma}, this is an exhaustive procedure for finding all $Y$-splittable vertices of $G_\mu$.
  \cref{thm_4_splittable_candidates} shows that we need to construct the auxiliary graph for at most four vertices, and \cref{thm_atmost2_splittable} guarantees that we can find at most two $Y$-splittable vertices.
\end{proof}

\begin{lemma}
  For a matrix $M\in\{0,1\}^{m\times n}$, an SPQR-tree $\T$ of the graph $G=(V,E)$ with the column edges $Y\subseteq E$ and a SPQR-tree node $\mu\in\V(\T)$,
  \label{thm_splittable_time_complexity}
  $\algoFindSplittableVertices(\T,\mu,Y)$ (\Cref{algo_findsplittablevertices}) runs in $\orderO(\alpha(|E_\mu|,m+n) \cdot |E_\mu|)$ time.
\end{lemma}

\begin{proof}
  The check in line~\ref{algo_findsplittablevertices_trivial} can be done in $\orderO(1)$ time.
  In line~\ref{algo_findsplittablevertices_star} we find the set of incident vertices, which can be done in $\orderO( \alpha(|\delta(v)|, m+n) \cdot |Y|)$ time.
  Note that the $\alpha(|\delta(v)|,m+n)$ factor comes from the fact that we need to perform \emph{find}-operations for every end-vertex of the edge.

  The intersection of all paths in line~\ref{algo_findsplittablevertices_intersect_path} is more complicated.
  We find the intersection of all paths in $Y$ using so-called \emph{lowest common ancestor} (\emph{LCA}) queries.
  First, we pick an arbitrary root vertex of $T_\mu$ and orient $T_\mu$ so that $T_\mu$ becomes a rooted tree.
  Then we can answer LCA queries on vertices of $T_\mu$ in $\orderO(1)$ by doing a $\orderO(|T_\mu|)$ preprocessing phase (see \cite{Harel1984,Bender2000}).
  Using a constant number of LCA queries, we can find the intersection of two paths $P_y(T_\mu)$ and $P_{y'}(T_\mu)$ in $\orderO(1)$.
  Since the intersection of two paths is again a path, we can repeat this procedure $(|Y|-1)$-times in total to find the intersection of all paths.
  Thus, the total time complexity for this step is $\orderO(|T_\mu| + |Y|)$.
  Note that by using the LCA queries, the final path is not given explicitly, but rather it is given by its two end-vertices in $T_\mu$.
  However, we can again efficiently check in $\orderO(1)$ time if a given vertex $v$ is on this path using LCA queries.
  In the above, we assumed constant time vertex queries from the edges; since we do not have this luxury, the time complexity becomes $\orderO(\alpha(|T| + |Y|, m+n) \cdot (|T|+|Y|))$.

  The articulation vertices of $G_\mu$ can be found in $\orderO(|E_\mu|+|V_\mu|)$  using a depth first search~\cite{Tarjan1972}.
  However, this is assuming that the end-vertices of an edge can be queried in time $\orderO(1)$.
  Since we need to perform an additional \emph{find}-operation for each query, we have a $\orderO(\alpha(|E_\mu|+|V_\mu|,m+n) \cdot (|E_\mu| + |V_\mu|))$ running time.
  Constructing $H^a_Y$, which involves finding the connected components of $G^a_Y$, can similarly be done in $\orderO(\alpha(|E_\mu|+|V_\mu|, m+n) \cdot (|E_\mu| + |V_\mu|))$ time, and determining if $H^a_Y$ is bipartite can also similarly be done using a depth first search over $G^a_Y$ in $\orderO(\alpha(|E_\mu|+|V_\mu|, m+n)|) \cdot (|E_\mu| + |V_\mu|))$ time.
  By \cref{thm_4_splittable_candidates}, we need to compute $H^a_Y$ for at most four candidates $a$, which establishes the total time complexity for this step.

  Since $\delta(v) \subseteq E_\mu$, $Y\subseteq E_\mu$ and $T_\mu\subseteq E_\mu$ hold, each step of \algoFindSplittableVertices runs in $\orderO(\alpha(|E_\mu|+|V_\mu|, m+n) \cdot (|E_\mu| + |V_\mu|))$ time.
  The claimed bound follows from $2$-connectivity since this implies $|V_\mu| \leq |E_\mu|$.
\end{proof}

Then, having shown the running time and validity of $\algoFindSplittableVertices$, we can show the running time and validity of  $\algoSplitSkeleton$, and conclude that it produces a $Y$-processed SPQR-tree.
\begin{lemma}
  \label{thm_update_single_3connected}
  Let $\T$ be a $Y$-reduced tree with $\V(\T) = \{\mu\}$ and let $\mu$ be of type~\eqref{node_R}.
  Let $(\T',X) \coloneqq \algoSplitSkeleton(\T,\mu,Y)$ be the output of \Cref{algo_splitskeleton}.
  If $X = \emptyset$, then $\T$ does not represent a graph $G$ that contains a $Y$-splittable vertex $v$. Otherwise, $X=\{v_1,v_2\}$ and $\T'$ is a $Y$-processed SPQR-tree with respect to $v_1$ and $v_2$.
\end{lemma}

\begin{proof}
  First, note that $\spqrVirtual(\T) = \emptyset$ implies that $\algoFindTreeSplittableVertices(\T,\mu,Y)$ (\Cref{algo_findtreesplittablevertices}) simply returns $\algoFindSplittableVertices(\T,\mu,Y)$ (\Cref{algo_findsplittablevertices}).
  By \cref{thm_findsplittable_correct}, the latter finds all $Y$-splittable vertices of $G_\mu$. By \cref{thm_atmost2_splittable}, there are at most two such $Y$-splittable vertices, and if there are two, they are adjacent to each other.
  Let $A$ denote the set of these $Y$-splittable vertices.
  Then we make a case distinction on $|A|$, as in the algorithm.

  If $|A| = 0$, i.e., $A = \emptyset$, then $G$ has no $Y$-splittable vertex.
  Hence, $\T$ does not represent any graph $G$ that contains a $Y$-splittable vertex.

  If $|A| = 1$, then there is a unique $Y$-splittable vertex $a$ for $G$.
  For this case we are exploiting a result that will be established in the next section due to its generality. \Cref{thm_triconnected_simple} shows that if $G$ has a unique $Y$-splittable vertex, that it remains $3$-connected and simple.
  In particular, \cref{thm_triconnected_simple} shows that the application of $\algoBipartiteSplit$ (\Cref{algo_bipartitesplit}), together with the newly added row edge, gives a new graph $G'$ that is still of type~\eqref{node_R}.
  Clearly, the updated SPQR-tree $\overline{\T'}$ obtained by adding this edge is minimal, as it contains a single \eqref{node_R}-node.

  Finally, if $|A|=2$, we know by \cref{thm_atmost2_splittable} that exactly one edge $e$ (since $G$ is simple) is incident to both vertices in $A$.

  Additionally, by combining \cref{thm_triconnected_proof_adjacent_split} and \cref{thm_SplitConstructionProof}, we observe that $\{e\}$ forms one side of the neighborhood partition for both $Y$-splittable vertices in $A$.
  If we then apply the construction of \cref{thm_SplitConstructionProof} to $G$, we end up with a graph $G'$ where $e$ is replaced by $e$ and the new row edge $r$, put in series.
  The order of $e$ and $r$ depends on the  vertex in $A$ to which \cref{def_SplitConstruction} was applied.
  Then we observe  that the SPQR-tree of $G$ then consists of $\mu$, where $e$ is replaced by a virtual edge which connects to a new node $\nu$ of type~\eqref{node_S}.
  Here, $E_\nu$ consists of a virtual edge connecting to $\mu$, $e$, and the new row edge.
  Since the structure of $G_\mu$ did not change, $G_\mu$ is still $3$-connected and simple.
  Moreover, $|E_\nu| = 3$ holds, so that $\nu$ is indeed of type~\eqref{node_S}.
  Clearly, $\overline{\T}'$ is minimal since it contains only a single \eqref{node_R}--\eqref{node_S} connection.

  For $|A| = 1$ and $|A| = 2$, we obtain a graphic realization $G'$ by splitting a $Y$-splittable vertex in $G$ by \cref{thm_SplitConstructionProof}.

  This shows that \cref{def_Y_processed_tree} \ref{def_Y_processed_tree_split} is satisfied.
  Additionally, we have shown that in both cases, $\overline{\T}'$ is a minimal SPQR-tree.
  Thus, $\T'$ is a $Y$-processed SPQR-tree with respect to $v_1$ and $v_2$ if $X\neq\emptyset$.
\end{proof}

\begin{proposition}
    \label{thm_splitskeleton_runtime}
  For a matrix $M\in\{0,1\}^{m\times n}$, a $Y$-reduced SPQR-tree $\T$ of the graph $G=(V,E)$ with the column edges $Y\subseteq E$ and an SPQR-tree node $\mu\in\V(\T)$,
  \label{thm_splitskeleton_time}
  $\algoSplitSkeleton(\T,\mu,Y)$ (\Cref{algo_splitskeleton}) runs in $\orderO(\alpha(|E_\mu|, m+n) \cdot |E_\mu|)$ time.
\end{proposition}

\begin{proof}
  First, we run \algoFindTreeSplittableVertices (\Cref{algo_findtreesplittablevertices}), which calls \algoFindSplittableVertices (\Cref{algo_findsplittablevertices}).
  This takes $\orderO(\alpha(|E_\mu|, m+n) \cdot |E_\mu|)$ time by \cref{thm_splittable_time_complexity}.
  Note that by $2$-connectivity at most $|V_\mu| = \orderO(|E_\mu|)$ vertices are returned.
  Then it intersects the resulting vertices with at most $|E_\mu|$ virtual edges' end-vertices.
  The intersection of two sets can be found in time linear in their sizes.
  For the first edge, this takes $\orderO(|E_\mu|)$ time. However, since each virtual edge has two vertices, the intersection with the remaining virtual edges takes $\orderO(1)$ time for each edge, since our set to intersect will have at most two vertices.
  Thus, the intersection can be computed in $\orderO(\alpha(|E_\mu|, m+n) \cdot |E_\mu|)$ time by considering the disjoint set data structure overhead.
  Together with the bound for $\algoFindSplittableVertices$, this yields a total running time of $\orderO(\alpha(|E_\mu|,m+n) \cdot |E_\mu|)$ for \algoFindTreeSplittableVertices.

  Then let us consider the other operations of \algoSplitSkeleton.
  In many cases, we execute \algoExtendSeries (\Cref{algo_extendseries}), which runs in constant time since we only split a single vertex with two edges.
  In all other cases,
  \algoBipartiteSplit (\Cref{algo_bipartitesplit}) runs in $\orderO(|\delta(v)|)$ since it reassigns $|\delta(v)|$ edges to new neighbors.

  Clearly, all steps run in $\orderO(\alpha(|E_\mu|, m+n) \cdot |E_\mu|)$ time, which proves the desired result.
\end{proof}

\begin{theorem}
  \label{thm_splitskeleton_yprocessed}
  Let $\T$ be a $Y$-reduced SPQR-tree with $\V(\T) = \{\mu\}$ and consider the output $(\T',X) \coloneqq \algoSplitSkeleton(\T,\mu,Y)$ of \Cref{algo_splitskeleton}.
  If $X = \emptyset$, then $\T$ does not represent a graph $G$ that contains a $Y$-splittable vertex $v$.
  Otherwise, $X = \{v_1,v_2\}$ holds and $\T'$ is a $Y$-processed SPQR-tree with respect to $v_1$ and $v_2$.
\end{theorem}

\begin{proof}
  By \cref{thm_structureReducedSimple}, $\mu$ must be of type \eqref{node_Q}, \eqref{node_S} or \eqref{node_R}.
  If $\mu$ is of type~\eqref{node_Q} or~\eqref{node_S}, then \cref{thm_update_single_cycle} shows that $X \neq \emptyset$ and that $\T'$ is a $Y$-processed tree with respect to $v_1$ and $v_2$.
  If $\mu$ is of type~\eqref{node_R}, then \cref{thm_update_single_3connected} shows the result.
\end{proof}

Note that the proof of \cref{thm_update_single_3connected} refers to \cref{thm_triconnected_simple}, which we have not yet presented. \cref{thm_triconnected_simple} provides a more general statement than needed for \cref{thm_update_single_3connected}, and will be explored in the next section, where we will discuss the more complicated case in which $|\V(\T)| \geq 2$ holds.

\section{Updating multiple skeletons by merging}
\label{sec_augmentation_merging}

In this section, we will consider what happens when the $Y$-reduced tree $\T$ has $|\V(\T)| \geq 2$.
Recall that our main motivation for considering the case $|\V(\T)| \geq 2$ separately, is that we observed in \cref{thm_shared_splittable} that only the boundary vertices of every 2-separation may be $Y$-splittable, and that their $Y$-splittability can be decomposed into the $Y$-splittability for the two graphs forming the 2-separation augmented with an edge.

\sharedsplit*

We start by using \cref{thm_shared_splittable} to show that it suffices to test $Y$-splittability of the individual skeleton nodes of the SPQR-tree.
In particular, we show in \cref{thm_skeleton_splittable_sufficient} that it is sufficient to check the $Y$-splittability for each skeleton separately, which we prove by repeatedly applying \cref{thm_shared_splittable}.

\begin{lemma}
  \label{thm_skeleton_splittable_sufficient}
  Let $\T$ be a $Y$-reduced SPQR-tree with at least two nodes, and let $G$ be represented by $\T$ using the mapping $\Phi_G$ in \cref{def_spqr_represented}, which maps the vertices from the skeletons of $\T$ to the vertices of $G$.
  Then $v \in V(G)$ is $Y$-splittable if and only if for each $\mu \in \V(\T)$, the skeleton $G_\mu$ contains a $(Y \cap E_\mu)$-splittable vertex $v_\mu$ such that $\Phi_G(v_\mu) = v$ holds.
\end{lemma}

\begin{proof}
  We define $\T_0$ as the SPQR-tree $\T$ in which the skeletons $\mu$ of type~\eqref{node_S} are reordered as in step~\ref{def_spqr_represented_permute} of \cref{def_spqr_represented}.
  For $i=1,2,\dotsc,k$, we define $\T_i$ as the tree obtained from $\T_{i-1}$ by carrying out a single step~\ref{def_spqr_represented_merge} from \cref{def_spqr_represented} using the mapping $\Phi_G$ of vertices for an arbitrary edge of $\T_{i-1}$.
  Consequently, $\T_k$ consists of a single node whose skeleton is equal to $G$.
  We consider the following statement for all $i=0,1,\dotsc,k$:
  \begin{equation}
    \text{for each $\mu \in \V(\T_i)$, the skeleton $G_\mu$ contains a $(Y \cap E_\mu)$-splittable vertex $v_\mu$ with $\Phi_G(v_\mu) = v$.}
    \setcounter{equivalenceproofcounter}{\value{equation}}
    \refstepcounter{equation}
    \tag*{(\theequation)$_i$} \label{eq_equivalenceproof}
  \end{equation}

  We now show that \refEquivalenceProofCounter{$i$} implies \refEquivalenceProofCounter{$i+1$}, where $i \in \{0,1,\dotsc,k-1\}$.
  Let $\{\mu, \nu\} \in \E(\T_i)$ be the unique edge of $\T_i$ that does not belong to $\T_{i+1}$, and whose skeletons $G_\mu$ and $G_\nu$ are merged into a new skeleton $G_\xi$.
  Let $v_\mu$ be the $(Y \cap E_\mu)$-splittable vertex with $\Phi_G(v_\mu) = v$ and let $v_\nu$ be the $(Y \cap E_\nu)$-splittable vertex with $\Phi_G(v_\nu) = v$.
  Since for each skeleton, $\Phi$ is injective, $v_\mu$ and $v_\nu$ must be identified to the same vertex in $G_\xi$, which we call $v_\xi$.
  This means that $v_\xi$ is one of the separating vertices of the 2-separation of $G_\xi$ that corresponds to the SPQR-tree edge $\{\mu, \nu\}$.
  \Cref{thm_shared_splittable_combine} implies that also $v_\xi$ is $(Y \cap E_\xi)$-splittable.
  The claim follows from $\Phi(v_\xi) = \Phi(v_\mu) = v$.

  By induction, we obtain that \refEquivalenceProofCounter{$0$} implies \refEquivalenceProofCounter{$k$}, which proves that $v \in V(G)$ being $Y$-splittable.
  Note that this is one direction of the lemma's statement since the premise is exactly \refEquivalenceProofCounter{$0$}.

  We now show the reverse direction, and thus assume that $v \in V(G)$ is $Y$-splittable.
  We consider any skeleton $\mu \in \V(\T)$ and show that its skeleton $G_\mu$ contains a $(Y \cap E_\mu)$-splittable vertex $v_\mu$ such that $\Phi(v_\mu) = v$ holds.
  Due to $|\V(\T_0)| \geq 2$ there exists a virtual edge $\{u,w\} \in E_\mu \cap \spqrVirtual(\T_0)$ that corresponds to a 2-separation of $G$ with separating vertices $\Phi_G(u)$ and $\Phi_G(w)$.
  Since $\T$ is $Y$-reduced, so is $\T_0$, and \cref{thm_shared_splittable} is applicable to this 2-separation of $G$, which implies $v \in \{ \Phi_G(u), \Phi_G(w) \}$.
  This shows that every skeleton $G_\mu$ of $\T_0$ contains a unique vertex $v_\mu$ with $\Phi_G(v_\mu) = v$.
  When applying step~\ref{def_spqr_represented_permute} of \cref{def_spqr_represented}, two such vertices are merged into one of the resulting skeleton.
  Hence, by induction on $i$, also every skeleton $G_\mu$ of $\T_i$ contains a unique vertex $v_\mu$ with $\Phi_G(v_\mu) = v$.

  We now show that \refEquivalenceProofCounter{$i+1$} implies \refEquivalenceProofCounter{$i$}, where $i \in \{0,1,\dotsc,k-1\}$.
  Let again $\{\mu, \nu\} \in \E(\T_i)$ be the unique edge of $\T_i$ that does not belong to $\T_{i+1}$, and whose skeletons $G_\mu$ and $G_\nu$ are merged into a new skeleton $G_\xi$.
  From the previous paragraph we know that $v_\xi$ must be one of the separating vertices of the 2-separation of $G_\xi$ that corresponds to the SPQR-tree edge $\{\mu, \nu\}$.
  Again, \cref{thm_shared_splittable_combine} implies that also $v_\mu$ is $(Y \cap E_\mu)$-splittable and that $v_\nu$ is $(Y \cap E_\nu)$-splittable.
  The claim follows.

  By induction, we obtain that \refEquivalenceProofCounter{$k$} implies \refEquivalenceProofCounter{$0$}.
  The latter statement, i.e., that each skeleton $G_\mu$ of $\T_0$ has a $(Y \cap E_\mu)$-splittable vertex $v_\mu$ with $\Phi_G(v_\mu) = v$ is clearly equivalent to the statement for the original SPQR-tree $\T$, i.e., that each skeleton $G_\mu$ of $\T$ has a $(Y \cap E_\mu)$-splittable vertex $v_\mu$ with $\Phi_G(v_\mu) = v$.
  This concludes the proof.
\end{proof}

\Cref{thm_skeleton_splittable_sufficient} shows that the structure of represented graphs $G$ that contain a $Y$-splittable vertex $v$ is quite restricted.
In particular, $v$ must be obtained by mapping all individual $Y\cap E_\mu$-splittable vertices into one.
Next, we show that this implies that each $v_\mu$ must be incident to all virtual edges within its skeleton.
The immediate algorithmic consequence of this characterization is that it can be checked locally in each skeleton.

\begin{lemma}
  \label{thm_splittable_all_virtual_adjacent}
  Let $\T$ be a $Y$-reduced SPQR-tree with at least two nodes.
  Then $\T$ represents a graph $G$ with a $Y$-splittable vertex if and only if each skeleton $\mu \in \V(\T)$ contains a $(Y \cap E_\mu)$-splittable vertex $v_\mu \in V(G_\mu)$ such that $v_\mu$ is incident to each virtual edge in $\spqrVirtual_\mu(\T)$.
\end{lemma}

\begin{proof}
  We consider any graph $G$ represented by $\T$ with a $Y$-splittable vertex $v$ and will show that each skeleton $\mu \in \V(\T)$ contains a $(Y \cap E_\mu)$-splittable vertex $v_\mu \in V(G_\mu)$ such that $v_\mu$ is incident to each virtual edge in $\spqrVirtual_\mu(\T)$.
  Applying \cref{thm_skeleton_splittable_sufficient}, each $\mu \in \V(\T)$ contains a $(Y \cap E_\mu)$-splittable vertex $v_\mu$ such that $\Phi_G(v_\mu) = v$.
  Since \cref{def_spqr_represented} only alters $\Phi_G$ for vertices that are merged, which are exactly the vertices that are incident to virtual edges, this implies that each $v_\mu$ must be incident to each virtual edge of $\spqrVirtual_\mu(\T)$.

  For the reverse direction, assume that each $\mu \in \V(\T)$ contains a $(Y \cap E_\mu)$-splittable vertex $v_\mu$ such that $v_\mu$ is incident to each virtual edge in $\spqrVirtual_\mu(\T)$.
  For a node $\mu$ of type~\eqref{node_S}, any vertex $v_\mu$ has two incident edges, which implies $|\spqrVirtual_\mu(\T)| \leq 2$.
  Since $\mu$ is of type~\eqref{node_S}, $|E_\mu| \geq 3$ holds, and thus \cref{thm_structureReducedSimple} implies $|\spqrVirtual_\mu(\T)| = 2$ and $|E_\mu| = 3$.
  We create a realization $G$ of $\T$ by mapping $v_\mu$ to $v_\nu$ for each 2-separation given by an edge $\{\mu,\nu\}$ of the SPQR-tree.
  We can ignore the permutation of series members in \cref{def_spqr_represented} since reorienting any series member with $|E_\mu| = 3$ does not change the realized graph.
  Since we map all the $Y\cap E_\mu$-splittable vertices into one vertex, it follows that $\Phi_G(v_\mu) = v$ holds for some vertex $v \in V(G)$ and for all $\mu \in \V(\T)$.
  The result follows by applying \cref{thm_skeleton_splittable_sufficient}.
\end{proof}

When proving the reverse direction in \cref{thm_splittable_all_virtual_adjacent}, we observed that all nodes of type~\eqref{node_S} had degree $2$.
We show that this is indeed necessary and that in this case the $Y$-splittable vertex must be unique.
Our main motivation for doing so is that uniqueness allows us to use \cref{thm_adjacent_splittable} to restrict the structure of the neighbourhood split.

\begin{corollary}
  \label{thm_series_implies_atmost_one}
  Let $\T$ be a $Y$-reduced SPQR-tree with a node $\mu \in \V(\T)$ of type~\eqref{node_S}.
  If $\mu$ has degree $2$ in $\T$, any graph $G$ represented by $\T$ has at most one $Y$-splittable vertex.
  Additionally, if $\mu$ has degree at least 3, then no graph $G$ represented by $\T$ has a $Y$-splittable vertex.
\end{corollary}

\begin{proof}
  We first consider the case where $\mu$ has degree at least $3$.
  Since every vertex of $G_\mu$ has degree $2$, but $|\spqrVirtual_\mu(\T)| \geq 3$ holds, there exists no vertex $v$ incident to each virtual edge in $\spqrVirtual_\mu(\T)$.
  Thus, \cref{thm_splittable_all_virtual_adjacent} implies that there does not exist a graph represented by $\T$ that contains a $Y$-splittable vertex $v$.

  Now consider the case where $\mu$ has degree $2$ and let $G$ be any graph represented by $\T$.
  Since $|\spqrVirtual_\mu(\T)| = 2$ holds, it follows from $|E_\mu| \geq 3$ that $|E_\mu| = 3$ holds by \cref{thm_structureReducedSimple}.
  Thus, there is exactly one vertex $v_\mu$ in $G_\mu$ that is incident to both virtual edges $\spqrVirtual_\mu(\T)$.
  Let $v$ be a $Y$-splittable vertex in $G$.
  By \cref{thm_skeleton_splittable_sufficient}, we have that $\Phi_G(v_\mu) = v$ holds.
  Since $v_\mu$ is the unique $(Y \cap E_\mu)$-splittable vertex in $G_\mu$ that is incident to each virtual edge in $\spqrVirtual_\mu(\T)$ and since $\Phi_G$ is surjective, $v$ must be unique as well.
\end{proof}

Next, we will argue that the previous results imply that minimal SPQR-trees $\T$ with $|\V(\T)| \geq 2$ admit at most one $Y$-splittable vertex.

\begin{lemma}
  \label{thm_uniqueSplittableNode}
  Let $\T$ be a minimal $Y$-reduced SPQR-tree with at least two nodes.
  Then every graph $G$ represented by $\T$ contains at most one $Y$-splittable vertex.
\end{lemma}

\begin{proof}
  Let $G$ be a graph represented by $\T$.
  If $\T$ contains an \eqref{node_S}-node, it has degree~2 or greater by \cref{thm_structureReducedSimple}. Then, $G$ contains at most one $Y$-splittable vertex by \cref{thm_series_implies_atmost_one}. Thus, we can assume that $\T$ does not have an \eqref{node_S}-node.

  By minimality of $\T$, it does not contain two adjacent \eqref{node_P}-nodes, and must thus contain some node of type \eqref{node_R}.
  Now, suppose that $\T$ has a leaf node $\mu$ of type~\eqref{node_R}.
  Let $\{u,w\}$ be the vertices of the 2-separation in $G$ given by the edges $\spqrNonvirtual_\mu(\T)$ and $\spqrNonvirtual(\T) \setminus\spqrNonvirtual_\mu(\T)$.
  Since \cref{thm_full_propagation_cotree} was not applicable, $Y \cap E_\mu \neq P^{-1}_{u,w}(G,T) \cap E_\mu$ holds.
  Then \cref{thm_full_propagation_3connected} implies that at most one of $u$ and $w$ can be $Y$-splittable.
  Consequently, \cref{thm_shared_splittable} implies that at most one vertex of $G$ can be $Y$-splittable, as $u$ and $w$ are the only possible $Y$-splittable vertices of $G$.

  It remains to show the statement for the case in which all leaves are of type \eqref{node_P}.
  Since $\T$ is minimal, \eqref{node_P}-nodes are not adjacent in $\T$.
  Hence, there must exist a non-leaf node $\nu$ of type~\eqref{node_R} with degree at least~2.
  This implies that $\nu$ has at least two virtual edges.
  Since $\nu$ is of type~\eqref{node_R}, $G_\nu$ is simple, and hence there is at most one vertex of $G_\nu$ that is incident to every virtual edge, since $G_\nu$ must have at least two virtual edges.
  By applying \cref{thm_shared_splittable} to each 2-separation of $G$ that corresponds to a virtual edge in $G_\nu$, only such a vertex may be $Y$-splittable.
  This concludes the proof.
\end{proof}

We now utilize the previous uniqueness results.
In particular, we show that if a graph $G$ represented by a $Y$-reduced SPQR-tree has a unique $Y$-splittable vertex, that its updated graph $G'$ is simple and $3$-connected.
This implies that in the SPQR-tree $\T$ is merged into a single node of type~\eqref{node_R} when we perform the neighborhood split.

\begin{theorem}
  \label{thm_triconnected_simple}
  Let $G$ be a $2$-connected multigraph with spanning tree $T \subseteq E(G)$ and a non-empty set of edges $Y \subseteq E(G) \setminus T$ with a unique $Y$-splittable vertex $v$.
  Let $G'$ be the graph constructed by splitting $v$ into $v_1$ and $v_2$, reassigning the edges according to \cref{thm_SplitConstructionProof}  and adding the new edge $r = \{v_1,v_2\}$.
  Then $G'$ has at least four edges.
  Additionally, if $G$ is irreducible then $G'$ is simple and $3$-connected.
\end{theorem}

\begin{proof}
  Assume that $|E(G')| < 4$, which implies that $|E(G)| < 3$.
  Since $Y$ is non-empty and $G$ contains a spanning tree $T$ with $T \cap Y = \emptyset$, we must have $|E(G)| = 2$.
  Since $G$ is 2-connected, $G$ consists of two parallel edges.
  Because  both vertices in $G$ are symmetric, $G$ cannot have a unique $Y$-splittable vertex, which is a contradiction.
  We conclude that $|E(G)| \geq 3$ and thus $|E(G')| \geq 4$ holds.

  Assume, for the sake of contradiction, that $G'$ is not $3$-connected or not simple.
  In either case, $G'$ has a $k$-separation $(E_1',E_2')$ for $k \in \{1,2\}$ with separating vertices $\{u',w'\}$ and the corresponding graphs $G_1'$ and $G_2'$, where we allow $u' = w'$ in case $k=1$.
  Without loss of generality we can assume $r \in E_1'$ and denote by $E_1 \coloneqq E_1' \setminus \{r\}$ and $E_2 \coloneqq E_2'$ a partition of the edge set of $G \coloneqq G' / r$.
  These edge sets belong to the graphs $G_1 \coloneqq G'_1 / r$ and $G_2 \coloneqq G_2'$, respectively.
  We denote by~$u$ and~$w$ the vertices $u'$ and $w'$ after contraction of~$r$, respectively.
  Let $T' \coloneqq T \cup \{r\}$ denote the spanning tree of $G'$ that is obtained from $T$ via \cref{thm_SplitConstructionProof}.

  First, observe that $G_2 = G_2'$ and $G_1 = G_1' / r$ together imply that $V(G_1) \cap V(G_2) = \{u,w\}$ holds, i.e., contraction of~$r$ cannot enlarge the set of vertices shared by the two subgraphs.

  Second, since $(E_1',E_2')$ is a $k$-separation of $G'$ we have $|E_2| = |E_2'| \geq k$.
  Similarly, $|E_1| = |E_1'| - 1 \geq k - 1$ follows.
  For the case $k = 1$, $|E_1| = k - 1 = 0$ would imply $E_1' = \{r\}$.
  Together with the fact that $u' = w'$ is an articulation vertex of $G'$ this would induce a degree~1 vertex in~$G$, which contradicts $2$-connectivity.
  For the case $k = 2$, $|E_1| = k - 1 = 1$ would imply $E_1 = \{e\}$ for some edge $e$.
  Since $u,w,v \in V(G_1)$ hold, these vertices cannot be distinct, say, $v = u$ as well as $e = \{u,w\}$, would hold.
  Moreover, we would have that $G_1'$ is a path of length~2 from $u' = v_1$ to $w'$ via edges $r$ and $e$ with inner vertex $v_2$.
  However, the neighborhood split of $v$ (in $G$) must have been of the form $\delta^I(v) = \{e\}$ and $\delta^J(v) = \delta(v) \setminus \{e\}$, implying $Y$-splittability of $w$ due to \cref{thm_adjacent_splittable}, which contradicts uniqueness of the $Y$-splittable vertex $v \neq w$.
  We conclude that $|E_1| = k - 1$ is not possible, and thus that $(E_1,E_2)$ form a $k$-separation of $G$.
  Clearly, $2$-connectivity of $G$ yields that $k = 1$ is impossible, and hence $k = 2$ must hold.

  Assume now that $G$ is irreducible.
  Hence, Reductions~\ref{def_reductions_A} and~\ref{def_reductions_B} are
  not applicable to the 2-separation $(E_1,E_2)$, which implies $Y \cap E_2 \neq \emptyset$.
  Let $y \in Y \cap E_2$ and note that $r \in P_y(T')$ holds.
  If $P_y(T) \subseteq E_2$ would hold, then $v_1$ and $v_2$ would both belong to $V(G'_2)$ and to $V(G'_1)$ (due to $r \in E'_1$), which would imply $\{v_1,v_2\} = \{u',w'\}$.
  However, the contraction of $r$ would yield an articulation vertex $u = w$ in this case, contradicting that $\{u,w\}$ is a 2-separation.
  We conclude that $P_y(T) \not\subseteq E_2$, i.e., that $P_{u,w}(T) \subseteq E_1$.
  Since this holds for any choice of $y \in Y \cap E_2$, we also have $Y \cap E_2 \subseteq P^{-1}_{u,w}(G,T) \cap E_2$.

  Since also Reduction~\ref{def_reductions_C}
  is not applicable to the 2-separation $(E_1,E_2)$, which implies that $Y \cap E_2 \subsetneqq P^{-1}_{u,w}(G,T) \cap E_2$ holds, i.e., there exists an edge $e \in (P^{-1}_{u,w}(G,T) \setminus Y) \cap E_2$.
  If $v \notin \{u,w\}$ would hold, then the fact that $(T \cap E_2) \cup \{e\}$ forms a spanning tree of $G_2$ implies that all vertices $V(G_2)$ would belong to the same connected component of $G^v_Y$.
  Moreover, each of the edges $y \in Y \cap E_2$ would induce a self-loop in the bipartite graph $H^v_Y$, contradicting that $v$ is $Y$-splittable.
  We conclude that $v \in \{u,w\}$ holds.

  Without loss of generality, we can assume that $v_1 = u'$ and that $v_2 \in V(G_1') \setminus V(G_2')$ holds.
  Let $f_1,f_2 \in \delta(v)$ be the unique edges of the cycle $P_e(T) \cup \{e\}$ that are incident to $v$. Since $P_{u,w}(T)\subseteq P_e(T)$ holds, exactly one of these two edges belongs to $P_{u,w}(T)$ and the other belongs to $E_2$.
  We assume without loss of generality that $f_1 \in P_{u,w(T)}$ and $f_2 \in E_2$.
  Note that by \cref{thm_SplitConstructionProof} we have $r \notin P_e(T')$ and hence $f_1$ and $f_2$ have $v_1$ in common in $G'$, so that $f_1$ is reassigned to $v_1$. This implies that $P_{u',w'}(T') = P_{u,w}(T)$ holds.

  Additionally, there must also exist an edge $y\in P^{-1}_{u,w}(G,T) \cap E_2$ since $Y\cap E_2$ is non-empty. However, since we have $r\in E_1$ and $r\not\in P_{u',w'}(T')$ (since $P_{u,w}(T) = P_{u',w'}(T')$), it follows that $r\notin P_y(T')$, which contradicts that $v$ was a $Y$-splittable vertex which was split using \cref{thm_SplitConstructionProof}.

  We conclude that $G'$ is indeed $3$-connected and simple.
\end{proof}

From the previous results we obtain a full characterization for the case where $|\V(\T)| \geq 2$. \Cref{thm_skeleton_splittable_sufficient}
shows that $\T$ represents a graph $G$ that contains a $Y$-splittable vertex $v$ if and only if we can find a $Y$-splittable vertex $v_\mu$ in the skeleton $G_\mu$ for each node $\mu\in\V(\T)$ such that we can map all $v_\mu$ into one vertex $v$. In \cref{thm_splittable_all_virtual_adjacent} we show that this is equivalent to finding a $Y$-splittable vertex that is incident to all virtual edges $\spqrVirtual_\mu(\T)$ for each $\mu\in\V(\T)$. \cref{thm_uniqueSplittableNode} shows that $v$ must be unique, and \cref{thm_triconnected_simple} then shows that the $Y$-processed SPQR-tree must consist of a single node of type~\eqref{node_R}.
This implies that all skeletons of $\T$ are merged into one big skeleton graph $G'$ in the $Y$-processed tree. We can find the specific realization $G$ where splitting a $Y$-splittable vertex leads to a 3-connected graph by mapping all splittable vertices into one vertex $v$, as in \cref{thm_skeleton_splittable_sufficient}, and then applying the neighborhood split to $v$.

Although we could apply the splitting to the unique $Y$-splittable vertex $v$ of the merged graph $G$ to obtain $G'$, this may be cumbersome as $G$ may be quite big.
Instead, we can already split $v_\mu$ locally in each skeleton, and map the resulting two vertices to the two vertices of $G'$ when performing the contraction outlined in \cref{thm_skeleton_splittable_sufficient}.
We prove this claim in \cref{thm_split_locally}.

\begin{corollary}
  \label{thm_split_locally}
  Let $G$ be a $2$-connected multigraph that has a 2-separation $(E_1,E_2)$ with separating vertices $\{u,w\}$, where $u$ is $Y$-splittable.
  For $i=1,2$, let $G_i$ denote the graph with vertex set $V(E_i)$ and edge set $E_i \cup \{e_i\}$ for a new edge $e_i \coloneqq \{u,w\}$, and let $Y_i \coloneqq Y \cap E_i$.
  Let $G'_i$ be the graph created by splitting $u$ into $u^i_1$ and $u^i_2$ according to the neighborhood split of $u$, which is given by $I_i,J_i\subseteq V(H^u_{Y_i}(G_i))$.
  We distinguish between $I_i$ and $J_i$ by assuming that the connected component of $H^u_{Y_i}(G_i)$ that contains $w$ is in $I_i$.
  Let $G'$ be the graph formed by merging $G'_1$ with $G'_2$, identifying $u^1_1$ with $u^2_1$, $u^1_2$ with $u^2_2$ and $w$ with itself and removing $e_1$ and $e_2$.
  Then $G'$ is equal to the graph obtained from $G$ by performing the neighborhood split on $u$.
\end{corollary}

\begin{proof}
  \Cref{thm_shared_splittable_combine} shows that $H^u_Y(G)$ is bipartite if and only if $H^{u}_{Y_i}(G_i)$ is bipartite for $i=1,2$.
  Let $h_w$ be the component of $H^{u}_Y(G)$ containing vertex $w$ and let, for $i=1,2$, $h^{i}_w$ be the component of $H^{u}_{Y_i}(G_i)$ containing vertex $w$.
  Then the proof of \cref{thm_shared_splittable_combine} shows that $I=I_1\cup I_2\cup\{h_w\} \setminus\{h^1_w,h^2_w\}$ and $J= J_1 \cup J_2$ hold.

  In particular, we observe that
  \[
    \delta^I(u) \coloneqq \{\{u,v\}\in \delta(u) \mid \text{ either } \{u,v\}\in Y \text{ or there exists } h \in I \text{ with } u \in h \}
  \]
  can be partitioned as $\delta^I(u) = \bigcup_{i=1,2} \delta^{I_i}(u) \setminus \{e_i\}$, where
  \[
    \delta^{I_i}(u) = \{\{u,v\}\in \delta(u)\cap E_i \mid \text{ either } \{u,v\}\in Y_i \text{ or there exists } h \in I_i \text{ with } u \in h \}.
  \]
  Similarly, we find that $\delta^J(u) = \delta^{J_1}(u)\cup \delta^{J_2}(u)$.
  Since $\delta^{I_i}$ and $\delta^{J_i}$ describe the neighborhood split of $G_i$, this shows that the neighborhood split of $G$ can be found by performing the neighborhood split on $G_1$ and $G_2$, identifying $u^{1}_i$ with $u^{2}_i$ for $i=1,2$ and $w$ with itself and removing $e_1$ and $e_2$.
  This concludes the proof.
\end{proof}

First, we show that \algoSplitSkeleton (\Cref{algo_splitskeleton}) correctly finds and splits the $Y$-splittable vertices that are incident to each virtual edge.

\begin{lemma}
  \label{thm_splitskeleton_merging_correct}
  Let $\T$ be a $Y$-reduced tree with $|\V(\T)| \geq 2$.
  For any $\mu \in \V(T)$, consider the output $(\T',X) \coloneqq \algoSplitSkeleton(\T,\mu,Y)$ of \Cref{algo_splitskeleton} and let $G_\mu$ and $G'_\mu$ be the skeleton of $\mu$ in $\T$ and $\T'$, respectively.
  Then the following hold:
  \begin{enumerate}
  \item
    $X = \emptyset$ holds if and only if $G_\mu$ does not contain a $(Y \cap E_\mu)$-splittable vertex that is incident to each virtual edge in $\spqrVirtual_\mu(\T)$.
  \item
    If $X \neq \emptyset$ holds, then $X = \{v_1,v_2\}$ and $G'_\mu$ is the graph obtained from the neighborhood split of a $(Y \cap E_\mu)$-splittable vertex $v_\mu$ that is incident to each virtual edge in $\spqrVirtual_\mu(\T)$.
  \end{enumerate}
\end{lemma}

\begin{proof}
  In $\algoSplitSkeleton$ (\Cref{algo_splitskeleton}), the sub-routine $\algoFindTreeSplittableVertices$ (\Cref{algo_findtreesplittablevertices}) first finds all $Y$-splittable vertices by \cref{thm_findsplittable_correct}, and then checks if any intersects all virtual edges $\spqrVirtual_\mu(\T)$.
  Note that $A \neq \emptyset$ holds if $\mu$ is of type~\eqref{node_P}, since the parallel structure implies that both end-vertices are always $Y$-splittable and incident to each (virtual) edge.
  Since $|\V(\T)| \geq 2$ holds, $\mu$ cannot be of type~\eqref{node_Q}.
  If $\mu$ is of type~\eqref{node_R} or~\eqref{node_S}, then we return $X = \emptyset$ when $A = \emptyset$ and return $X \neq \emptyset$ otherwise.
  This shows the first statement.

  For the second point, let us assume $X \neq \emptyset$.
  If $\mu$ is of type~\eqref{node_S}, then it has at least one virtual edge, which implies that the case $A = V(G_\mu)$ does not occur due to $|E_\mu| \geq 3$.
  If $\mu$ is of type~\eqref{node_R}, then the case $|A| = 2$ cannot occur; if $\mu$ is a leaf of $\T$, then \cref{thm_full_propagation_cotree} was not applicable, and \cref{thm_full_propagation_3connected} then shows that at most one vertex incident to a virtual edge can be $Y$-splittable, as otherwise $\mu$ could have been reduced.
  If $\mu$ is not a leaf of $\T$, then there is at most one vertex in $G_\mu$ that is incident to all virtual edges $\spqrVirtual_\mu(\T)$ since $G_\mu$ is simple.
  Since $A \neq \emptyset$, we must have $|A| = 1$.
  Finally, if $\mu$ is of type~\eqref{node_P}, then both vertices in $G_\mu$ are trivially $Y$-splittable and incident to each virtual edge in $\spqrVirtual_\mu(\T)$.
  Hence, in all cases $\algoSplitSkeleton$ runs $\algoBipartiteSplit(\T, \mu, Y, v_\mu)$ (\Cref{algo_bipartitesplit}) on some $(Y \cap E_\mu)$-splittable vertex $v_\mu$ that is incident to each virtual edge in $\spqrVirtual_\mu(\T)$, which shows the second statement.
\end{proof}

\begin{figure}[htpb]
    \begin{subfigure}[t]{0.3\textwidth}
        \centering
       \usetikzlibrary{positioning}
        \begin{tikzpicture}
                    \draw[behindrect] (-1.7,0.5) rectangle (2.25, -3.0);
    \node[main] (8) {};
    \node[main] (9) [below = of 8]{};
    \node[main] (10) [left = of 9]{};
    \draw[virtualtree] (8) -- (9);
    \draw[virtualtree] (9) -- (10);
    \draw[marked] (10) -- (8);

    \node[main] (13) [right = of 8]{};
    \node[main] (14) [below = of 13]{};
    \draw[marked] (13)  to [bend left] (14) ;
    \draw[tree] (13) -- (14);

    \draw[virtualcotree] (13) to [bend right] (14);

    \node[main] (15) [below = of 10]{};
    \node[main] (16) [right = of 15]{};
    \draw[marked] (15) to [bend right] (16);
    \draw[tree] (15) -- (16) ;
    \draw[virtualcotree] (15) to [bend left] (16);

    \draw[black] ($0.5*(9.south)+0.5*(10.south) + (0.0,-0.3)$) -- ($0.5*(15.north)+0.5*(16.north)+ (0.0,0.45)$);

    \node[black] (Box3) at ($(8.north) + 0.5*(-0.1,0.3)$) {$\mu_3$ ~\bf{S}};
    \node[black] (Box5) at ($(13.north) + (0.5,0.15)$) {$\mu_5$~\bf{P}};
    \node[black] (Box6) at ($(16.north) + 0.5*(0.0,0.5)$) {$\mu_6$~\bf{P}};

    \draw[black] ($(8.north)+(-1.5,0.35)$) rectangle ($(9.south)+(0.4,-0.3)$);

    \draw[black] ($0.5*(8.north)+0.5*(9.south) + (0.4,0.0)$) -- ($0.5*(13.north)+0.5*(14.south)+ (-0.4,0.0)$) ;

    \draw[black] ($(13.north)+(-0.4,0.35)$) rectangle ($(14.south)+(0.9,-0.3)$);

    \draw[black] ($(15.north)+(-0.4,0.45)$) rectangle ($(16.south)+(0.4,-0.4)$);
    \end{tikzpicture}
    \subcaption{A $Y$-reduced SPQR-tree $\T$.}
    \end{subfigure}\hfill
    \begin{subfigure}[t]{0.3\textwidth}
        \centering
       \usetikzlibrary{positioning}
        \begin{tikzpicture}
                    \draw[behindrect] (-1.7,0.5) rectangle (2.25, -3.0);
    \node[main] (8) {};
    \node[main] (9) [below = of 8]{};
    \node[main] (10) [left = of 9]{};
    \draw[virtualtree] (8) -- (9);
    \draw[virtualtree] (9) -- (10);
    \draw[marked] (10) -- (8);

    \node[main] (13) [right = of 8]{};
    \node[main] (14) [below = of 13]{};
    \draw[marked] (13)  to [bend left] (14) ;
    \draw[tree] (13) -- (14);

    \draw[virtualcotree] (13) to [bend right] (14);

    \node[main] (15) [below = of 10]{};
    \node[main] (16) [right = of 15]{};
    \node[main] (17) [right = of 16]{};
    \draw[marked] (15) to [bend right] (17) ;
    \draw[tree] (15) -- (16) ;
    \draw[virtualcotree] (15) to [bend left] (16);

    \draw[black] ($0.5*(9.south)+0.5*(10.south) + (0.0,-0.3)$) -- ($0.5*(15.north)+0.5*(16.north)+ (0.0,0.45)$);

    \node[black] (Box3) at ($(8.north) + 0.5*(-0.1,0.3)$) {$\mu_3$ ~\bf{S}};
    \node[black] (Box5) at ($(13.north) + (0.5,0.15)$) {$\mu_5$~\bf{P}};
    \node[black] (Box6) at ($(17.north) + 0.5*(0.0,0.5)$) {$\mu_6$~\bf{P}};

    \draw[black] ($(8.north)+(-1.5,0.35)$) rectangle ($(9.south)+(0.4,-0.3)$);

    \draw[black] ($0.5*(8.north)+0.5*(9.south) + (0.4,0.0)$) -- ($0.5*(13.north)+0.5*(14.south)+ (-0.4,0.0)$) ;

    \draw[black] ($(13.north)+(-0.4,0.35)$) rectangle ($(14.south)+(0.9,-0.3)$);

    \draw[black] ($(15.north)+(-0.4,0.45)$) rectangle ($(17.south)+(0.4,-0.4)$);

    \end{tikzpicture}
        \subcaption{$\T'_1$ obtained from\\ $\algoSplitSkeleton(\T,\mu_6,Y)$.}
    \end{subfigure}\hfill
    \begin{subfigure}[t]{0.3\textwidth}
        \centering
       \usetikzlibrary{positioning}
        \begin{tikzpicture}
                    \draw[behindrect] (-2.7,0.5) rectangle (2.25, -3.0);
    \node[main] (8) {};
    \node[main] (9) [below = of 8]{};
    \node[main] (10) [left = of 9]{};
    \node[main] (11) [left = of 10]{};
    \draw[virtualtree] (8) -- (9);
    \draw[virtualtree] (10) -- (11);
    \draw[marked] (11) -- (8);

    \node[main] (13) [right = of 8]{};
    \node[main] (14) [below = of 13]{};
    \draw[marked] (13)  to [bend left] (14) ;
    \draw[tree] (13) -- (14);

    \draw[virtualcotree] (13) to [bend right] (14);

    \node[main] (15) [below = of 11]{};
    \node[main] (16) [right = of 15]{};
    \node[main] (17) [right = of 16]{};
    \draw[marked] (15) to [bend right] (17) ;
    \draw[tree] (15) -- (16) ;
    \draw[virtualcotree] (15) to [bend left] (16);

    \draw[black] ($0.5*(10.south)+0.5*(11.south) + (0.0,-0.3)$) -- ($0.5*(15.north)+0.5*(16.north)+ (0.0,0.45)$);

    \node[black] (Box3) at ($(8.north) + 0.5*(-0.1,0.3)$) {$\mu_3$ ~\bf{S}};
    \node[black] (Box5) at ($(13.north) + (0.5,0.15)$) {$\mu_5$~\bf{P}};
    \node[black] (Box6) at ($(17.north) + 0.5*(0.0,0.5)$) {$\mu_6$~\bf{P}};

    \draw[black] ($(8.north)+(-2.5,0.35)$) rectangle ($(9.south)+(0.4,-0.3)$);

    \draw[black] ($0.5*(8.north)+0.5*(9.south) + (0.4,0.0)$) -- ($0.5*(13.north)+0.5*(14.south)+ (-0.4,0.0)$) ;

    \draw[black] ($(13.north)+(-0.4,0.35)$) rectangle ($(14.south)+(0.9,-0.3)$);

    \draw[black] ($(15.north)+(-0.2,0.45)$) rectangle ($(17.south)+(0.52,-0.4)$);

    \end{tikzpicture}
            \subcaption{$\T'_2$ obtained from\\ $\algoSplitSkeleton(\T'_1,\mu_3,Y)$.}
    \end{subfigure}

    \begin{subfigure}[t]{0.33\textwidth}
       \usetikzlibrary{positioning}
       \centering
        \begin{tikzpicture}
                    \draw[behindrect] (-2.7,0.5) rectangle (2.25, -1.7);
    \node[main] (8) {};
    \node[main] (9) [below = of 8]{};
    \node[main] (10) [left = of 9]{};
    \node[main] (11) [left of = 10]{};
    \draw[virtualtree] (8) -- (9);
    \draw[tree] (10) -- (11);
    \draw[marked] (9) to [bend left] (11);
    \draw[marked] (11) -- (8);

    \node[main] (13) [right = of 8]{};
    \node[main] (14) [below = of 13]{};
    \draw[marked] (13)  to [bend left] (14) ;
    \draw[tree] (13) -- (14);

    \draw[virtualcotree] (13) to [bend right] (14);

    \node[black] (Box3) at ($(8.north) + 0.5*(0.0,0.3)$) {$\mu_6$ ~\bf{P}};
    \node[black] (Box5) at ($(13.north) + (0.5,0.15)$) {$\mu_5$~\bf{P}};

    \draw[black] ($(8.north)+(-2.5,0.35)$) rectangle ($(9.south)+(0.5,-0.35)$);

    \draw[black] ($0.5*(8.north)+0.5*(9.south) + (0.5,0.0)$) -- ($0.5*(13.north)+0.5*(14.south)+ (-0.4,0.0)$) ;

    \draw[black] ($(13.north)+(-0.4,0.35)$) rectangle ($(14.south)+(0.9,-0.35)$);

    \end{tikzpicture}
                \subcaption{$\T'_3$ obtained from $\T'_2$ by \\ merging $\mu_3$ into $\mu_6$. }
    \end{subfigure}\hfill
    \begin{subfigure}[t]{0.39\textwidth}
        \centering
       \usetikzlibrary{positioning}
        \begin{tikzpicture}
                    \draw[behindrect] (-2.7,0.5) rectangle (3.35, -1.7);
    \node[main] (8) {};
    \node[main] (9) [below = of 8]{};
    \node[main] (10) [left = of 9]{};
    \node[main] (11) [left = of 10]{};
    \draw[virtualtree] (8) -- (9);
    \draw[tree] (10) -- (11);
    \draw[marked] (9) to [bend left] (11);
    \draw[marked] (11) -- (8);

    \node[main] (15) [right = of 9]{};
    \node[main] (14) [right = of 15]{};
    \node[main] (13) [above = of 14]{};

    \draw[marked] (13)  to (15) ;
    \draw[tree] (13) -- (14);

    \draw[virtualcotree] (13) to [bend right] (14);

    \node[black] (Box3) at ($(8.north) + 0.5*(0.0,0.3)$) {$\mu_6$ ~\bf{P}};
    \node[black] (Box5) at ($(13.north) + (0.5,0.15)$) {$\mu_5$~\bf{P}};

    \draw[black] ($(8.north)+(-2.5,0.35)$) rectangle ($(9.south)+(0.5,-0.35)$);

    \draw[black] ($0.5*(8.north)+0.5*(9.south) + (0.5,0.0)$) -- ($0.5*(13.north)+0.5*(14.south)+ (-1.4,0.0)$) ;

    \draw[black] ($(13.north)+(-1.4,0.35)$) rectangle ($(14.south)+(0.9,-0.35)$);

    \end{tikzpicture}
                    \subcaption{$\T'_4$ obtained from \\ $\algoSplitSkeleton(\T'_3,\mu_5,Y)$.}
    \end{subfigure}\hfill
    \begin{subfigure}[t]{0.25\textwidth}
       \usetikzlibrary{positioning}
       \centering
        \begin{tikzpicture}
                    \draw[behindrect] (-2.7,0.5) rectangle (0.65, -1.7);
    \node[main] (8) {};
    \node[main] (9) [below = of 8]{};
    \node[main] (10) [left = of 9]{};
    \node[main] (11) [left = of 10]{};
    \draw[tree] (8) -- (9);
    \draw[tree] (10) -- (11);
    \draw[marked] (9) to [bend left] (11);
    \draw[marked] (11) -- (8);
    \draw[marked] (8)--(10);

    \node[black] (Box3) at ($(8.north) + 0.5*(0.0,0.3)$) {$\mu_6$ ~\bf{R}};

    \draw[black] ($(8.north)+(-2.5,0.35)$) rectangle ($(9.south)+(0.5,-0.35)$);

    \end{tikzpicture}
                    \subcaption{The $Y$-processed tree $\T'_5$ obtained from $\T'_4$ by merging $\mu_5$ into $\mu_6$ and marking $\mu_6$ as type~\eqref{node_R}. }
                    \label{fig_mergetree_example_final}
    \end{subfigure}
    \caption{%
      A run of \algoMergeTree (\Cref{algo_mergetree}) on the $Y$-reduced tree from \Cref{fig_reducetreeexample_final}.
      Virtual edges in the SPQR-tree are given by dashed edges, tree edges are marked in red and bold and all other edges are marked in blue. Edges in $Y$ are marked by two stripes.
    }%
    \label{fig_mergetree_example}
\end{figure}
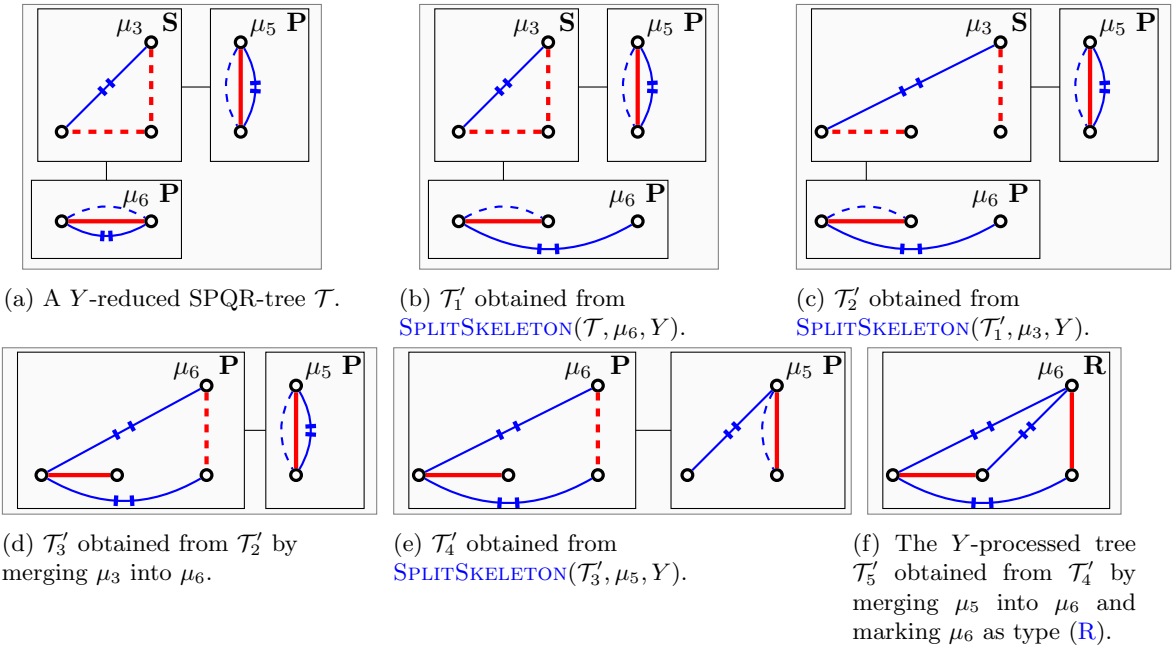

We are ready to state the full algorithm for the case where the $Y$-reduced tree $\T$ has $|\V(\T)| >1$.
$\algoMergeTree$ (\Cref{algo_mergetree}) contains the proposed merging algorithm.
\Cref{fig_mergetree_example} contains an example of the execution of \algoMergeTree on the $Y$-reduced tree from \cref{fig_reducetreeexample_final}.

\begin{algorithm}[H]
  \label{algo_mergetree}
  \footnotesize
  \caption{Merging $\T_R$ into a single node of type~\eqref{node_R}}
    \LinesNumbered
    \TitleOfAlgo{MergeTree$(\T, \T_R, Y_R)$}
    \KwIn{Reduced SPQR-tree $\T$, edges $Y$}
    \KwOut{$Y$-processed SPQR-tree $\T$, vertex set $X'$ such that $X'=\{v_1,v_2\}$ or $X=\emptyset$}
    Let $\mu_1,\mu_2,\dotsc,\mu_n$ be an ordering of $\V(\T)$ such that $\mu_1$ is a leaf and every other $\mu_i$ is adjacent to exactly one node $\mu_j$ with $j < i$. \;
    Let $(\T',X') \coloneqq \algoSplitSkeleton(\T,\mu_1,Y)$.\;
    \lIf{$X' = \emptyset$}{\Return $\T,\emptyset$}
    \For{$i=2,\dotsc,n$}{
        Let $(\T',X) \coloneqq \algoSplitSkeleton(\T',\mu_i,Y$).\;
        \lIf{ $X = \emptyset$}{\Return $\T,\emptyset$}
        Let $e,f$ denote the virtual edge pair connecting $\mu_1$ with $\mu_i$.\;
        Merge $\mu_i$ into $\mu_1$, identifying $X'\cup f$ with $X\cup e$ such that $f\cap X'$ is identified with $e\cap X$, $f\setminus X'$ is identified with $e\setminus X$ and $X'\setminus f$ is identified with $X\setminus e$.\;

        Remove the virtual edges $e$ and $f$ from $\T'$, $T_{\mu_1}$ and $T_{\mu_i}$. \;
        Update $T_{\mu_1} \coloneqq  T_{\mu_1} \cup T_{\mu_i}$.
    }
    Change $\mu_1$ to type~\eqref{node_R}. \;
    \Return $(\T',X)$
\end{algorithm}

Finally, we show that \algoMergeTree (\Cref{algo_mergetree}) indeed returns a $Y$-processed tree $\T$, if one exists.

\begin{theorem}
  \label{thm_mergetree_yprocessed}
  Let $\T$ be a $Y$-reduced SPQR-tree with $|\V(\T)| \geq 2$, and consider the pair $(\T',X' ) \coloneqq \algoMergeTree(\T,Y)$.
  If $X' = \emptyset$, then $\T$ does not represent a graph that contains a $Y$-splittable vertex.
  Otherwise, there exist distinct vertices $v_1$ and $v_2$ such that $X' = \{v_1,v_2\}$ holds, and $\T'$ is a $Y$-processed SPQR-tree of $\T$ with respect to $v_1$ and $v_2$.
\end{theorem}

\begin{proof}
  Let $\overline{\T}'$ denote $\T'$ with additional edge $e'$ between $v_1$ and $v_2$.
  We note that, by the structure of \algoMergeTree (\Cref{algo_mergetree}), either $X' = \emptyset$ or $\V(\T') = \{\mu\}$ must hold, where $\mu$ is of type~\eqref{node_R} after adding $e'$ between $v_1$ and $v_2$.

  If $X' = \emptyset$, this can only occur if $\algoSplitSkeleton(\T,\mu,Y)$ (\Cref{algo_splitskeleton}) returns $X=\emptyset$ for some $\mu \in \V(\T)$.
  \Cref{thm_splitskeleton_merging_correct} implies that $G_\mu$ does not contain a $(Y \cap E_\mu)$-splittable vertex that is incident to each virtual edge in $\spqrVirtual_\mu(\T)$.
  Thus, \cref{thm_splittable_all_virtual_adjacent} shows that $\T$ does not represent any graph that contains a $Y$-splittable vertex.

  If $X'\neq \emptyset$, then applying \cref{thm_splitskeleton_merging_correct} shows that, for each $\mu \in \V(\T)$, the skeleton $G_\mu$ contains a $(Y \cap E_\mu)$-splittable vertex that is incident to each virtual edge in $\spqrVirtual_\mu(\T)$.
  Then \cref{thm_splittable_all_virtual_adjacent} shows that $\T$ represents a graph $G$ that contains a $Y$-splittable vertex $v$.
  \Cref{thm_uniqueSplittableNode} shows that this $Y$-splittable vertex must be unique in $G$, and then it follows from \cref{thm_triconnected_simple} that the graph $G'$ obtained after performing the neighborhood split and adding the new edge must indeed be a skeleton of type~\eqref{node_R} in $\overline{\T}'$.
  In particular, we  have a graph-tree pair $(G,T)$ that is represented by $\T$ with a $Y$-splittable vertex $v$.
  Then, we could simply apply the split construction from \cref{thm_SplitConstructionProof} to obtain the updated graph-tree pair $(G',T')$.
  Then, in order to apply \cref{thm_Y_processed_single_sufficient} to show property \ref{def_Y_processed_tree_split} of \cref{def_Y_processed_tree}, we only need to show that the final graph $G_{\mu_1}$ obtained by $\algoMergeTree$ is equal to such a $G'$.

  \Cref{thm_splitskeleton_merging_correct} shows that $\algoSplitSkeleton(\T, \mu, Y)$ returns a graph $G'_\mu$ obtained from the neighborhood split of a $(Y \cap E_\mu)$-splittable vertex $v_\mu$ that is incident to each virtual edge in $G_\mu$.
  In \algoMergeTree we exploit \cref{thm_split_locally} to combine these neighborhood splits.
  Note that $X \cap e$ and $X' \cap f$ are well-defined and have size $1$ because each virtual edge in $\spqrVirtual_\mu(\T)$ must be incident to $v_\mu$.
  Since the neighborhood split of a vertex moves each virtual edge to be incident to exactly one vertex of $\{v_1,v_2\}$, $X'\setminus f$ and $X \setminus e$ also are single-vertex sets.
  It follows that $e \setminus X$ and $f \setminus X'$ are also single-vertex sets and well-defined.
  By identifying $X \cap e$ with $X' \cap f$, we exactly pick the realization $G$ that maps all $Y\cap E_\mu$-splittable vertices $v_\mu$ into one vertex.
  By using a connected ordering of $\V(\T)$, we ensure that the conditions of \cref{thm_split_locally} hold in subsequent steps, by maintaining one neighborhood split in $\mu_1$.
  Then, after \algoMergeTree terminates, $G_{\mu_1}$ must contain the neighborhood split of the graph obtained by merging all $v_\mu$ into a single vertex $v$. Thus, $G'= G_{\mu_1}$ holds and $\overline{\T}'$ does indeed represent $G'$, which shows property \ref{def_Y_processed_tree_split} of \cref{def_Y_processed_tree}.

  Clearly, $\overline{\T}'$ is a minimal SPQR-tree since it consists of a single node of type~\eqref{node_R} as shown by \cref{thm_triconnected_simple}. Thus, $\T'$ is a $Y$-processed tree with respect to $v_1$ and $v_2$.
\end{proof}

\begin{lemma}
  \label{thm_mergetree_complexity}
    For a matrix $M\in\{0,1\}^{m\times n}$ and $Y$-reduced SPQR-tree $\T$ of the graph $G=(V,E)$ with $Y\subseteq E$,
  $\algoMergeTree(\T,Y)$ (\Cref{algo_mergetree}) runs in $\orderO(\alpha(|E|, m+n) \cdot |E|)$ time.
\end{lemma}

\begin{proof}
  For each skeleton, we execute \algoSplitSkeleton in $\orderO(\alpha(|E_\mu|,m+n) \cdot |E_\mu|)$ time by \cref{thm_splitskeleton_time}.
  Summing over all $\mu\in \V(\T)$, we then find using \cref{thm_SPQRTreeEdgeBound} that this takes $\orderO(\alpha(|E|, m+n) \cdot |E|)$ time in total.
  Merging two skeletons into one includes identification of a constant number of vertices and removal of the virtual edges.
  We do this at most $|\V(\T)|-1$ times, so that we obtain a $\orderO(\alpha(|\V(\T)|,m+n) \cdot |\V(\T)|)$ bound.
  By \cref{thm_SPQRTreeNodeAndSkeletonBound}, this is equivalent to $\orderO(\alpha(|E|,m+n) \cdot |E|)$.
  Because $X$ has size 2, all other set operations in the merging loop have constant time complexity and thus take $\orderO(|\V(\T)|) = \orderO(|E|)$ time in total.
  Thus, all steps can be done in $\orderO(\alpha(|E|, m+n) \cdot |E|)$ time, which concludes our proof.
\end{proof}

\section{Overall algorithm}
\label{sec_algorithm}

In the previous sections we discussed how, given a new row $b$ to add to the graphic matrix $M$, one can reduce and update an SPQR-tree $\T$ in order to represent the matrix $M'=\begin{bmatrix}
    M\\
    b^T
\end{bmatrix}$. Since this is a procedure that we would like to be able to repeat, the SPQR-tree $\T'$ that we find must be minimal and it must represent $M'$.
In our algorithm, the SPQR-tree $\T'$ obtained by performing $\algoReduceTree$ and \algoSplitSkeleton (\Cref{algo_splitskeleton}) or \algoMergeTree (\Cref{algo_mergetree}) might not explicitly represent $M'$, because of the reductions performed in \algoReduceTree (\Cref{algo_reducetree}).
In particular, we have only shown that $\T'$ is $Y$-processed with respect to $\T_R$, the reduced tree, whereas our actual goal is to obtain a $Y$-processed tree with respect to the original tree $\T$.
In order to do so, we can reverse the reductions that we used to derive $\T_R$.
We will use $\widehat{\T}$ to denote the SPQR-tree where these reversions have been performed.

Since reversing the reductions is equivalent to reversing the SPQR-tree operations of $\algoReduceTree$, we will not describe them in full detail.
\Cref{thm_reductions_overview} shows that the reductions preserved splittability of vertices of any graph represented by $\T$ in $\T'$.
Thus, if we perform \cref{thm_SplitConstructionProof}  on some $Y$-splittable vertex $v$ of a graph $G'$ represented by $\T'$ and reverse the reductions, we obtain a graph $\widehat{G}$ with a set of edges $\widehat{Y}$ that are elongated, where $\widehat{Y}$ is precisely the original set of marked edges $Y$.
In particular, $\widehat{G}$ is realized by $\widehat{\T}$.
This is shown in detail by the second statements of \cref{thm_empty_propagation_cotree,thm_empty_propagation_tree1,thm_full_propagation_cotree}, which precisely argue that the sequence of performing the reduction, splitting a $Y$-splittable vertex and undoing the reduction yields the same graph as when we split the $Y$-splittable vertex in the original graph.

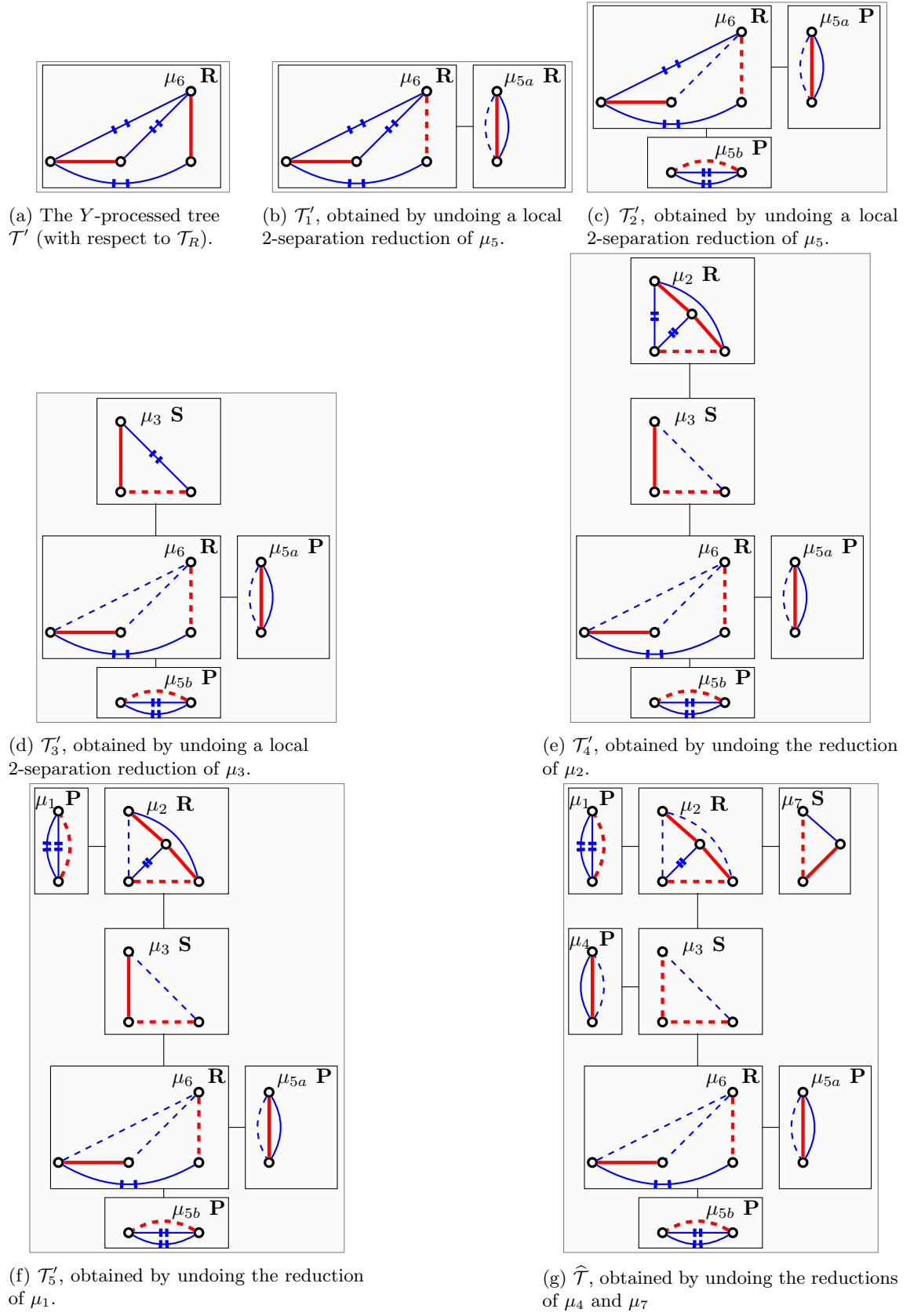
\begin{figure}[htpb]
    \begin{subfigure}[t]{0.28\textwidth}
        \centering
       \usetikzlibrary{positioning}
        \begin{tikzpicture}
                    \draw[behindrect] (-2.6,0.5) rectangle (0.65, -1.7);
    \node[main] (8) {};
    \node[main] (9) [below = of 8]{};
    \node[main] (10) [left = of 9]{};
    \node[main] (11) [left = of 10]{};
    \draw[tree] (8) -- (9);
    \draw[tree] (10) -- (11);
    \draw[marked] (9) to [bend left] (11);
    \draw[marked] (11) -- (8);
    \draw[marked] (8)--(10);

    \node[black] (Box3) at ($(8.north) + 0.5*(0.0,0.3)$) {$\mu_6$ ~\bf{R}};

    \draw[black] ($(8.north)+(-2.5,0.35)$) rectangle ($(9.south)+(0.5,-0.35)$);

    \end{tikzpicture}

    \subcaption{The $Y$-processed tree\\ $\T' $ (with respect to $\T_R$).}
    \end{subfigure}\hfill
    \begin{subfigure}[t]{0.36\textwidth}
        \centering
       \usetikzlibrary{positioning}
        \begin{tikzpicture}
                            \draw[behindrect] (-2.6,0.5) rectangle (2.45, -1.7);
    \node[main] (8) {};
    \node[main] (9) [below = of 8]{};
    \node[main] (10) [left = of 9]{};
    \node[main] (11) [left = of 10]{};
    \draw[virtualtree] (8) -- (9);
    \draw[tree] (10) -- (11);
    \draw[marked] (9) to [bend left] (11);
    \draw[marked] (11) -- (8);
    \draw[marked] (8)--(10);

    \node[main] (12) [right = of 8]{};
    \node[main] (13) [below = of 12]{};
    \draw[virtualcotree] (12) to [bend right] (13);
    \draw[tree] (12) -- (13);
    \draw[cotree] (12) to [bend left] (13);

    \node[black] (Box3) at ($(8.north) + 0.5*(0.0,0.3)$) {$\mu_6$ ~\bf{R}};

    \draw[black] ($(8.north)+(-2.5,0.35)$) rectangle ($(9.south)+(0.5,-0.35)$);

    \node[black] (Box5a) at ($(12.north) + (0.6,0.15)$) {$\mu_{5a}$\bf{ R}};

    \draw[black] ($(12.north)+(-0.4,0.35)$) rectangle ($(13.south)+(1.15,-0.35)$);

    \draw[black] ($0.5*(8.south)+0.5*(9.north) + (0.5,0.0)$) -- ($0.5*(12.north)+0.5*(13.south)+ (-0.4,0.0)$) ;

    \end{tikzpicture}
    \subcaption{$\T'_1$, obtained by undoing a local\\ 2-separation reduction of $\mu_5$.}
    \end{subfigure}\hfill
    \begin{subfigure}[t]{0.35\textwidth}
       \usetikzlibrary{positioning}
        \begin{tikzpicture}
\draw[behindrect] (-2.6,0.5) rectangle (2.45, -2.7);
    \node[main] (8) {};
    \node[main] (9) [below = of 8]{};
    \node[main] (10) [left = of 9]{};
    \node[main] (11) [left = of 10]{};
    \draw[virtualtree] (8) -- (9);
    \draw[tree] (10) -- (11);
    \draw[marked] (9) to [bend left] (11);
    \draw[marked] (11) -- (8);
    \draw[virtualcotree] (8)--(10);

    \node[main] (12) [right = of 8]{};
    \node[main] (13) [below = of 12]{};
    \draw[virtualcotree] (12) to [bend right] (13);
    \draw[tree] (12) -- (13);
    \draw[cotree] (12) to [bend left] (13);

    \node[main] (14) [below = of 9] {};
    \node[main] (15) [below = of 10] {};
    \draw[virtualtree] (14) to [bend right] (15);
    \draw[marked] (14) -- (15);
    \draw[marked] (14) to [bend left] (15);

    \node[black] (Box3) at ($(8.north) + 0.5*(0.0,0.3)$) {$\mu_6$ ~\bf{R}};

    \draw[black] ($(8.north)+(-2.5,0.35)$) rectangle ($(9.south)+(0.5,-0.35)$);

    \node[black] (Box5a) at ($(12.north) + (0.6,0.15)$) {$\mu_{5a}$\bf{ P}};

    \draw[black] ($(12.north)+(-0.4,0.35)$) rectangle ($(13.south)+(1.15,-0.35)$);

    \draw[black] ($0.5*(8.south)+0.5*(9.north) + (0.5,0.0)$) -- ($0.5*(12.north)+0.5*(13.south)+ (-0.4,0.0)$) ;

    \node[black] (Box5b) at ($(14.north) + (0.0,0.3)$) {$\mu_{5b}$\bf{ P}};

    \draw[black] ($(15.north)+(-0.4,0.5)$) rectangle ($(14.north)+(0.5,-0.35)$);

    \draw[black] ($0.5*(9.south)+0.5*(10.south) + (0.0,-0.35)$) -- ($0.5*(14.north)+0.5*(15.north)+ (0.0,0.5)$) ;
    \end{tikzpicture}
    \subcaption{$\T'_2$, obtained by undoing a local 2-separation reduction of $\mu_5$.}    \end{subfigure}

    \begin{subfigure}[t]{0.4\textwidth}
        \centering
       \usetikzlibrary{positioning}
        \begin{tikzpicture}
\draw[behindrect] (-2.6,2.85) rectangle (2.45, -2.7);
    \node[main] (8) {};
    \node[main] (9) [below = of 8]{};
    \node[main] (10) [left = of 9]{};
    \node[main] (11) [left = of 10]{};
    \draw[virtualtree] (8) -- (9);
    \draw[tree] (10) -- (11);
    \draw[marked] (9) to [bend left] (11);
    \draw[virtualcotree] (11) -- (8);
    \draw[virtualcotree] (8)--(10);

    \node[main] (12) [right = of 8]{};
    \node[main] (13) [below = of 12]{};
    \draw[virtualcotree] (12) to [bend right] (13);
    \draw[tree] (12) -- (13);
    \draw[cotree] (12) to [bend left] (13);

    \node[main] (14) [below = of 9] {};
    \node[main] (15) [below = of 10] {};
    \draw[virtualtree] (14) to [bend right] (15);
    \draw[marked] (14) -- (15);
    \draw[marked] (14) to [bend left] (15);

    \node[main] (16) [above = of 8] {};
    \node[main] (17) [left = of 16] {};
    \node[main] (18) [above = of 17] {};
    \draw[virtualtree] (16) -- (17);
    \draw[tree] (17) -- (18);
    \draw[marked] (16) -- (18);

    \node[black] (Box3) at ($(8.north) + 0.5*(0.0,0.3)$) {$\mu_6$ ~\bf{R}};

    \draw[black] ($(8.north)+(-2.5,0.35)$) rectangle ($(9.south)+(0.5,-0.35)$);

    \node[black] (Box5a) at ($(12.north) + (0.6,0.15)$) {$\mu_{5a}$\bf{ P}};

    \draw[black] ($(12.north)+(-0.4,0.35)$) rectangle ($(13.south)+(1.15,-0.35)$);

    \draw[black] ($0.5*(8.south)+0.5*(9.north) + (0.5,0.0)$) -- ($0.5*(12.north)+0.5*(13.south)+ (-0.4,0.0)$) ;

    \node[black] (Box5b) at ($(14.north) + (0.0,0.3)$) {$\mu_{5b}$\bf{ P}};

    \draw[black] ($(15.north)+(-0.4,0.5)$) rectangle ($(14.north)+(0.5,-0.35)$);

    \draw[black] ($0.5*(9.south)+0.5*(10.south) + (0.0,-0.35)$) -- ($0.5*(14.north)+0.5*(15.north)+ (0.0,0.5)$) ;

     \node[black] (Box53a) at ($(18.north) + (0.7,0.0)$) {$\mu_{3}$\bf{ S}};

    \draw[black] ($(18.north)+(-0.4,0.3)$) rectangle ($(16.north)+(0.5,-0.3)$);

    \draw[black] ($0.5*(16.south)+0.5*(17.south) + (0.0,-0.1)$) -- ($0.5*(16.south)+0.5*(17.south) + (0.0,-0.65)$) ;
    \end{tikzpicture}
    \subcaption{$\T'_3$, obtained by undoing a local \\ 2-separation reduction of $\mu_3$.}
    \end{subfigure}\hfill
    \begin{subfigure}[t]{0.4\textwidth}
        \centering
       \usetikzlibrary{positioning}
        \begin{tikzpicture}

\draw[behindrect] (-2.6,5.2) rectangle (2.45, -2.7);
    \node[main] (8) {};
    \node[main] (9) [below = of 8]{};
    \node[main] (10) [left = of 9]{};
    \node[main] (11) [left = of 10]{};
    \draw[virtualtree] (8) -- (9);
    \draw[tree] (10) -- (11);
    \draw[marked] (9) to [bend left] (11);
    \draw[virtualcotree] (11) -- (8);
    \draw[virtualcotree] (8)--(10);

    \node[main] (12) [right = of 8]{};
    \node[main] (13) [below = of 12]{};
    \draw[virtualcotree] (12) to [bend right] (13);
    \draw[tree] (12) -- (13);
    \draw[cotree] (12) to [bend left] (13);

    \node[main] (14) [below = of 9] {};
    \node[main] (15) [below = of 10] {};
    \draw[virtualtree] (14) to [bend right] (15);
    \draw[marked] (14) -- (15);
    \draw[marked] (14) to [bend left] (15);

    \node[main] (16) [above = of 8] {};
    \node[main] (17) [left = of 16] {};
    \node[main] (18) [above = of 17] {};
    \draw[virtualtree] (16) -- (17);
    \draw[tree] (17) -- (18);
    \draw[virtualcotree] (16) -- (18);

    \node[main] (19) [above = of 18]{};
    \node[main] (20) [right = of 19]{};
    \node[main] (21) [above = of 19]{};
    \node[main] (22) [above right = 0.707cm of 19]{};
    \draw[virtualtree] (19) -- (20);
    \draw[marked] (19) -- (21);
    \draw[marked] (19) -- (22);
    \draw[tree] (20) -- (22);
    \draw[tree] (21) -- (22);
    \draw[cotree] (21) to [bend left] (20);

    \node[black] (Box3) at ($(8.north) + 0.5*(0.0,0.3)$) {$\mu_6$ ~\bf{R}};

    \draw[black] ($(8.north)+(-2.5,0.35)$) rectangle ($(9.south)+(0.5,-0.35)$);

    \node[black] (Box5a) at ($(12.north) + (0.6,0.15)$) {$\mu_{5a}$\bf{ P}};

    \draw[black] ($(12.north)+(-0.4,0.35)$) rectangle ($(13.south)+(1.15,-0.35)$);

    \draw[black] ($0.5*(8.south)+0.5*(9.north) + (0.5,0.0)$) -- ($0.5*(12.north)+0.5*(13.south)+ (-0.4,0.0)$) ;

    \node[black] (Box5b) at ($(14.north) + (0.0,0.3)$) {$\mu_{5b}$\bf{ P}};

    \draw[black] ($(15.north)+(-0.4,0.5)$) rectangle ($(14.north)+(0.5,-0.35)$);

    \draw[black] ($0.5*(9.south)+0.5*(10.south) + (0.0,-0.35)$) -- ($0.5*(14.north)+0.5*(15.north)+ (0.0,0.5)$) ;

     \node[black] (Box3a) at ($(18.north) + (0.7,0.0)$) {$\mu_{3}$\bf{ S}};

    \draw[black] ($(18.north)+(-0.4,0.3)$) rectangle ($(16.north)+(0.5,-0.3)$);

    \draw[black] ($0.5*(16.south)+0.5*(17.south) + (0.0,-0.1)$) -- ($0.5*(16.south)+0.5*(17.south) + (0.0,-0.65)$) ;

         \node[black] (Box2) at ($(21.north) + (0.7,0.0)$) {$\mu_{2}$\bf{ R}};

    \draw[black] ($(21.north)+(-0.4,0.3)$) rectangle ($(20.north)+(0.5,-0.3)$);

    \draw[black] ($0.5*(19.south)+0.5*(20.south) + (0.0,-0.1)$) -- ($0.5*(19.south)+0.5*(20.south) + (0.0,-0.7)$) ;

    \end{tikzpicture}
    \subcaption{$\T'_4$, obtained by undoing the reduction of $\mu_2$.}
    \end{subfigure}

    \begin{subfigure}[t]{0.4\textwidth}
        \centering
       \usetikzlibrary{positioning}
        \begin{tikzpicture}
        \draw[behindrect] (-2.85,5.2) rectangle (2.45, -2.7);

    \node[main] (8) {};
    \node[main] (9) [below = of 8]{};
    \node[main] (10) [left = of 9]{};
    \node[main] (11) [left = of 10]{};
    \draw[virtualtree] (8) -- (9);
    \draw[tree] (10) -- (11);
    \draw[marked] (9) to [bend left] (11);
    \draw[virtualcotree] (11) -- (8);
    \draw[virtualcotree] (8)--(10);

    \node[main] (12) [right = of 8]{};
    \node[main] (13) [below = of 12]{};
    \draw[virtualcotree] (12) to [bend right] (13);
    \draw[tree] (12) -- (13);
    \draw[cotree] (12) to [bend left] (13);

    \node[main] (14) [below = of 9] {};
    \node[main] (15) [below = of 10] {};
    \draw[virtualtree] (14) to [bend right] (15);
    \draw[marked] (14) -- (15);
    \draw[marked] (14) to [bend left] (15);

    \node[main] (16) [above = of 8] {};
    \node[main] (17) [left = of 16] {};
    \node[main] (18) [above = of 17] {};
    \draw[virtualtree] (16) -- (17);
    \draw[tree] (17) -- (18);
    \draw[virtualcotree] (16) -- (18);

    \node[main] (19) [above = of 18]{};
    \node[main] (20) [right = of 19]{};
    \node[main] (21) [above = of 19]{};
    \node[main] (22) [above right = 0.707cm of 19]{};
    \draw[virtualtree] (19) -- (20);
    \draw[virtualcotree] (19) -- (21);
    \draw[marked] (19) -- (22);
    \draw[tree] (20) -- (22);
    \draw[tree] (21) -- (22);
    \draw[cotree] (21) to [bend left] (20);

    \node[main] (23) [left = of 19]{};
    \node[main] (24) [above = of 23]{};
    \draw[virtualtree] (23) to [bend right] (24);
    \draw[marked] (23) to (24);
    \draw[marked] (23) to [bend left] (24);

    \node[black] (Box3) at ($(8.north) + 0.5*(0.0,0.3)$) {$\mu_6$ ~\bf{R}};

    \draw[black] ($(8.north)+(-2.5,0.35)$) rectangle ($(9.south)+(0.5,-0.35)$);

    \node[black] (Box5a) at ($(12.north) + (0.6,0.15)$) {$\mu_{5a}$\bf{ P}};

    \draw[black] ($(12.north)+(-0.4,0.35)$) rectangle ($(13.south)+(1.15,-0.35)$);

    \draw[black] ($0.5*(8.south)+0.5*(9.north) + (0.5,0.0)$) -- ($0.5*(12.north)+0.5*(13.south)+ (-0.4,0.0)$) ;

    \node[black] (Box5b) at ($(14.north) + (0.0,0.3)$) {$\mu_{5b}$\bf{ P}};

    \draw[black] ($(15.north)+(-0.4,0.5)$) rectangle ($(14.north)+(0.5,-0.35)$);

    \draw[black] ($0.5*(9.south)+0.5*(10.south) + (0.0,-0.35)$) -- ($0.5*(14.north)+0.5*(15.north)+ (0.0,0.5)$) ;

     \node[black] (Box3a) at ($(18.north) + (0.7,0.0)$) {$\mu_{3}$\bf{ S}};

    \draw[black] ($(18.north)+(-0.4,0.3)$) rectangle ($(16.north)+(0.5,-0.3)$);

    \draw[black] ($0.5*(16.south)+0.5*(17.south) + (0.0,-0.1)$) -- ($0.5*(16.south)+0.5*(17.south) + (0.0,-0.65)$) ;

         \node[black] (Box2) at ($(21.north) + (0.7,0.0)$) {$\mu_{2}$\bf{ R}};

    \draw[black] ($(21.north)+(-0.4,0.3)$) rectangle ($(20.north)+(0.5,-0.3)$);

    \draw[black] ($0.5*(24.south)+0.5*(23.north) + (0.5,0.0)$) -- ($0.5*(23.south)+0.5*(24.north) + (0.8,0.0)$) ;

         \node[black] (Box1) at ($(24.north) + (0.,0.1)$) {$\mu_{1}$\bf{ P}};

    \draw[black] ($(24.north)+(-0.4,0.3)$) rectangle ($(23.north)+(0.5,-0.3)$);

    \draw[black] ($0.5*(19.south)+0.5*(20.south) + (0.0,-0.1)$) -- ($0.5*(19.south)+0.5*(20.south) + (0.0,-0.7)$) ;

    \end{tikzpicture}
    \subcaption{$\T'_5$, obtained by undoing the reduction of $\mu_1$.}    \end{subfigure}\hfill
    \begin{subfigure}[t]{0.4\textwidth}
        \centering
       \usetikzlibrary{positioning}
        \begin{tikzpicture}
        \draw[behindrect] (-2.85,5.2) rectangle (2.45, -2.7);
    \node[main] (8) {};
    \node[main] (9) [below = of 8]{};
    \node[main] (10) [left = of 9]{};
    \node[main] (11) [left = of 10]{};
    \draw[virtualtree] (8) -- (9);
    \draw[tree] (10) -- (11);
    \draw[marked] (9) to [bend left] (11);
    \draw[virtualcotree] (11) -- (8);
    \draw[virtualcotree] (8)--(10);

    \node[main] (12) [right = of 8]{};
    \node[main] (13) [below = of 12]{};
    \draw[virtualcotree] (12) to [bend right] (13);
    \draw[tree] (12) -- (13);
    \draw[cotree] (12) to [bend left] (13);

    \node[main] (14) [below = of 9] {};
    \node[main] (15) [below = of 10] {};
    \draw[virtualtree] (14) to [bend right] (15);
    \draw[marked] (14) -- (15);
    \draw[marked] (14) to [bend left] (15);

    \node[main] (16) [above = of 8] {};
    \node[main] (17) [left = of 16] {};
    \node[main] (18) [above = of 17] {};
    \draw[virtualtree] (16) -- (17);
    \draw[virtualtree] (17) -- (18);
    \draw[virtualcotree] (16) -- (18);

    \node[main] (19) [above = of 18]{};
    \node[main] (20) [right = of 19]{};
    \node[main] (21) [above = of 19]{};
    \node[main] (22) [above right = 0.707cm of 19]{};
    \draw[virtualtree] (19) -- (20);
    \draw[virtualcotree] (19) -- (21);
    \draw[marked] (19) -- (22);
    \draw[tree] (20) -- (22);
    \draw[tree] (21) -- (22);
    \draw[virtualcotree] (21) to [bend left] (20);

    \node[main] (23) [left = of 19]{};
    \node[main] (24) [above = of 23]{};
    \draw[virtualtree] (23) to [bend right] (24);
    \draw[marked] (23) to (24);
    \draw[marked] (23) to [bend left] (24);

    \node[main] (25) [left = of 17] {};
    \node[main] (26) [above = of 25]{};
    \draw[virtualcotree] (25) to [bend right] (26);
    \draw[tree] (25) to (26);
    \draw[cotree] (25) to [bend left] (26);

    \node[main] (27) [right = of 20] {};
    \node[main] (28) [above = of 27] {};
    \node[main] (29) [above right = 0.707cm of 27]{};
    \draw[virtualtree] (27) -- (28);
    \draw[cotree] (28) -- (29);
    \draw[tree] (27) -- (29);

    \node[black] (Box3) at ($(8.north) + 0.5*(0.0,0.3)$) {$\mu_6$ ~\bf{R}};

    \draw[black] ($(8.north)+(-2.5,0.35)$) rectangle ($(9.south)+(0.5,-0.35)$);

    \node[black] (Box5a) at ($(12.north) + (0.6,0.15)$) {$\mu_{5a}$\bf{ P}};

    \draw[black] ($(12.north)+(-0.4,0.35)$) rectangle ($(13.south)+(1.15,-0.35)$);

    \draw[black] ($0.5*(8.south)+0.5*(9.north) + (0.5,0.0)$) -- ($0.5*(12.north)+0.5*(13.south)+ (-0.4,0.0)$) ;

    \node[black] (Box5b) at ($(14.north) + (0.0,0.3)$) {$\mu_{5b}$\bf{ P}};

    \draw[black] ($(15.north)+(-0.4,0.5)$) rectangle ($(14.north)+(0.5,-0.35)$);

    \draw[black] ($0.5*(9.south)+0.5*(10.south) + (0.0,-0.35)$) -- ($0.5*(14.north)+0.5*(15.north)+ (0.0,0.5)$) ;

     \node[black] (Box3a) at ($(18.north) + (0.7,0.0)$) {$\mu_{3}$\bf{ S}};

    \draw[black] ($(18.north)+(-0.4,0.3)$) rectangle ($(16.north)+(0.5,-0.3)$);

    \draw[black] ($0.5*(16.south)+0.5*(17.south) + (0.0,-0.1)$) -- ($0.5*(16.south)+0.5*(17.south) + (0.0,-0.65)$) ;

         \node[black] (Box2) at ($(21.north) + (0.7,0.0)$) {$\mu_{2}$\bf{ R}};

    \draw[black] ($(21.north)+(-0.4,0.3)$) rectangle ($(20.north)+(0.5,-0.3)$);

    \draw[black] ($0.5*(24.south)+0.5*(23.north) + (0.5,0.0)$) -- ($0.5*(23.south)+0.5*(24.north) + (0.79,0.0)$) ;

         \node[black] (Box1) at ($(24.north) + (0.,0.1)$) {$\mu_{1}$\bf{ P}};

    \draw[black] ($(24.north)+(-0.4,0.3)$) rectangle ($(23.north)+(0.5,-0.3)$);

    \draw[black] ($0.5*(19.south)+0.5*(20.south) + (0.0,-0.1)$) -- ($0.5*(19.south)+0.5*(20.south) + (0.0,-0.7)$) ;

    \node[black] (Box4) at ($(26.north) + (0,0.1)$) {$\mu_{4}$\bf{ P}};

    \draw[black] ($(26.north)+(-0.4,0.3)$) rectangle ($(25.north)+(0.5,-0.3)$);

    \draw[black] ($0.5*(26.south)+0.5*(25.north) + (0.5,0.0)$) -- ($0.5*(26.south)+0.5*(25.north) + (0.79,0.0)$) ;

        \node[black] (Box7) at ($(28.north) + (0,0.1)$) {$\mu_{7}$\bf{ S}};

    \draw[black] ($(28.north)+(-0.4,0.3)$) rectangle ($(27.north)+(0.8,-0.3)$);

    \draw[black] ($0.5*(28.south)+0.5*(27.north) + (-0.4,0.0)$) -- ($0.5*(28.south)+0.5*(27.north) + (-0.69,0.0)$) ;

    \end{tikzpicture}

    \subcaption{$\widehat{\T}$, obtained by undoing the reductions\\
    of $\mu_4$ and $\mu_7$}
    \end{subfigure}
    \caption{%
      Reversing the reductions from \Cref{fig:reducetreeexample} on the $Y$-processed SPQR-tree from  \Cref{fig_mergetree_example_final}.
      Virtual edges in the SPQR-tree are given by dashed edges, tree edges are marked in red and bold and all other edges are marked in blue.
      Edges in $Y$ are marked by two stripes.
    }
    \label{fig_reduction_reversal}
\end{figure}

We will sketch how $\widehat{\T}$ can be obtained by reversing the reductions, undoing the operations from \algoReduceTree.
An example of the reversal of reductions is provided in \Cref{fig_reduction_reversal}.
First, we reverse the local 2-separations that are removed by $\algoReduceSeries$ (\Cref{algo_reduceseries}) and $\algoReduceParallel$ (\Cref{algo_reduceParallel}).
In $\widehat{\T}$, these reductions are reversed by replacing the single edges by virtual edges pointing to a skeleton of type~\eqref{node_S} and type~\eqref{node_P} for each reduction performed in \algoReduceSeries and \algoReduceParallel, respectively.

Secondly, we undo the reductions that removed leaf nodes from $\T$ by simply making the relevant edge $e$ virtual again and adding back the removed node and its skeleton.
Since $\T'$ is an SPQR-tree, it is clear that $\widehat{\T}$ is one, too.
However, although we have shown that $\T'$ is minimal, $\widehat{\T}$ is not necessarily minimal, as reversing the reductions may create new \eqref{node_S}--\eqref{node_S} or \eqref{node_P}--\eqref{node_P} connections.
If this occurs, the adjacent nodes can be merged into a single node of type~\eqref{node_S} or~\eqref{node_P}, instead.
In \cref{thm_few_merges_recovery}, we argue that we need at most one such merge.

\begin{lemma}
  \label{thm_few_merges_recovery}
  Let $\T$ be a minimal SPQR-tree with a set of $Y$-reduced edges that represents a graph $G$ that contains a $Y$-splittable vertex $v$.
  Let $(\T_R,Y_R) \coloneqq \algoReduceTree(\T,Y)$.
  If $\V(\T) = \{\mu\}$, let $(\T',X) \coloneqq \algoSplitSkeleton(\T_R,\mu,Y_R)$ be the output of \Cref{algo_splitskeleton}, otherwise let $(\T',X) \coloneqq \algoMergeTree(\T_R,Y_R)$ be the output of \Cref{algo_mergetree}.
  Let $\widehat{\T}$ be the SPQR-tree obtained by reversing the reductions from $\algoReduceTree$ in $\T'$.
  Then $\widehat{\T}$ has at most one pair of adjacent \eqref{node_S}--\eqref{node_S} or \eqref{node_P}--\eqref{node_P} nodes.
\end{lemma}

\begin{proof}
First, note that
the minimality of $\T$ implies minimality of $\T_R$ by \cref{thm_reductions_overview}, and that minimality of $\T_R$ implies minimality of $\T'$ by \cref{thm_mergetree_yprocessed,thm_splitskeleton_yprocessed}.

First, we consider the case where $|\V(\T_R)| > 1$. Then, $\T'$ consists of a single node $\mu$ of type~\eqref{node_R}.
Assume that $\widehat{\T}$ is not minimal, such that it contains an edge $\{\nu,\omega\}$ where $\nu$ and $\omega$ are either both of type~\eqref{node_S} or~\eqref{node_P}. Since neither $\nu$ nor $\omega$ are equal to $\mu$ (which is of type~\eqref{node_R}), both nodes $\nu$ and $\omega$ as well as their common edge were already part of $\T$, which contradicts minimality of $\T$.
Thus, $\widehat{\T}$ must be minimal if $|\V(\T_R)| >1$ holds.

Second, we consider the case where $\V(\T_R) = \{\mu\}$. We consider different cases based on the type of $\mu$. Note that $\mu$ cannot be of type~\eqref{node_P} by \cref{thm_nodelabelsvalid_reduction}.

First, consider the case where $\mu$ has type~\eqref{node_Q} in $\T_R$. By \cref{thm_singlecycle}, $\mu$ becomes a node of type~\eqref{node_S} in $\T'$.
If $\mu$ is of type~\eqref{node_Q} in $\T_R$, then $\mu$ must be of type~\eqref{node_Q} or~\eqref{node_P} in $\T$. If $\mu$ was of type~\eqref{node_Q} in $\T$, there are no 2-separations that can be reversed, so then $\widehat{\T}$ is clearly minimal, too, since it consists of a single~\eqref{node_S} node.

If $\mu$ was of type~\eqref{node_P} in $\T$, at least one of two types of local 2-separations in $\algoReduceParallel$ (\Cref{algo_reduceParallel}) must have been performed,  since we know that $\spqrVirtual_\mu(\T_R) = \emptyset$ and $|E^{\T}_\mu|\geq 3$ hold but $|E^{\T_R}_\mu| = 2$.
Such a local 2-separation splits off a set of edges $E'$ from $\mu$ into a new~\eqref{node_P} node $\nu$.
Since any virtual edge in $E'$ also exists in $\T$ and since $\T$ was minimal, no virtual edge in $E'$ does not connect to nodes of type~\eqref{node_P} in $\widehat{\T}$.
Because $\mu$ is converted to a node of type~\eqref{node_S} in $\widehat{T}$, we create a new ~\eqref{node_S}--\eqref{node_P} connection this way between $\mu$ and $\nu$, and $\widehat{\T}$ can thus not have any \eqref{node_P}--\eqref{node_P} connections by reversing the reductions. If both local 2-separations in lines \ref{algo_reduceparallel_reduction_1} and \ref{algo_reduceparallel_reduction_2} of \algoReduceParallel were performed, then the minimality of $\widehat{\T}$ follows from the minimality of $\T$. Otherwise, exactly one reduction was performed, and there can be one edge $e'\in E^{\widehat{\T}}_\mu$ that could be a virtual edge to a node of type~\eqref{node_S} after reversing the reductions. This implies that $\widehat{\T}$ contains at most one \eqref{node_S}--\eqref{node_S} connection.

Secondly, we consider the case where $\mu$ has type~\eqref{node_S} in $\T_R$. Then, \cref{thm_singlecycle} shows that $\mu$ stays a node of type~\eqref{node_S} in $\T'$ and $\widehat{\T}$. Then, since \algoReduceSeries (\Cref{algo_reduceseries}) does not perform local 2-separations when $\T_R$ is a single node of type~\eqref{node_S}, the node sets $\V(\T)$ and $\V(\widehat{\T})$ are the same and have identical types. Thus, $\widehat{\T}$ is minimal because $\T$ is minimal.

Lastly, consider the case where $\mu$ has type~\eqref{node_R} in $\T_R$. If $|A|=1$ in $\algoSplitSkeleton$, $\T'$ consists of a single node of type~\eqref{node_R} by \cref{thm_triconnected_simple}, and we must again have that $\V(\T) = \V(\widehat{\T})$ holds with identical types, since no local reductions were performed because $\mu$ is of type~\eqref{node_R}. This implies that $\widehat{\T}$ is minimal.
In the other case, we have $|A|=2$, where an edge $e$ connects two $Y$-splittable vertices in $\T_R$. Then, $\T'$ consists of $\mu$ and a new node $\nu$ of type~\eqref{node_S}, where $e$ is put in series with the new row edge.

Since we create a new~\eqref{node_S}-node, we may create a new~\eqref{node_S}--\eqref{node_S} connection in $\widehat{\T}$. In particular, this case occurs if $e$ is a virtual edge pointing to a node of type~\eqref{node_S} in $\T$. Since $e$ is the only edge in $\nu$ that can be a virtual edge that is not pointing to $\mu$, at most one~\eqref{node_S}--\eqref{node_S} connection can be created in this way.
\end{proof}

In $\algoProcessTree$, we present the complete algorithm for updating a single SPQR-tree.
In the last step of $\algoProcessTree$, we reverse the reductions performed in $\algoReduceTree$ and ensure that $\widehat{\T}$ remains minimal, by merging two adjacent \eqref{node_S} or \eqref{node_P} nodes if necessary.

\bigskip
\begin{algorithm}[H]
  \label{algo_processtree}
  \footnotesize
  \caption{Processing a single SPQR-tree $\T$}
    \LinesNumbered
    \TitleOfAlgo{ProcessTree$(\T, Y)$}
    \KwIn{Minimal SPQR-tree $\T$, edges $Y\neq\emptyset$}
    \KwOut{$Y$-processed SPQR-tree $\widehat{\T}$, vertex set $\widehat{X}$ such that $\widehat{X} = \{v_1,v_2\}$ or $\widehat{X} = \emptyset$}
    Let $(\T_R,Y_R) \coloneqq \algoReduceTree(\T,Y)$.\;
    \lIf{$\V(\T_R) = \{\mu\}$}{%
    let $(\T', X) \coloneqq \algoSplitSkeleton(\T_R,\mu,Y_R)$.
    }\lElse{%
    let $(\T', X) \coloneqq \algoMergeTree(\T_R,Y_R)$.
    }
    \lIf{$X=\emptyset$}{%
     \Return $(\T,\emptyset)$
    }
    Let $(\widehat{\T}, \widehat{X})$ be obtained by reversing the reductions that reduced $\T$ to $\T_R$.\;
    \Return $(\widehat{\T}, \widehat{X}) $
\end{algorithm}
\bigskip

\begin{corollary}
\label{thm_processed_notin_P}
  Let $\T'$ be a $Y$-processed tree with respect to vertices $v_1$ and $v_2$ obtained by \algoProcessTree (\Cref{algo_processtree}).
  Then $v_1$ and $v_2$ never lie in a node of type~\eqref{node_P}.
\end{corollary}

\begin{proof}
  First, we note that $\T'$ was obtained using some $Y$-reduced tree $\T_R$.
  By \cref{thm_reductions_overview}, we cannot have $\V(\T_R) = \{\mu\}$ where $\mu$ is of type~\eqref{node_P}.
  If $|\V(\T_R)| > 1$, $\algoMergeTree$ (\Cref{algo_mergetree}) ensures that $v_1$ and $v_2$ must be in a skeleton of type~\eqref{node_R}.
  Otherwise, $\algoSplitSkeleton$ (\Cref{algo_splitskeleton}) only places $v_1$ and $v_2$ in nodes of type~\eqref{node_S} or~\eqref{node_R}.
\end{proof}

In \cref{sec_separations_connectivity} we argued that it is sufficient to consider the connected components of $M$ separately, as we could easily connect them in a realization by identifying the new row edges of each component with one another.
Since we have an SPQR-tree for every connected component of $M$, we actually maintain an SPQR-forest consisting of the individual trees for each component.
Interpreting \cref{thm_combine_blocks} in terms of SPQR-trees, we observe that combining multiple connected components in $M$ can be thought of as creating a new node $\mu$ of type~\eqref{node_P}, that connects $\mu$ with a virtual edges in between the vertices of $X=\{v_1,v_2\}$ for each $Y$-processed SPQR-tree.
Additionally, it may be the case that we encounter a nonzero in a column for the first time.
Clearly, the resulting edge of such a column must always be placed in parallel with the row edge.
These cases are handled by our main algorithm, \algoGraphicRowAugmentation (\Cref{algo_graphicrowaugmentation}).

\bigskip

\begin{algorithm}[H]
  \label{algo_graphicrowaugmentation}
  \footnotesize
  \caption{Processing a single row}
  \LinesNumbered
  \TitleOfAlgo{GraphicRowAugmentation$(M,b,\mathcal{F})$}
  \KwIn{Graphic Matrix $M$, new row $b$, SPQR-forest $\mathcal{F}$}
  \KwOut{SPQR-forest $\mathcal{F}'$ representing $M' = \begin{bmatrix} M \\
  b^T\end{bmatrix}$ or FALSE if $M'$ is not graphic}
  Let $Y \coloneqq \supp(b)\cap \spqrNonvirtual(\mathcal{F})$ and let $Y'\coloneqq\supp(b)\setminus \spqrNonvirtual(\mathcal{F})$.\;
  Let $Y_{\T}\coloneqq Y \cap \spqrNonvirtual(\T)$ for each SPQR-tree $\T\in\mathcal{F}$.\;
  Let $\mathcal{F}_Y\coloneqq\{\T\in\mathcal{F} \mid Y_{\T} \neq \emptyset\}$.\;

  \uIf{$\mathcal{F}_Y = \{\T\}$}{%
    Let $(\T',X) \coloneqq \algoProcessTree(\T,Y_{\T})$.\;
    \lIf{$X=\emptyset$}{%
      \Return FALSE
    }
    \uIf{$Y'=\emptyset$}{%
      Add $b$ between $v_1$ and $v_2$.\;
    }\Else{%
      Create a new skeleton $\mu$ of type~\eqref{node_P} with edges $Y'\cup\{b\}$ in $\T'$. \;
      Connect $\mu$ and $\{v_1,v_2\}$ in $\T'$ using a virtual edge pair.\;
    }
    $\mathcal{F}' \coloneqq (\mathcal{F} \cup\{\T'\})\setminus\{\T\}$\;
  }
  \uElseIf{$|\mathcal{F}_Y| > 1$}{
    Let $\T_{\mathrm{new}}$ be an SPQR-tree with a single member $\mu$ of type~\eqref{node_P} with edges $Y'\cup\{b\}$.\;
    \For{$\T\in\mathcal{F}_Y$}{
        Let $(\T',X) \coloneqq \algoProcessTree(\T,Y_{\T})$.\;
        \lIf{$X=\emptyset$}{\Return FALSE}
        Update $\T_{\mathrm{new}}$ by connecting $\T_{\mathrm{new}}$ with $\T'$, by connecting $\mu$ and $v_1$ and $v_2$ using a virtual edge pair.
    }
    Let $\mathcal{F}'\coloneqq ( \mathcal{F}\cup\{\T_{\mathrm{new}}\})\setminus\mathcal{F}_Y$.\;
  }
  \Else{%
      Create an SPQR-tree $\T$ with a single skeleton $\mu$ of type~\eqref{node_Q} (if $|Y'| = 1$) or~\eqref{node_P} with edges $Y'\cup \{b\}$.\;
      $\mathcal{F}'\coloneqq \mathcal{F} \cup \{T\}$.
  }
  \Return $\mathcal{F}'$
\end{algorithm}
\medskip

\begin{corollary}
  Let $\mathcal{F}$ be a minimal SPQR-forest.
  Then every SPQR-tree in $\mathcal{F}'$ returned by \linebreak \algoGraphicRowAugmentation (\Cref{algo_graphicrowaugmentation}) is minimal.
\end{corollary}

\begin{proof}
  In the case where $\mathcal{F}_Y\neq\emptyset$ we always obtain a new SPQR-tree $\T_{\mathrm{new}}$ from a $Y$-processed SPQR-tree $\T'$ by adding an edge between the two vertices $v_1$ and $v_2$ in $\T'$ where the new edge may be added.
  In particular, we either connect $v_1$ and $v_2$ with row edge or we connect $v_1$ and $v_2$ using a virtual edge pointing to a new node $\mu$ of type~\eqref{node_P}.
  By \cref{thm_processed_notin_P}, $v_1$ and $v_2$ cannot lie in a node of~\eqref{node_P}, and we can thus not create a new~\eqref{node_P}--\eqref{node_P} connection in this manner. Then, since $\T'$ is $Y$-processed, the SPQR-tree $\T_{\mathrm{new}}$ obtained  by these operations is minimal. In the case where $|\mathcal{F}_Y| > 1$, we identify these minimal SPQR-trees in $\mu$, which clearly yields another minimal SPQR-tree.
  In the case where $\mathcal{F}_Y=\emptyset$, we simply add a new SPQR-tree to $\mathcal{F}'$ with a single node of type~\eqref{node_Q} or~\eqref{node_P}, which is clearly minimal.
\end{proof}

Although our algorithmic description both detects whether the matrix augmented with the new row is graphic and updates the SPQR-tree to reflect the new row at the same time, we note that these steps are easy to separate within an implementation.
Similarly, although the SPQR-tree is frequently copied throughout our pseudocode, it is easy to avoid these copies in an implementation. In particular, both  the reductions from \algoReduceTree (\Cref{algo_reducetree}) and the update steps in \algoSplitSkeleton (\Cref{algo_splitskeleton}) and \algoMergeTree (\Cref{algo_mergetree}) can be performed on a single SPQR-tree by editing a sub SPQR-tree in-place, and the reversals of the reductions are then performed implicitly.
We have intentionally omitted these details from the algorithmic description to keep it simpler.

\begin{lemma}
  \label{thm_processtree_complexity}
  For a graphic matrix $M\in\{0,1\}^{m\times n}$, an SPQR-tree $\T$ of the graph $G=(V,E)$ and the column edges $Y\subseteq E$, $\algoProcessTree(\T,Y)$ (\Cref{algo_processtree}) runs in $\orderO(\alpha(|E|, m+n) \cdot |E|)$ time.
\end{lemma}

\begin{proof}
  First, \algoReduceTree (\Cref{algo_reducetree}) runs in $\orderO(\alpha(|E|,m+n) \cdot |E|)$ time by \cref{thm_reducetree_complexity} plus the time for running \algoFindSplittableVertices for each skeleton $G_\mu$ of type~\eqref{node_R}.
  By \cref{thm_splittable_time_complexity} this can be done in time $\orderO(\alpha(|E_\mu|, m+n) \cdot |E_\mu|)$. Summing over all skeletons yields a worst-case runtime of $\orderO(\alpha(|E|, m+n) \cdot |E|)$.
  We obtain an SPQR-tree $\T_R$ that represents a graph $G_R=(V_R,E_R)$ where $|V_R|\leq |V|$ and $|E_R|\leq |E|$.
  In the case $|\V(\T_R) | = 1$ holds, $\algoSplitSkeleton$ (\Cref{algo_splitskeleton}) runs in $\orderO(\alpha(|E_\mu|, m+n) \cdot (|E_\mu|))$ time by \cref{thm_splitskeleton_runtime}, which is dominated by $\orderO(\alpha(|E_R|, m+n) \cdot |E_R|)$.
  Otherwise, $\algoMergeTree$ (\Cref{algo_mergetree}) runs in $\orderO(\alpha(|E_R|, m+n) \cdot |E_R|)$ time by \cref{thm_mergetree_complexity}.
  Then finally, we can reverse the reductions from $\algoReduceTree$ in the same time as it took to perform them. To restore minimality, \cref{thm_few_merges_recovery} shows we need at most one merge of adjacent nodes in $\T'$ which takes amortized $\orderO(\alpha(1,m+n))$ time.
  Clearly, all steps are dominated by the given time complexity $\orderO(\alpha(|E|, m+n) \cdot |E|)$.
\end{proof}

Considering the time bounds, we can roughly distinguish three cases in which the worst case time complexity is attained.
The first occurs when $\T$ consists of a single large node (of type~\eqref{node_R}), where \algoFindSplittableVertices (\Cref{algo_findsplittablevertices}) is responsible for the time bound.
The second case occurs when the $Y$-reduced tree $\T$ has a large number of nodes that need to be merged into one single large node, where the time bound is given by the identification of these nodes in $\algoMergeTree$ (\Cref{algo_mergetree}).
Finally, if the initial SPQR-tree is large and contains many reductions, $\algoReduceTree$ (\Cref{algo_reducetree}) attains the worst case time complexity.

\begin{theorem}
  For a matrix $M\in\{0,1\}^{m\times n}$, \algoGraphicRowAugmentation (\Cref{algo_graphicrowaugmentation}) runs in $\orderO((m+n) \cdot \alpha(m+n, m+n))$ time and $\orderO(m+n)$ space.
\end{theorem}
\begin{proof}
First, we distribute $\supp{b}$ over $Y$ and $Y'$, by checking if the associated column edge exists, which takes $\orderO(|\supp{b}|) = \orderO(n)$ time.
Then, for each $y\in Y$, we find the corresponding SPQR-tree by first finding the corresponding SPQR-node and then finding the corresponding SPQR-tree.
Using the amortized union find operations, this step takes $\orderO(n  \cdot \alpha(n, m+n))$ time.

Now, consider various cases based on the size of $\mathcal{F}_Y$. If $\mathcal{F}_Y = \emptyset$, then we simply create a new SPQR-tree with $|Y'|+1$ edges. Clearly, this can be done in $\orderO(n)$ time.

If $\mathcal{F}_Y = \{\T\}$, then we call \algoProcessTree (\Cref{algo_processtree}).
Since $\T$ is a member of the SPQR-forest that represents $M$, any realization of $\T$ has at most $m+n$ edges. Then, $\algoProcessTree$ runs in $\orderO((m+n) \cdot \alpha(m+n, m+n))$ time by \cref{thm_processtree_complexity}.
For the other steps in this branch, adding $b$ and the edges $Y'$, and initializing the data structures for the new SPQR-tree takes at most $\orderO(n)$ time.

If $|\mathcal{F}_Y| > 1$, then we first create a new skeleton with $|Y'|+1$ edges, which can be done in $\orderO(n)$ time. For all $\T\in \mathcal{F}_Y$, let $G_{\T}$ be any graph represented by $\T$.
Because $\mathcal{F}_Y$ is a subset of the SPQR-forest representing $M$, we obtain $\sum_{\T\in \mathcal{F}_Y} |E(G_{\T})|\leq m+n$. Thus, if we run $\algoProcessTree$ for each $\T\in\mathcal{F}_Y$, the sum of their time complexities is of order $\orderO((m+n) \cdot \alpha(m+n, m+n))$.
In the loop, we additionally connect $\T_{\mathrm{new}}$ with $\T$. We do this at most $n$ times, which gives us a time complexity of $\orderO(n \cdot \alpha(n,m+n))$, which is clearly dominated by the time complexity of running $\algoProcessTree$ for each tree.

We obtain that the first 3 lines and every branch of $\algoGraphicRowAugmentation$ (\Cref{algo_graphicrowaugmentation}) can be done in $\orderO((m+n) \cdot \alpha(m+n,m+n))$ time, which concludes the proof
\end{proof}

\begin{corollary}
  For a matrix $M\in\{0,1\}^{m\times n}$, we can determine its graphicness using repeated calls of $\algoGraphicRowAugmentation$ (\Cref{algo_graphicrowaugmentation}) in $\orderO((m^2 + mn)\cdot \alpha(m+n, m+n))$ time.
  \label{thm_total_time_complexity}
\end{corollary}

\begin{proof}
  We perform $m$ calls of $\algoGraphicRowAugmentation$, which gives us a total time  complexity of $\orderO((m^2 + mn)\cdot \alpha(m+n, m+n))$.
\end{proof}

Note that the time complexity we obtain in \cref{thm_total_time_complexity} is strictly worse than the $\orderO(k \alpha(k,m))$ running time that is achieved by Bixby and Wagner, where $k$ is the number of nonzeros of $M$.

\section{Discussion}
\label{sec_discussion}

In this paper, we formulated an algorithm for solving the graphic row augmentation problem in $\orderO(\alpha(m+n,m+n) \cdot (m+n))$ time.
By adding all rows sequentially we obtain a $\orderO((m^2+mn) \cdot \alpha(m+n, m+n))$ algorithm to detect whether a matrix $M$ is graphic.
Note that this is strictly worse than the $\orderO(k  \cdot \alpha(k,m))$ running time achieved by Bixby and Wagner.
However, we do suspect that the $\orderO((m+n) \cdot \alpha(m+n, m+n))$ running time can be improved.
In particular, we suspect one can use (modifications of) the dynamic connectivity data structures from~\cite{Kapron2013} and the dynamic LCA data structure from \cite{Sleator1981} to speed up \algoFindSplittableVertices, which attains the worst case bound if we have one big component of type~\eqref{node_R}.
Additionally, for the purpose of detecting graphicness of the entire matrix, one can use \cref{thm_SPQRTreeNodeAndSkeletonBound} to show that at most $\orderO(m+n)$ pairs of skeletons in $\T$ are merged over multiple graphic row additions.
This hints that the other worst case where the $Y$-reduced SPQR-tree $\T$ has many nodes is somewhat rare when one sequentially adds all rows of a matrix. We hypothesize that using dynamic data structures, an implementation of the described algorithms that is polynomial in $k$ is possible.
However, using dynamic data structures comes at the cost of both space complexity and a more complex algorithmic implementation, which are the reasons why we did not pursue this direction of research further in this work.

Although our method does not reach the best known running time for finding graphic matrices, it is, to the best of our knowledge, the first complete algorithm for the graphic row augmentation problem.
The proposed algorithm can be combined with that of Bixby and Wagner~\cite{BixbyWagner1988} in order to find inclusion-wise maximal graphic submatrices.
This can be advantageous for applications in mixed-integer linear programming, where knowledge of (transposed) network submatrices in the problem may be useful to derive stronger cutting planes, primal heuristics or integrality properties.
An interesting new research direction would be to investigate the presence of (transposed) network submatrices in mixed-integer linear programming problems, and to examine their properties.
Our prototype implementations of the row-wise and column-wise algorithms for graphic matrices are publicly available in~\cite{matrecgithub}.
As this work primarily investigates the validity of the row-wise algorithm, we consider computational results to be outside its scope. In future work, we plan to experimentally compare both methods and investigate the presence of graphic/network submatrices in mixed-integer programming problems.

Another promising future research direction is to generalize the current results from graphic matrices and undirected graphs to  network matrices and directed graphs.
In particular, we suspect that there exists an SPQR-tree type data structure that uniquely represents directed graphs with the same set of directed cycles.
Moreover, we hypothesize that both Bixby and Wagner's column-wise algorithm and our proposed row-wise algorithm can be modified to work in this generalized setting.

Furthermore, because graphic and network matrices are closely related to regular matroids and totally unimodular matrices, this work, together with that of Bixby and Wagner in \cite{BixbyWagner1988}, can be a starting point for the development of matrix augmentation algorithms maintaining regularity or total unimodularity.

\paragraph{Acknowledgments.}
We thank two anonymous reviewers for their valuable suggestions and constructive feedback that led to an improved manuscript.
Both authors acknowledge funding support from the Dutch Research Council (NWO) on grant number OCENW.M20.151.

\paragraph{Competing interests.}
After submission the first author became affiliated with Mosek ApS, which produces proprietary optimization software. The second author has no competing interests to declare.

\bibliographystyle{plainurl}
\bibliography{network-row-augmentation}

\end{document}